\documentclass[11pt,a4paper]{article}
\usepackage{graphicx}
\usepackage{amssymb}
\usepackage{hyperref}
\pdfoutput=1
\def\bra#1{\left\langle#1\right|}
\def\ket#1{\left|#1\right\rangle}
\usepackage[active]{srcltx}

\begin{document}
\renewcommand{\thefootnote}{\fnsymbol{footnote}}

\begin{center}
{\LARGE\bf\sf Elliptic recursion for 4-point superconformal blocks and bootstrap in N=1 SLFT} \\
\end{center}

\begin{center}

\vspace*{1cm}

    {\large\bf\sf
        Paulina Suchanek \footnote{\emph{e-mail}: paulina@ift.uni.wroc.pl}
    }
     \\

\vskip 3mm
     Institute for Theoretical Physics,
    University of Wroc{\l}aw \\
    pl. M. Borna 9, 50-204~Wroc{\l}aw, Poland. \\
\end{center}

\vspace*{5mm}

\begin{abstract}
All types of 4-point spheric conformal blocks in both sectors of $N=1$ superconformal field theory
are introduced and analyzed. The elliptic recurrence formulae are derived for all the types
of blocks not previously discussed in the literature.
The results are used for numerical verification of the crossing symmetry of some 4-point functions
in the  $N=1$ superconformal Liouville field theory.
\end{abstract}
PACS: 11.25.Hf, 11.30.Pb


\renewcommand{\thefootnote}{\arabic{footnote}}

\section*{Introduction}

The Liouville field theory (LFT) is one of the most important examples of
 2 dimensional non-rational  conformal field theories. Its numerous applications range from the
 2-dim quantum gravity and matrix models \cite{sei, Ginsparg:1993is} through  D-brane dynamics
 in string theory \cite{Nakayama:2004vk}
to the recently discovered AGT relation \cite{Alday:2009aq}.
The exact analytical expression for the Liouville structure constants was first
proposed by Dorn and Otto \cite{DO} and independently
 by Zamolodchikovs \cite{Zamolodchikov:1995aa}. The proposal was motivated by
 an analytic continuation of the 3-point functions perturbatively calculated within the Coulomb gas approach.
 Another derivation of the DOZZ formula
based on functional relations for structure constants
 was presented by Teschner in \cite{Teschner:1995yf}.

 In principle any n-point function in the Liouville theory is given by the DOZZ structure constants  and
conformal blocks \cite{Belavin:1984vu}. Such representation however is not unique and
the consistency of the theory requires various decompositions of the same correlator to yield the same result.
In case of a CFT on closed Riemman surfaces  the consistency conditions for all the correlators are satisfied
if and only if
the 4-point spheric  functions
are crossing symmetric   and
 the 1-point toric functions are modular invariant  \cite{Sonoda:1988fq}.
The first numerical check of the crossing symmetry of the 4-point functions in LFT was done in
\cite{Zamolodchikov:1995aa}.
It was based on Al. Zamolodchikov's effective recursive relations for  4-point blocks worked out
 in a series of papers \cite{Zamolodchikov:ie,Zamolodchikov:2,Zamolodchikov:3}.
An analytical proof of the crossing symmetry was derived by Ponsot and Teschner \cite{PT1,PT2}
 and Teschner \cite{T.G, T.V}. It was recently shown \cite{Hadasz:2009sw}
 that the  modular invariance  follows from the relations between the toric 1-point and the spheric
4-point blocks  \cite{Poghossian:2009mk, Hadasz:2009db}.
 Let us note that the proof of these relations
 is based on the recursive representations of both types of blocks \cite{Hadasz:2009db}.

 The $N=1$ supersymmetric generalization of the Liouville theory is much less developed.
 Although the structure constants on closed surfaces  have been known for a long time 
 \cite{Poghosian:1996dw, Rashkov:1996jx}
 and certain couplings on bordered surfaces were successfully derived \cite{Fukuda:2002bv}
 the computation of all 4-point functions is still an open question.
  The main  reason is that the 4-point superconformal blocks
   are much more complex objects than the bosonic ones.

The first complication arises already in the Neveu-Schwarz (NS) sector
where the superconformal Ward identities determine the 3-point
functions up to 2 (rather than 1) structure constants \cite{Zamolodchikov:1988nm}.
This leads to eight types of NS 4-point blocks
\cite{Hadasz:2006sb}, \cite{Belavin:2007zz}.
The second difficulty comes from the Ward identities in the Ramond (R) sector which are
considerably more involved \cite{Friedan:1984rv}.
A method of finding a proper basis for 3-point blocks  and an appropriate
 representation of the R fields in terms of chiral vertex
operators was presented in \cite{R}. It was also shown that the correlation functions of 4 R fields decompose into
 eight types of corresponding 4-point blocks.

The recursive relations for  the NS 4-point  blocks were derived by suitable modification of Zamolodchikov's method
\cite{Hadasz:2006sb}, \cite{Belavin:2007zz}.
 The more efficient  elliptic recursion was proposed by Belavins, Neveu and Zamolodchikov
  \cite{Belavin:2007gz,Belavin:2007eq}  and  derived in full generality in \cite{Hadasz:2007nt}.
  With the help of this recursion the bootstrap equations for 4-point functions
  in the NS sector of $N=1$ SLFT were numerically verified \cite{Belavin:2007gz,Belavin:2007eq}.
An analytical check of the crossing symmetry in the NS sector based on the braiding and the fusion properties of
the NS blocks was worked out in \cite{Hadasz:2007wi, Chorazkiewicz:2008es}.
The recursive
representations of the eight 4-point blocks corresponding to correlation functions of 4 R fields were
derived in \cite{R}.

The aim of the present paper is to extend the analysis of \cite{R}
to all types of $N=1$ superconformal blocks.
This concerns in particular the conformal blocks corresponding to the correlation functions
of 2 NS and 2 R fields. We apply the techniques developed in \cite{R} in order to find a diagonal representation
of an NS superprimary field in terms of vertex operators acting in the R sector. Such a representation suggests
a convenient basis for the corresponding  3-point blocks. These new 3-point blocks together with those of
 \cite{Hadasz:2006sb}, \cite{R} constitute a complete set indispensable for
defining all types of 4-point superconformal blocks corresponding to the correlation functions of
the R primaries and the  NS superprimaries. These blocks include in particular the 4-point blocks with R
intermediate states,
which have not
been investigated so far. In all new cases the elliptic recursive relations are derived.

Using recursive representations for the 4-point blocks we numerically check
the crossing symmetry of correlators of 4 R fields and  2 R and 2 NS fields in $N=1$ SLFT.
The crossing symmetry of a correlator of 4 R fields can be seen as a verification of two
structure constants in the R sector \cite{Poghosian:1996dw}.
We preform two more checks involving all four SLFT structure constants
and  4-point blocks with NS and R intermediate states. These checks not only test all the
 structure constants  but also give a strong verification of the  definitions and
recursive representations of all the types of the 4-point blocks involved.
This makes the considerations of the present paper a firm starting point for deriving
an analytic proof of the bootstrap equations in $N=1$ SLFT
which was one of the motivations of the present work.
Another one is  a study of 1-point functions on a torus in $N=1$ SLFT and their
 modular invariance.
 The recursive representation of the corresponding 1-point superconformal blocks
can be found using the techniques developed in the present paper.
It would be very interesting to check if there exists a relation
between supersymmetric 1-point toric and 4-point spheric blocks  similar to that found in the bosonic case
\cite{Poghossian:2009mk, Hadasz:2009db}.

The organization of the paper is as follows. In the first section we make a brief review of $N=1$ SLFT and
introduce our notation. Section 2 is devoted to diagonal representations of NS superprimary and R primary fields
in terms of chiral vertex operators. We collect earlier results from \cite{Hadasz:2006sb},
\cite{R}\footnote{We also correct the formula for chiral decomposition of R primary fields in R-NS sector
introduced in \cite{R}.}
and derive a formula for NS superprimary field in R-R sector. In section 3 the 4-point blocks corresponding to
correlators of 2 NS and 2 R fields factorized on NS and R states are defined
and their recursion representations are derived. These are the main  results of the present paper.
 In the last section we present the results of numerical checks
 of bootstrap equations in three different cases: the correlator of 4 R fields, the correlator of 2 NS and 2 R fields
 factorized on R states and finally the equation combining correlators of 2 NS and 2 R fields factorized
 on R and NS states.

\section{N=1 Supersymmetric Liouville Theory}

The $N=1$ supersymmetric Liouville theory is defined by the action:
\begin{equation}
\label{SLFT}
{\cal S}_{\rm\scriptscriptstyle SL} =
\int d^2z \, \left(\frac{1}{2\pi}\left|\partial\phi_{\rm \scriptscriptstyle SL}\right|^2
+ \frac{1}{2\pi}\left(\psi_{\rm \scriptscriptstyle SL}\bar\partial\psi_{\rm\scriptscriptstyle SL}
 + \bar\psi_{\rm\scriptscriptstyle SL}\partial\bar\psi_{\rm\scriptscriptstyle SL}\right)
+ 2i\mu b^2\bar\psi_{\scriptscriptstyle\rm SL}\psi_{\scriptscriptstyle\rm SL} {\rm e}^{b\phi_{\scriptscriptstyle\rm SL}}
+ 2\pi b^2\mu^2{\rm e}^{2b\phi_{\scriptscriptstyle\rm SL}}\right)
\end{equation}
where $b$ is the dimensionless coupling constant and $\mu$ is the scale parameter. The central charge of the theory
is $c = \frac32 + 3 Q^2$, where $  Q=b + \frac1b$ is the background charge.

The $N=1$ superconformal symmetry is generated by the holomorphic bosonic and fermionic generators
$T(z),S(z) $ (and their antiholomorphic counterparts $\bar T(\bar z),\bar S(\bar z)$) fulfilling the OPEs
\cite{Friedan:1984rv}, \cite{Zamolodchikov:1988nm}:
\begin{eqnarray*}
T(z)T(0) & = & \frac{c}{2z^4} + \frac{2}{z^2}T(z) + \frac{1}{z}\partial T(0) + \ldots,
\\
T(z)S(0) & = & \frac{3}{2z^2}S(0) + \frac{1}{z}\partial S(0)+ \ldots,
\\
S(z)S(0) & = & \frac{2c}{3z^3} + \frac{2}{z}T(0) + \ldots\,.
\end{eqnarray*}
The generators
have definite parity with respect to the common (left and right) parity operator:
$$
(-1)^F \, T(z) = T(z) \, (-1)^F , \qquad
(-1)^F \, S(z) = - S(z) \, (-1)^F.
$$
The space of fields contains two types of fields:
the \emph{Neveu-Schwarz} fields $\varphi(z_i,\bar z_i)$ local with respect to $S(z)$
and  the \emph{Ramond} fields $R(z_i,\bar z_i)$  ``half-local'' with respect to $S(z)$.
  The ``half-locality'' of Ramond fields means that any correlation function containing product
$ S(z) R(z_i,\bar z_i)$ changes the sing upon analytic continuation in $z$ around the point $z=z_i.$
The locality properties determine the form of the  OPEs:
$$
S(z)\varphi(0,0) \; = \; \sum\limits_{k\in {\mathbb Z} + \frac12} z^{k-\frac32} \, S_{-k}\varphi(0,0),
\qquad
T(z)\varphi(0,0) \; = \; \sum\limits_{n\in {\mathbb Z} } z^{n-2} \,L_{-n}\varphi(0,0)
$$
$$
S(z)R(0,0) \; = \; \sum\limits_{m\in {\mathbb Z} } z^{m-\frac32} \, S_{-m}R(0,0),
\qquad
T(z)R(0,0) \; = \; \sum\limits_{n\in {\mathbb Z} } z^{n-2} \,L_{-n}R(0,0)
$$
The fermionic modes together with the Virasoro generators $L_n$
form two independent copies of the Neveu-Schwarz (for $p,q$ half-integer) or Ramond (for $p,q$ integer) algebra:
\begin{eqnarray*}
\left[L_m,L_n\right] & = & (m-n)L_{m+n} +\frac{c}{12}m\left(m^2-1\right)\delta_{m+n},
\\
\left[L_m,S_p\right] & = &\frac{m-2p}{2}S_{m+p},
\\
\left\{S_p,S_q\right\} & = & 2L_{p+q} + \frac{c}{3}\left(p^2 -\frac14\right)\delta_{p+q},
\\
\left[ L_n , \bar{L}_m \right] &=& 0\, , \qquad [ L_n , \bar{S}_p ] = 0\, , \qquad \left\{S_p,\bar S_q\right\} = 0.
\end{eqnarray*}

Each Liouville superprimary NS field $\phi_a$ is represented by an exponential:
\begin{equation}\label{superprimary}
\phi_a = {\rm e}^{a\phi_{\rm \scriptscriptstyle SL}},
\end{equation}
with the conformal dimension $\Delta_a=\bar\Delta_a={a(Q-a)\over 2} $. We shall also use the
parametrization in terms of the momentum $p$:
\begin{equation}\label{momentum}
    \Delta_a = \frac{Q^2}{8} + \frac{p^2}{2}, \qquad \mathrm{where } \quad a = \frac{Q}{2} + i p.
\end{equation}
The superconformal family of $\phi_a$ corresponds to the tensor product
 $\mathcal{V}_{\Delta_a} \otimes \bar{\mathcal{V}}_{\bar \Delta_a}$ of
 ${1\over 2}\mathbb{Z}$-graded representations of the left and the right NS algebra. It contains
four Virasoro primaries, $\phi_a$ itself and three descendants:
\begin{eqnarray}\label{NSprimaries}
\psi_a &=& \left[S_{-1/2},\phi_a\right]
    = -ia\psi_{\rm \scriptscriptstyle SL}{\rm e}^{a\phi_{\rm \scriptscriptstyle SL}},
\qquad
\bar\psi_a = \left[\bar S_{-1/2},\phi_a\right]
    = -i a\bar \psi_{\rm \scriptscriptstyle SL}{\rm e}^{a \phi_{\rm \scriptscriptstyle SL}},
\\[5pt]  \nonumber
\widetilde \phi_a &=& \left\{S_{-1/2},\left[\bar S_{-1/2},\phi_a\right]
\right\} = a^2\psi_{\rm \scriptscriptstyle SL}\bar\psi_{\rm \scriptscriptstyle SL}
    {\rm e}^{a\phi_{\rm \scriptscriptstyle SL}} -2i\pi\mu b a {\rm e}^{(a+b)\phi_{\rm \scriptscriptstyle SL}}
\end{eqnarray}

There are two Virasoro primary fields in the Ramond sector  represented by
\begin{equation}\label{Rprimary}
R^{\pm}_{a} = \sigma^{\pm} \, \mathrm{e}^{a \phi_{\rm SL}},
\end{equation}
where $\sigma^{\pm}$ are the twist operators with the conformal weight $\frac{1}{16}$.
The weights of  primary  fields $R^{\pm}_{a}$ read:
$$ \Delta_{[a]} = \bar\Delta_{[a]} = \frac{1}{16}+\frac{a(Q-a)}{2}.
$$
The OPEs of the R primary fields with the fermionic current take the following form:
\begin{eqnarray*}
S(z) R_{a}^{\pm}(z_i,\bar z_i)
& \sim &
{i \beta {\rm e}^{\mp i\frac{\pi}{4}} \over (z-z_i)^{\frac{3}{2}}} R_{a}^{\mp}(z_i,\bar z_i)+ \ldots
,
\\[6pt]
\bar S(\bar z) R_{a}^{\pm}(z_i,\bar z_i)
& \sim &
{-i\beta {\rm e}^{\pm i\frac{\pi}{4}} \over (\bar z-\bar z_i)^{3\over 2}}R_{a}^{\mp}(z_i,\bar z_i) + \ldots,
\end{eqnarray*}
where $\beta$ is related to the conformal weight by
\begin{equation}\label{beta}
\Delta_{[a]} = {c\over 24}-\beta^2, \qquad \mathrm{and} \quad \beta = \frac{1}{\sqrt{2}}\left( \frac{Q}{2} - a\right) .
\end{equation}
We assume that the superprimary fields $\phi_a$ and primaries $R^+_{a}$ are even with respect
to the common parity operator.

We define the R supermodule ${\cal W}_\Delta\,$ as a highest weight  representation of the
R  algebra extended by the chiral parity operator $(-1)^{F_L}$:
$$
[(-1)^{F_L}, L_m]=\{(-1)^{F_L},S_n\}=0\;\;\;,\;\;\;m,n\in \mathbb{Z}.
$$
The tensor product ${\cal W}_\Delta \otimes \bar{\cal W}_{\bar\Delta}$ of the left and the right
R supermodules provides a representation of the direct sum $R\oplus \bar R$ of the left and the right
extended Ramond algebras.
In  the Liouville theory however
we need an extension of the left and the right R algebras only by the common parity operator
$
(-1)^F =(-1)^{F_L}(-1)^{F_R}.
$
This can be achieved by reducing the representation ${\cal W}_\Delta \otimes \bar{\cal W}_{\bar\Delta}$
 to the invariant subspace ${\cal W}_{\Delta, \bar\Delta}
\subset {\cal W}_\Delta \otimes \bar{\cal W}_{\bar\Delta}$. For $\Delta \neq {c\over 24}$ it is
 generated by the vectors
\begin{equation}
\label{basis0}
\begin{array}{llllll}
w^+_{\Delta,\bar\Delta}&=&
 {1\over \sqrt{2}}\left( w^+_\Delta \otimes  w^+_{\bar\Delta} -i\,
 w^-_\Delta \otimes  w^-_{\bar\Delta}\right)\ ,
\\ [4pt]
 w^-_{\Delta,\bar\Delta}&=&
 {1\over \sqrt{2}}\left(
 w^+_\Delta \otimes  w^-_{\bar\Delta}\; +\;
 w^-_\Delta \otimes  w^+_{\bar\Delta}\right)\ .
\end{array}
\end{equation}
where $ w^+_{\Delta}, w^+_{\bar \Delta}$ are the even highest weight states in
$ {\cal W}_\Delta, \bar{\cal W}_{\bar\Delta}$ and
$
w^-_\Delta=  {{\rm e}^{i{\pi\over 4}}\over i \beta}S_0  w^+_{\Delta}\;\;,\;\;
w^-_{\bar \Delta}=  {{\rm e}^{-i{\pi\over 4}}\over - i  \beta}\bar S_0  w^+_{\bar \Delta}$.
This is so called ``small representation'' \cite{R}.

Correlation functions in $N=1$ SLFT are determined by the superconformal Ward identities
 up to 3-point structure constants of the four independent types:
\begin{eqnarray}\label{constantNS}
\nonumber
C_{\scriptscriptstyle 321} & = & \Big\langle\phi_3(\infty,\infty)\phi_2(1,1)\phi_1(0,0)\Big\rangle,
\\[-8pt]
\\[-8pt] \nonumber
\widetilde C_{\scriptscriptstyle 321}
    & = & \Big\langle\phi_3(\infty,\infty)\widetilde\phi_2(1,1)\phi_1(0,0)\Big\rangle,
\\ \label{constantR}
C^{\pm}_{\scriptscriptstyle 3[2][1]} &=& \Big\langle\phi_3(\infty,\infty)R^{\pm}_2(1,1) R^{\pm}_1(0,0)\Big\rangle,
\end{eqnarray}
They were derived in  \cite{Poghosian:1996dw, Rashkov:1996jx} and read:
\begin{eqnarray} \label{structureC_NS}
C_{\scriptscriptstyle 321} &=& \frac12
\left( \frac{\pi \mu}{2} \gamma\left(\frac{Qb}{2}\right) b^{2-2b^2}\right)^{Q-a \over b}
\\ \nonumber
&\times&
\frac{\Upsilon_0 \,\Upsilon_{\rm NS}(2a_1) \Upsilon_{\rm NS}(2a_2) \Upsilon_{\rm NS}(2a_3)}{
    \Upsilon_{\rm NS}(a-Q) \Upsilon_{\rm NS}(a_1+a_2-a_3)\Upsilon_{\rm NS}(a_2+a_3-a_1)\Upsilon_{\rm NS}(a_3+a_1-a_2)}
    \\[6pt] \label{structureCt_NS}
\tilde C_{\scriptscriptstyle 321} &=& i
\left( \frac{\pi \mu}{2} \gamma\left(\frac{Qb}{2}\right) b^{2-2b^2}\right)^{Q-a \over b}
\\ \nonumber &\times&
\frac{\Upsilon_0 \,\Upsilon_{\rm NS}(2a_1) \Upsilon_{\rm NS}(2a_2) \Upsilon_{\rm NS}(2a_3)}{
    \Upsilon_{\rm R}(a-Q) \Upsilon_{\rm R}(a_1+a_2-a_3)\Upsilon_{\rm R}(a_2+a_3-a_1)\Upsilon_{\rm R}(a_3+a_1-a_2)}
\\[6pt]\label{structureC_R}
 C^{\epsilon}_{\scriptscriptstyle 3[2][1]} &=& \frac12
\left( \frac{\pi \mu}{2} \gamma\left(\frac{Qb}{2}\right) b^{2-2b^2}\right)^{Q-a \over b}
\\ \nonumber &\times& \Bigg[
\frac{\Upsilon_0 \,\Upsilon_{\rm R}(2a_1) \Upsilon_{\rm R}(2a_2) \Upsilon_{\rm NS}(2a_3)}{
    \Upsilon_{\rm R}(a-Q) \Upsilon_{\rm R}(a_1+a_2-a_3)\Upsilon_{\rm NS}(a_2+a_3-a_1)\Upsilon_{\rm NS}(a_3+a_1-a_2)}
    \\ [4pt] \nonumber
&+& \epsilon \,
 \frac{\Upsilon_0 \,\Upsilon_{\rm R}(2a_1) \Upsilon_{\rm R}(2a_2) \Upsilon_{\rm NS}(2a_3)}{
    \Upsilon_{\rm NS}(a-Q) \Upsilon_{\rm NS}(a_1+a_2-a_3)\Upsilon_{\rm R}(a_2+a_3-a_1)\Upsilon_{\rm R}(a_3+a_1-a_2)}
    \Bigg]
\end{eqnarray}
where $a= \sum_{i=1}^{3} a_i$, $\Upsilon_0 = \frac{\mathrm{d}  \Upsilon_b(x)}{\mathrm{d}x}|_{x=0}$
and $\Upsilon_{\rm NS/R} $  \cite{FH} denote:
\begin{eqnarray*}
\Upsilon_{\rm NS}(x)&=& \Upsilon_b\left(\frac{x}{2}\right) \, \Upsilon_b\left(\frac{x+Q}{2}\right)
\\
\Upsilon_{\rm R}(x)&=& \Upsilon_b\left(\frac{x+b}{2}\right) \, \Upsilon_b\left(\frac{x+ b^{-1}}{2}\right).
\end{eqnarray*}
The special function $\Upsilon_b(x)$ was introduced by Zamolodchikovs in \cite{Zamolodchikov:1995aa}.
In the strip  $ 0<Re(x)<Q$  it has the integral representation:
\begin{eqnarray}\label{Ups}
\log \Upsilon_b(x) = \int_0^{\infty} \frac{\mathrm{d} t}{t}
\left\{
\left( \frac{Q}{2} - x\right)^2 \mathrm{e}^{ -t} -
{ \sinh^2\left[ \left( \frac{Q}{2} - x\right)\frac{t}{2}\right] \over
\sinh\left( \frac{bt}{2}\right) \, \sinh\left( \frac{t}{2b}\right)
}
\right\}.
\end{eqnarray}

In this paper we are interested in 4-point functions of the R fields and the bootstrap equations they should satisfy.
 Restricting ourselves to the $R^+$ fields and the NS superprimaries  we have the following crossing symmetry conditions:
\begin{eqnarray}
\label{RRRRcs1}
&& \hspace{-30pt}
\left\langle R^+_4(\infty,\infty) R^+_3(1,1) R^+_2(z,\bar z)  R_1^+(0,0)
\right\rangle
=
\left\langle R^+_4(\infty,\infty) R^+_1(1,1) R^+_2(1-z,1-\bar z)  R_3^+(0,0)
\right\rangle
 \\[4pt]
\label{NRNRcs1}
&& \hspace{-30pt}
\left\langle \, \phi_4(\infty,\infty) R^+_3(1,1) \phi_2(z,\bar z) \, R_1^+(0,0)
\right\rangle
=
\left\langle \, \phi_4(\infty,\infty) R^+_1(1,1) \phi_2(1-z,1-\bar z) \, R_3^+(0,0)
\right\rangle
\\[4pt] \label{NNRRcs1}
&& \hspace{-30pt}
\left\langle \, \phi_4(\infty,\infty) \phi_3(1,1) R^+_2(z,\bar z) \, R_1^+(0,0)
\right\rangle
=
\left\langle \, \phi_4(\infty,\infty) R^+_1(1,1) R^+_2(1-z,1-\bar z) \, \phi_3^+(0,0)
\right\rangle
\end{eqnarray}
 and
\begin{eqnarray}
\label{RRRRcs2}
&&
\left\langle R^+_4(\infty,\infty)\, R^+_3(1,1) R^+_2(z,\bar z) \, R_1^+(0,0)
\right\rangle
\\ \nonumber
&&\hspace{2cm} =
 \left[(1-z)(1- \bar z)\right]^{-2\Delta_2}
\left\langle R^+_3(\infty,\infty)\, R^+_4(1,1) R^+_2(\frac{z}{z-1},\frac{\bar z}{\bar z-1}) \, R_1^+(0,0)
\right\rangle
\\ \label{NRNRcs2}
&&
\left\langle \, \phi_4(\infty,\infty) R^+_3(1,1) \phi_2(z,\bar z) \, R_1^+(0,0)
\right\rangle
\\ \nonumber
&&\hspace{2cm} =
 \left[(1-z)(1- \bar z)\right]^{-2\Delta_2}
\left\langle \, R^+_3(\infty,\infty) \phi_4(1,1)  \phi_2(\frac{z}{z-1},\frac{\bar z}{\bar z-1}) \, R_1^+(0,0)
\right\rangle
\\  \label{NNRRcs2}
&&\left\langle \, \phi_4(\infty,\infty) \phi_3(1,1) R^+_2(z,\bar z) \, R_1^+(0,0) \right\rangle
\\ \nonumber
&&\hspace{2cm} =
 \left[(1-z)(1- \bar z)\right]^{-2\Delta_2}
\left\langle \, \phi_3(\infty,\infty) \phi_4(1,1) R^+_2(\frac{z}{z-1},\frac{\bar z}{\bar z-1})  \, R_1^+(0,0)
\right\rangle.
\end{eqnarray}

\section{The 3-point blocks}

The aim of this section is to collect the formulae for the NS superprimary and the $R^+$ primary fields written in terms
of normalized chiral vertex operators.
The strategy is to express  3-point functions by the structure constants and suitably normalized 3-point blocks.
In order to find such decompositions we will define the chiral 3-forms using the
Ward identities for correlation functions of three fields with an arbitrary number of holomorphic generators.
The  NS and the R operators have the following block structure:
\begin{eqnarray*}
\varphi = \left[\begin{array}{c|c}
\varphi_{\scriptscriptstyle \rm{NN}} & 0\\
\hline
0 & \varphi_{\scriptscriptstyle \rm RR}\;
\end{array}\right]
\;\;\;,
\;\;\;
R = \left[\begin{array}{c|c}
0 & R_{\scriptscriptstyle \rm{NR}}\\
\hline
R_{\scriptscriptstyle \rm{RN}} & 0
\end{array}\right]
\end{eqnarray*}
with respect to the direct sum decomposition
 $ {\cal H}= {\cal H}_{\scriptscriptstyle \rm NS}\oplus  {\cal H}_{\scriptscriptstyle\rm R} $ of the space of states.
 Since for each block there are different Ward identities,
  it is convenient to investigate these four cases separately.

The simplest case is that of pure NS sector \cite{Hadasz:2006sb}.
The Ward identities for a 3-point function suggest the definition of the chiral 3-form
(anti-linear in the left argument and
linear in the central and the right ones)
$ \varrho_{\scriptscriptstyle \rm NN}(\xi_3, \xi_2, \xi_1|z)\ : \
\mathcal{V}_{\Delta_3} \times \mathcal{V}_{\Delta_2} \times \mathcal{V}_{\Delta_1} \rightarrow \mathbb{C} $:
\begin{eqnarray*}
\varrho_{\scriptscriptstyle \rm{NN}}( \xi_3,S_{k}\xi_2,\xi_1|z) &=&
 \sum\limits_{m=0}^{k+{1\over 2}}
 \left(
\begin{array}{c}
\scriptstyle k+{1\over 2}\\[-6pt]
\scriptstyle m
\end{array}
\right)
 (-z)^{m}
\left(\varrho_{\scriptscriptstyle \rm{NN}}( S_{m-k}\xi_3,\xi_2,\xi_1|z)\right.
\\
 &&\hspace{20pt} -\ (-1)^{|\xi_1|+|\xi_3|}\; \left.
\varrho_{\scriptscriptstyle \rm{NN}}(\xi_3,\xi_2,S_{k-m}\xi_1|z) \right),
\hskip 5mm k\geqslant-\scriptstyle {1\over 2},
\\[10pt]
\varrho_{\scriptscriptstyle \rm{NN}}(\xi_3,S_{-k}\xi_2,\xi_1|z) &=&
 \sum\limits_{m=0}^{\infty}
 \left(
\begin{array}{c}
\scriptstyle k-{3\over 2}+m\\[-6pt]
\scriptstyle m
\end{array}
\right)
z^{m}
\varrho_{\scriptscriptstyle \rm{NN}}(S_{k+m} \xi_3,\xi_2,\xi_1|z)
\\
 && \hspace{-95pt} -\;
(-1)^{|\xi_1|+|\xi_3|+k+{1\over 2}}
\sum\limits_{m=0}^{\infty}
 \left(
\begin{array}{c}
\scriptstyle k-{3\over 2}+m\\[-6pt]
\scriptstyle m
\end{array}
\right)
 z^{-k-m+{1\over 2}}
\varrho_{\scriptscriptstyle \rm{NN}}(\xi_3,\xi_2, S_{m-{1\over 2}}\xi_1|z), \hskip 5mm k>\scriptstyle {1\over 2}.
\\ [10pt]
\varrho_{\scriptscriptstyle \rm{NN}}(S_{-k} \xi_3,\xi_2,\xi_1|z)
&=&
(-1)^{|\xi_1|+|\xi_3|+1}
\varrho_{\scriptscriptstyle \rm{NN}}( \xi_3,\xi_2,S_k\xi_1|z)
\\
 &+&
\sum\limits_{m=-1}^{l(k-{1\over 2})}
 \left(
\begin{array}{c}
\scriptstyle k+{1\over 2}\\[-6pt]
\scriptstyle m+1
\end{array}
\right)
  z^{k-{1\over 2} -m}
  \varrho_{\scriptscriptstyle \rm{NN}}(\xi_3,S_{m+{1\over 2}}\xi_2,\xi_1|z)
\end{eqnarray*}
\begin{eqnarray}
\label{translationNS}
\varrho_{\scriptscriptstyle \rm{NN}}( \xi_3,L_{-1}\xi_2,\xi_1|z)
 &=&
 \partial_z
 \varrho_{\scriptscriptstyle \rm{NN}}( \xi_3,\xi_2,\xi_1|z),
\\[5pt]
\label{formaL2aNS}
\varrho_{\scriptscriptstyle \rm{NN}}( \xi_3,L_{n}\xi_2,\xi_1|z) &=&
 \sum\limits_{m=0}^{n+1} \left(\,_{\;\;m}^{n+1}\right) (-z)^{m}
\Big( \varrho_{\scriptscriptstyle \rm{NN}}( L_{m-n}\xi_3,\xi_2,\xi_1|z)
\\ \nonumber &&\hspace{70pt} -\;
\varrho_{\scriptscriptstyle \rm{NN}}( \xi_3,\xi_2,L_{n-m}\xi_1|z) \Big), \quad n>-1,
\\ [8pt]
\label{formaL2bNS}
\varrho_{\scriptscriptstyle \rm{NN}}( \xi_3,L_{-n}\xi_2,\xi_1|z) &=&
 \sum\limits_{m=0}^{\infty} \left(\,_{\;\;n-2}^{n-2+m}\right)
z^{m}
\varrho_{\scriptscriptstyle \rm{NN}}(L_{n+m} \xi_3,\xi_2,\xi_1|z)
\\
\nonumber && \hspace{-40pt} +\;(-1)^n \sum\limits_{m=0}^{\infty}
\left(\,_{\;\;n-2}^{n-2+m}\right) z^{-n+1-m}
\varrho_{\scriptscriptstyle \rm{NN}}(\xi_3,\xi_2, L_{m-1}\xi_1|z), \hskip 5mm n>1,
\\ [4pt]
\label{formaL3NS}
\varrho_{\scriptscriptstyle \rm{NN}}(L_{-n} \xi_3,\xi_2,\xi_1|z)
&=&
\varrho_{\scriptscriptstyle \rm{NN}}( \xi_3,\xi_2,L_n\xi_1|z)
+
\sum\limits_{m=-1}^{l(n)}
{\textstyle { n+1\choose  m+1}}
  z^{n-m}
  \varrho_{\scriptscriptstyle \rm{NN}}(\xi_3,L_m\xi_2,\xi_1|z),
\end{eqnarray}
where  $|\xi_i|=0$ for an even state and $|\xi_i|=1$ for an odd state.
The 3-form is set by the definition up to two independent constants:
\begin{equation} \label{varrhoNN}
 \varrho_{\scriptscriptstyle \rm NN}(\nu_3, \nu_2, \nu_1|1),  \qquad
 \varrho_{\scriptscriptstyle \rm NN}(\nu_3, S_{-\frac12}\nu_2, \nu_1|1).
\end{equation}
  For the  highest weight state $\nu_2\in \mathcal{V}_{\Delta_2} $ we define the 3-point blocks by:
 \begin{eqnarray*}
&& \hspace{-0.5cm}
\varrho_{\scriptscriptstyle \rm NN}(\xi_3, \nu_2, \xi_1|z)
 = 
\rho_{\scriptscriptstyle \rm NN,e}(\xi_3, \nu_2, \xi_1|z)
 \varrho_{\scriptscriptstyle \rm NN}(\nu_3, \nu_2, \nu_1|1)
+ \ \rho_{\scriptscriptstyle \rm NN,o}(\xi_3, \nu_2, \xi_1|z)
 \varrho_{\scriptscriptstyle \rm NN}(\nu_3, S_{-\frac12}\nu_2, \nu_1|1),
 \\[4pt]
 && \hspace{-0.5cm} \rho_{\scriptscriptstyle \rm NN,e/o}(\xi_3, \nu_2, \xi_1|z)  =
z^{\Delta_3(\xi_3)- \Delta_2(\nu_2)-\Delta_1(\xi_1)}  \rho_{\scriptscriptstyle \rm NN,e/o}(\xi_3, \nu_2, \xi_1)
\end{eqnarray*}
where the indices ${\rm e,o}$ denote even and odd parity of the 3-point block respectively.
The even part of the block vanishes when $\xi_3, \xi_1$ are states of different parity,
 while the odd part vanishes for $\xi_3, \xi_1$ of the same parity.

The even-even and the odd-odd products of the left and the right constants (\ref{varrhoNN})
 yield the two structure constants (\ref{constantNS})
\begin{eqnarray*}
C_{\scriptscriptstyle 321} &=& \varrho_{\scriptscriptstyle \rm NN}(\nu_3, \nu_2, \nu_1|1) \,
 \bar \varrho_{\scriptscriptstyle \rm NN}(\bar\nu_3, \bar\nu_2, \bar\nu_1|1), \\
\widetilde C_{\scriptscriptstyle 321} &=&
\varrho_{\scriptscriptstyle \rm NN}(\nu_3, S_{-\frac12} \nu_2, \nu_1|1) \,
 \bar \varrho_{\scriptscriptstyle \rm NN}(\bar\nu_3, \bar S_{-\frac12}\bar\nu_2, \bar\nu_1|1).
\end{eqnarray*}
 An arbitrary 3-point function of the NS fields with a definite parity is determined
by the Ward identities up to one of the two structure constants:
\begin{eqnarray*}\bra{\xi_3 \otimes \xi_3} \phi_{\scriptscriptstyle NN}(z, \bar{z}) \ket{\xi_1 \otimes \xi_1}
&=& C_{\scriptscriptstyle 321} \, \rho_{\scriptscriptstyle \rm NN,e}(\xi_3, \nu_2, \xi_1|z) \,
\rho_{\scriptscriptstyle \rm NN,e}(\bar \xi_3, \bar \nu_2, \bar \xi_1|\bar z)
\\
&-& (-1)^{|\xi_1|} \widetilde C_{\scriptscriptstyle 321} \, \rho_{\scriptscriptstyle \rm NN,o}(\xi_3, \nu_2, \xi_1|z) \,
\rho_{\scriptscriptstyle \rm NN,o}(\bar \xi_3, \bar \nu_2, \bar \xi_1|\bar z)
\end{eqnarray*}
Thus the decomposition of the superprimary field $\phi_{NN}$ in terms of normalized vertex operators
$$ \bra{\xi_3} V_{\scriptscriptstyle \rm NN,f}(\nu_2|z) \ket{\xi_1} =
    \rho_{\scriptscriptstyle \rm NN,f}(\xi_3, \nu_2, \xi_1|z)
$$
reads \footnote{The vertex decomposition of the other 3 primary NS  fields (\ref{NSprimaries}) is presented in \cite{Hadasz:2006sb}}:
\begin{equation}\label{phiNN_vertex}
\phi_{\scriptscriptstyle NN}(z, \bar{z}) = 
    C_{321} V_{\scriptscriptstyle \rm NN, e}(\nu|z)\otimes
    V_{\scriptscriptstyle \rm NN, e}(\bar \nu|\bar z)
    -\tilde C_{321} V_{ \scriptscriptstyle \rm NN, o}(\nu|z)\otimes
    V_{\scriptscriptstyle \rm NN, o}(\bar \nu|\bar z).
\end{equation}

Due to the complicated form of the Ward identities
the chiral decomposition of R primary fields is considerably more involved \cite{R}.
In contrast to NS sector any 3-point function involving R fields with definite parity
depends on both R structure constants (\ref{constantR}).
The chiral Ward identities defining  3-forms
$$ \varrho_{\scriptscriptstyle \rm NR}(\xi_3, \eta_2, \eta_1|z)\ : \
\mathcal{V}_{\Delta_3} \times \mathcal{W}_{\Delta_2} \times \mathcal{W}_{\Delta_1} \rightarrow \mathbb{C},
$$
$$ \varrho_{\scriptscriptstyle \rm RN}(\eta_3, \eta_2, \nu_1|z)\ : \
\mathcal{W}_{\Delta_3} \times \mathcal{W}_{\Delta_2}  \times \mathcal{V}_{\Delta_1}\rightarrow \mathbb{C}
$$
have the following form\footnote{The Ward identities for the Virasoro generators
$L_n$ (\ref{translationNS})-(\ref{formaL3NS}) are the same in all sectors. }:
\begin{eqnarray}\label{Ward_rhoNR}
\nonumber
&&\hspace{-70pt} \sum_{p=0}^{\infty} \left(^{n+\frac12}_p \right) \ z^{n+
\frac12 -p} \
\varrho_{\scriptscriptstyle \rm NR}
(\xi_3,S_{p} \eta_2,\eta_1 |z)
= \sum_{p=0}^{\infty} \left(^{\frac12}_p \right) (-z)^p \
\varrho_{\scriptscriptstyle \rm NR}
(S_{p-n - \frac12} \xi_3, \eta_2,\eta_1 |z)
 \\
&& - i (-1)^{|\xi_3|+|\eta_1|+1}  \sum_{p=0}^{\infty} \left(^{\frac12}_p \right) (-1)^p \ z^{\frac12 -p}
\varrho_{\scriptscriptstyle \rm NR}
(\xi_3, \eta_2,S_{n+p}\eta_1|z ) ,
\\  \nonumber
&& \hspace{-70pt} \sum_{p=0}^{\infty} \left(^{\frac12}_p \right) \ z^{\frac12 -p} \
\varrho_{\scriptscriptstyle \rm NR}
(\xi_3,S_{p-n} \eta_2,\eta_1 |z)
=
 \sum_{p=0}^{\infty} \left(^{-n+ \frac12}_p \right) (-z)^p \
 \varrho_{\scriptscriptstyle \rm NR}
(S_{p+n - \frac12} \xi_3, \eta_2,\eta_1 |z)
 \\  \nonumber
 &&- i (-1)^{|\xi_3|+|\eta_1|+1 } \sum_{p=0}^{\infty} \left(^{-n+\frac12}_p \right)(-1)^{n+p}  z^{ \frac12-n -p}
 \varrho_{\scriptscriptstyle \rm NR}
(\xi_3,\eta_2,S_{p}\eta_1 |z)
 ,
\end{eqnarray}
\begin{eqnarray}\label{Ward_rhoRN}
 \nonumber
&&\hspace{-70pt} \sum_{p=0}^{\infty} \left(^{-n}_{\;p} \right) \ z^{-p-n} \
\varrho_{\scriptscriptstyle \rm RN}
(\eta_3,S_{p} \eta_2,\xi_1|z )
=\sum_{p=0}^{\infty} \left(^{\frac12}_p \right) (-z)^p \
\varrho_{\scriptscriptstyle \rm RN}
(S_{n+p} \eta_3, \eta_2,\xi_1|z )
 \\
&& - i (-1)^{|\eta_3|+|\xi_1|+1}  \sum_{p=0}^{\infty} \left(^{\frac12}_p \right) (-1)^p \ z^{\frac12 -p}
\varrho_{\scriptscriptstyle \rm RN}
(\eta_3, \eta_2,S_{p-n-{1\over 2}}\xi_1 |z) ,
\\  \nonumber
&& \hspace{-70pt}
\varrho_{\scriptscriptstyle \rm RN}
(\eta_3,S_{-n} \eta_2,\xi_1|z)
= \sum_{p=0}^{\infty} \left(^{-n+ \frac12}_{\;\;\;\;p} \right) (-z)^p \
\varrho_{\scriptscriptstyle \rm RN}
(S_{p+n } \eta_3, \eta_2,\xi_1|z)
 \\  \nonumber
 && - i (-1)^{|\eta_3|+|\xi_1|+1 } \sum_{p=0}^{\infty} \left(^{-n+\frac12}_{\;\;\;\;p} \right)
 (-1)^{n+p}\,z^{ \frac12 -n-p}
 \varrho_{\scriptscriptstyle \rm RN}
(\eta_3,\eta_2,S_{p-\frac12 }\xi_1|z ).
\end{eqnarray}
They determine each 3-form  up to four rather than two constants:
\begin{eqnarray*}
\varrho_{\scriptscriptstyle \rm NR}(\xi_3,\eta_2,\eta_1|z)
& =& 
 \rho^{++}_{\scriptscriptstyle \rm NR}(\xi_3,\eta_2,\eta_1|z)
 \varrho_{\scriptscriptstyle \rm NR}(\nu_3, w^+_2, w^+_1|1) \\
&+& \ \rho^{+-}_{\scriptscriptstyle \rm NR}(\xi_3,\eta_2,\eta_1|z)
 \varrho_{\scriptscriptstyle \rm NR}(\nu_3, w^+_2, w^-_1|1)\\
&+& \ \rho^{-+}_{\scriptscriptstyle \rm NR}(\xi_3,\eta_2,\eta_1|z)
 \varrho_{\scriptscriptstyle \rm NR}(\nu_3, w^-_2, w^+_1|1)\\
&+& \ \rho^{--}_{\scriptscriptstyle \rm NR}(\xi_3,\eta_2,\eta_1|z)
 \varrho_{\scriptscriptstyle \rm NR}(\nu_3, w^-_2, w^-_1|1)
\\[6pt]
\varrho_{\scriptscriptstyle \rm RN}(\xi_3,\eta_2,\eta_1|z)
&=& 
 \rho^{++}_{\scriptscriptstyle \rm RN}(\eta_3,\eta_2,\xi_1|z)
 \varrho_{\scriptscriptstyle \rm RN}(w^+_3, w^+_2, \nu_1|1)\\
&+&\ \rho^{+-}_{\scriptscriptstyle \rm RN}(\eta_3,\eta_2,\xi_1|z)
 \varrho_{\scriptscriptstyle \rm RN}(w^+_3, w^-_2, \nu_1|1 )
 \\
&+&\ \rho^{-+}_{\scriptscriptstyle \rm RN}(\eta_3,\eta_2,\xi_1|z)
 \varrho_{\scriptscriptstyle \rm RN}(w^-_3, w^+_2, \nu_1|1)\\
&+&\ \rho^{--}_{\scriptscriptstyle \rm RN}(\eta_3,\eta_2,\xi_1|z)
\varrho_{\scriptscriptstyle \rm RN}(w^-_3, w^-_2, \nu_1|1).
\end{eqnarray*}
Since the R fields correspond to states from the ``small representation'' (\ref{basis0})
 eight even products of the left and the right
$\varrho_{\scriptscriptstyle \rm NR/RN}$ constants  reduce to the two
 structure constants \cite{R}
 \begin{equation}\label{strCpm}
    C^{(\pm)}_{\scriptscriptstyle 3[2][1]}
 ={ C^+_{\scriptscriptstyle 3[2][1]} \pm C^-_{\scriptscriptstyle 3[2][1]}\over 2},
\end{equation}
where
$C^{\pm}_{\scriptscriptstyle 3[2][1]}$ are 3-point correlators of primary fields (\ref{constantR}).
The mechanism of constants' number reduction together with the properties of
$\rho^{\imath\jmath}_{\scriptscriptstyle \rm NR/RN}$ suggest the convenient basis for the 3-point blocks
in each sector:
\begin{eqnarray}\label{3ptBNR}
\nonumber
\rho^{(\pm)}_{\scriptscriptstyle \rm NR, e}
&=&
\rho^{++}_{\scriptscriptstyle \rm NR} \pm
        \rho^{--}_{\scriptscriptstyle \rm NR}
\;\;\;,\;\;\;
\bar \rho^{(\pm)}_{\scriptscriptstyle \rm NR, e}
\;=\;
\bar \rho^{++}_{\scriptscriptstyle \rm NR} \pm
        \bar \rho^{--}_{\scriptscriptstyle \rm NR}
\\[-6pt]
\\[-6pt] \nonumber
\rho^{(\pm)}_{\scriptscriptstyle \rm NR, o}
&=&
\rho^{+-}_{\scriptscriptstyle \rm NR} \pm i
        \rho^{-+}_{\scriptscriptstyle \rm NR}
\;\;\;,\;\;\;
\bar \rho^{(\pm)}_{\scriptscriptstyle \rm NR, o}
\;=\;
\bar \rho^{+-}_{\scriptscriptstyle \rm NR} \mp i
        \bar \rho^{-+}_{\scriptscriptstyle \rm NR}
\end{eqnarray}
\begin{eqnarray}\label{3ptBRN}
\nonumber
\rho^{(\pm)}_{\scriptscriptstyle \rm RN, e}
&=&
\rho^{++}_{\scriptscriptstyle \rm RN} \pm
        \rho^{--}_{\scriptscriptstyle \rm RN}
\;\;\;,\;\;\; \quad
\bar \rho^{(\pm)}_{\scriptscriptstyle \rm RN, e}
\;=\;
\bar \rho^{++}_{\scriptscriptstyle \rm RN} \pm
        \bar \rho^{--}_{\scriptscriptstyle \rm RN}
\\[-6pt]
\\[-6pt] \nonumber
\rho^{(\pm)}_{\scriptscriptstyle \rm RN, o}
&=&
 \rho^{-+}_{\scriptscriptstyle \rm RN} \pm i
        \rho^{+-}_{\scriptscriptstyle \rm RN}
\;\;\;,\;\;\;
\bar \rho^{(\pm)}_{\scriptscriptstyle \rm RN, o}
\;=\;
\bar \rho^{-+}_{\scriptscriptstyle \rm RN} \mp i
        \bar \rho^{+-}_{\scriptscriptstyle \rm RN}
\end{eqnarray}
Using these bases we define the chiral vertex operators:
$$
\begin{array}{ccccclllllll}
\langle \xi_3|V^{(\pm)}_{\scriptscriptstyle \rm NR, f}(z)|\eta_1\rangle &=&
\rho^{(\pm)}_{\scriptscriptstyle \rm  NR, f}(\xi_3,w^+,\eta_1|z),
\end{array}
\qquad
\begin{array}{ccccclllllll}
\langle \eta_3|V^{(\pm)}_{\scriptscriptstyle \rm RN, f}(z)|\xi_1\rangle &=&
\rho^{(\pm)}_{\scriptscriptstyle \rm  RN, f}(\eta_3,w^+,\xi_1|z).
\end{array}
$$
In terms of these operators the fields $R^+_{\scriptscriptstyle \rm NR}$
and $R^+_{\scriptscriptstyle \rm RN}$ have the following diagonal representation \cite{R}:
\begin{eqnarray}\label{R_NRvertex}
\nonumber
  R_{\scriptscriptstyle \rm NR}^+ = \,
{C^{(+)} \over \sqrt{2}}
  \left( V_{\scriptscriptstyle \rm NRe}^{(+)} \otimes
  \bar V_{\scriptscriptstyle \rm NRe}^{(+)}
            \,-\, i \,
V_{\scriptscriptstyle \rm NRo}^{(+)} \otimes
\bar V_{\scriptscriptstyle \rm NRo}^{(+)} \right)
+
{C^{(-)} \over \sqrt{2}}
\left( V_{\scriptscriptstyle \rm NRe}^{(-)} \otimes
\bar V_{\scriptscriptstyle \rm NRe}^{(-)}
            \,-\, i \,
V_{\scriptscriptstyle \rm NRo}^{(-)} \otimes
\bar V_{\scriptscriptstyle \rm NRo}^{(-)}\right),
\\[-6pt]
\\[-2pt] \nonumber
  R_{\scriptscriptstyle \rm RN}^+ = \,
{C^{(+)} \over \sqrt{2}}
  \left( V_{\scriptscriptstyle \rm RNe}^{(+)} \otimes
  \bar V_{\scriptscriptstyle \rm RNe}^{(+)}
            \,-\, i \,
V_{\scriptscriptstyle \rm RNo}^{(+)} \otimes
\bar V_{\scriptscriptstyle \rm RNo}^{(+)} \right)
+
{C^{(-)} \over \sqrt{2}}
\left( V_{\scriptscriptstyle \rm RNe}^{(-)} \otimes
\bar V_{\scriptscriptstyle \rm RNe}^{(-)}
            \,-\, i \,
V_{\scriptscriptstyle \rm RNo}^{(-)} \otimes
\bar V_{\scriptscriptstyle \rm RNo}^{(-)}\right).
\end{eqnarray}
Let us stressed that $R_{\scriptscriptstyle \rm RN} $ are expressed in terms of the vertex operators
$ V_{\scriptscriptstyle \rm RN,f}^{(\pm)}$
 \footnote{ In the original paper \cite{R} it was erroneously assumed that the
 3-point blocks $ \rho^{(\pm)}_{\scriptscriptstyle \rm  RN, f}$
are conjugated to  $\rho^{(\pm)}_{\scriptscriptstyle \rm  NR, f} $.
For this reason the recursive relation for the 4-point blocks (5.8),(5.9) in \cite{R} needs correction.
The correct formula for this recursion is given by the equation (\ref{H_RRRR_rek}).
}.

The case of the NS superprimary field in the R-R sector has not been investigated before.
The chiral  Ward identities take the form:
\begin{eqnarray*}
\varrho_{\scriptscriptstyle \rm RR}
(S_{-n} \eta_3,\xi_2,\eta_1|z)
&=&
(-1)^{|\eta_1|+|\eta_3|+1}
\varrho_{\scriptscriptstyle \rm RR}
( \eta_3,\xi_2,S_n\eta_1|z)
\\
\nonumber &+&
\sum\limits_{k=-\frac12}^{\infty}
 \Big(\!\!
\begin{array}[c]{c}
\scriptstyle n+{1\over 2}\\[-7pt]
\scriptstyle k+\frac12
\end{array}
\!\!\Big)
  z^{n-k}
 \varrho_{\scriptscriptstyle \rm RR}(\eta_3,S_{k}\xi_2,\eta_1|z),
\\[4pt]
\sum_{p=0}^{\infty} \left(^{\frac12}_p \right) \ z^{\frac12 -p} \
\varrho_{\scriptscriptstyle \rm RR}(\eta_3,S_{p-k} \xi_2,\eta_1|z )
&=&
 \sum_{p=0}^{\infty} \left(^{ \frac12-k}_{\;\;\;p} \right) (-z)^p \
\varrho_{\scriptscriptstyle \rm RR}(S_{p+k - \frac12} \eta_3, \xi_2,\eta_1|z )
 \\
 &&\hspace{-70pt} - (-1)^{|\eta_3|+|\eta_1|+1 } \sum_{p=0}^{\infty}
 \left(^{\frac12-k}_{\;\;\;p} \right)(-z)^{ \frac12 -k-p}
 \varrho_{\scriptscriptstyle \rm RR}(\eta_3,\xi_2,S_{p}\eta_1 |z).
\end{eqnarray*}
As in the previous two cases the Ward identities determine the 3-form
 $$ \varrho_{\scriptscriptstyle \rm RR}(\eta_3, \xi_2, \eta_1|z)\ : \
\mathcal{W}_{\Delta_3} \times \mathcal{V}_{\Delta_2} \times \mathcal{W}_{\Delta_1} \rightarrow \mathbb{C}
$$
up to four constants:
 \begin{eqnarray*}
\varrho_{\scriptscriptstyle \rm RR}(\eta_3,\xi_2,\eta_1|z)
& =& 
 \rho^{++}_{\scriptscriptstyle \rm RR}(\eta_3,\xi_2,\eta_1|z)
 \varrho_{\scriptscriptstyle \rm RR}(w^+_3, \nu_2, w^+_1) \\
&+& \ \rho^{+-}_{\scriptscriptstyle \rm RR}(\eta_3,\xi_2,\eta_1|z)
 \varrho_{\scriptscriptstyle \rm RR}( w^+_3, \nu_2, w^-_1)\\
&+&\ \rho^{-+}_{\scriptscriptstyle \rm RR}(\eta_3,\xi_2,\eta_1|z)
 \varrho_{\scriptscriptstyle \rm RR}( w^-_3, \nu_2, w^+_1)\\
&+& \ \rho^{--}_{\scriptscriptstyle \rm RR}(\eta_3,\xi_2,\eta_1|z)
 \varrho_{\scriptscriptstyle \rm RR}(w^-_3,\nu_2, w^-_1)\, , 
 \\[4pt]
 \rho^{\imath\jmath}_{\scriptscriptstyle \rm RR}(\eta_3,\xi_2,\eta_1|z)
 &=& z^{\Delta_3(\eta_3)- \Delta_2(\xi_2)-\Delta_1(\eta_1)}
 \rho^{\imath\jmath}_{\scriptscriptstyle \rm RR}(\eta_3,\xi_2,\eta_1)
\end{eqnarray*}
where $ \varrho_{\scriptscriptstyle \rm RR}(w^{\pm}_3,\nu_2, w^{\pm}_1) \equiv
\varrho_{\scriptscriptstyle \rm RR}(w^{\pm}_3,\nu_2, w^{\pm}_1|1) $.
We write $\phi_{\scriptscriptstyle RR}$ in terms of the
non normalized chiral vertex operators
\begin{eqnarray*}
 \bra{\eta_3} V_{\scriptscriptstyle \rm RR, e}(\nu|z) \ket{\eta_1}
&=& \varrho_{\scriptscriptstyle \rm RR}(\eta_3,\nu,\eta_1|z),
	\qquad 	\eta_3, \eta_1 \quad \mathrm{of \ equal \ parities}
\\
\bra{\eta_3} V_{\scriptscriptstyle \rm RR, o}(\nu|z) \ket{\eta_1}
&=& \varrho_{\scriptscriptstyle \rm RR}(\eta_3,\nu,\eta_1|z),
	\qquad 	\eta_3, \eta_1 \quad \mathrm{of \ opposite \ parities}
\end{eqnarray*}
in the following form:
$$
\phi_{\scriptscriptstyle \rm RR}(z, \bar z)
= A \, V_{\scriptscriptstyle \rm RR, e}(\nu|z)\otimes  V_{\scriptscriptstyle \rm RR, e}(\bar \nu| \bar z) +
B \,  V_{\scriptscriptstyle \rm RR, o}(\nu|z)\otimes     V_{\scriptscriptstyle \rm RR, o}(\bar \nu | \bar z)
$$
Considering the matrix elements of $\phi_{\scriptscriptstyle RR}$ between primary
states $w^{\pm}_{\Delta, \bar \Delta}$
 from the ``small representation'' (\ref{basis0}) one can see that
 eight  even products of the left and the right
$\varrho_{\scriptscriptstyle \rm RR}$ constants  reduce to the two structure constants (\ref{strCpm}):
\begin{eqnarray}\label{C+RNR}
C^{(\pm)}_{\scriptscriptstyle  [3]2[1]} \! & = & \!
	\frac12 A \, \left( \varrho_{\scriptscriptstyle \rm RR}(w^+_3, \nu_2, w^+_1)
		\pm  \varrho_{\scriptscriptstyle \rm RR}(w^-_3, \nu_2, w^-_1) \right)
\left( \bar \varrho_{\scriptscriptstyle \rm RR}(\bar w^+_3, \bar \nu_2, \bar w^+_1)
		\pm \bar \varrho_{\scriptscriptstyle \rm RR}(\bar w^-_3, \bar \nu_2, \bar w^-_1) \right)
\nonumber
\\[-4pt]
\\[-4pt] \nonumber
&+& \!\frac{i}{2} B \,
\left( \varrho_{\scriptscriptstyle \rm RR}(w^+_3, \nu_2, w^-_1)
		\mp i \varrho_{\scriptscriptstyle \rm RR}(w^-_3, \nu_2, w^+_1) \right)
\left( \bar \varrho_{\scriptscriptstyle \rm RR}(\bar w^+_3, \bar \nu_2, \bar w^-_1)
		\pm \bar \varrho_{\scriptscriptstyle \rm RR}(\bar w^-_3, \bar \nu_2, \bar w^+_1) \right).
\end{eqnarray}
In order to express an arbitrary 3-point correlation function  in terms
of these structure constants one needs several relations between the
normalized 3-forms $\rho^{\imath\jmath}_{\scriptscriptstyle RR}$.
From the Ward identities it follows that the 3-forms of the same parity satisfy:
\begin{eqnarray} \label{e-e,o-o:p}
\nonumber
\rho^{++}_{\scriptscriptstyle \rm RR,e}(S_{M}w^+, \nu, S_{N} w^+)
	&=& \rho^{--}_{\scriptscriptstyle \rm RR,e}(S_{M}w^-, \nu, S_{N} w^-)
\\ \nonumber
\rho^{--}_{\scriptscriptstyle \rm RR,e}(S_{M}w^+, \nu, S_{N} w^+)
	&=&\rho^{++}_{\scriptscriptstyle \rm RR,e}(S_{M}w^-, \nu, S_{N} w^-)
\\[-4pt]
\\[-4pt] \nonumber
 \rho^{+-}_{\scriptscriptstyle \rm RR,o}(S_{M}w^+, \nu, S_{N} w^-)
	&=& \rho^{-+}_{\scriptscriptstyle \rm RR,o}(S_{M}w^-, \nu, S_{N} w^+)
\\ \nonumber
 \rho^{-+}_{\scriptscriptstyle \rm RR,o}(S_{M}w^+, \nu,S_{N} w^-)
	&=& - \rho^{+-}_{\scriptscriptstyle \rm RR,o}(S_{M}w^-, \nu, S_{N} w^+)
\end{eqnarray}
for the even number of fermionic operators in both strings: $\# M +\, \# N = 2 \mathbb{N}$
 and
\begin{eqnarray} \label{e-e,o-o:np}
\nonumber
\rho^{+-}_{\scriptscriptstyle \rm RR,o}(S_{K}w^-, \nu,S_{N} w^-)
	&=& - i \rho^{-+}_{\scriptscriptstyle \rm RR,o}(S_{K}w^+, \nu, S_{N}w^+)
\\ \nonumber
\rho^{-+}_{\scriptscriptstyle \rm RR,o}(S_{K}w^-, \nu, S_{N}w^-)
	 &=& i \rho^{+-}_{\scriptscriptstyle \rm RR,o}(S_{K}w^+, \nu,S_{N} w^+)
\\[-4pt]
\\[-4pt] \nonumber
\rho^{++}_{\scriptscriptstyle \rm RR,e}(S_{M}w^-, \nu, S_{N}w^+)
	&=& -i  \rho^{--}_{\scriptscriptstyle \rm RR,e}(S_{K}w^+, \nu, S_{N} w^-)
\\ \nonumber
\rho^{--}_{\scriptscriptstyle \rm RR,e}(S_{K}w^-, \nu,S_{N} w^+)
	&=& -i \rho^{++}_{\scriptscriptstyle \rm RR,e}(S_{K}w^+, \nu, S_{N}w^-)
\end{eqnarray}
for  $\# M +\, \# N = 2 \mathbb{N}+1$.
The relations between the 3-forms of different parity have the following form:
\begin{eqnarray} \label{e-o}
\nonumber
\rho^{++}_{\scriptscriptstyle \rm RR,e}(S_{I}w^+, \nu, S_{J} w^+)
	&=& (-1)^{\# J} \rho^{+-}_{\scriptscriptstyle \rm RR,o}(S_{I}w^+, \nu, S_{J} w^-)
\\ \nonumber
\rho^{--}_{\scriptscriptstyle \rm RR,e}(S_{I}w^+, \nu, S_{J} w^+)
	&=& (-1)^{\# J}\,  i \, \rho^{-+}_{\scriptscriptstyle \rm RR,o}(S_{I}w^+, \nu, S_{J} w^-)
\\ [-4pt]
\\[-4pt]\nonumber
\rho^{-+}_{\scriptscriptstyle \rm RR,o}(S_{I}w^+, \nu, S_{J} w^+)
	&=& (-1)^{\# J} \rho^{--}_{\scriptscriptstyle \rm RR,e}(S_{I}w^+, \nu, S_{J} w^-)
\\ \nonumber
\rho^{+-}_{\scriptscriptstyle \rm RR,o}(S_{I}w^+, \nu, S_{J} w^+)
&=& (-1)^{\# J}\,  i \, \rho^{++}_{\scriptscriptstyle \rm RR,e}(S_{I}w^+, \nu, S_{J} w^-).
\end{eqnarray}
These identities together with the constants' number reduction (\ref{C+RNR})
 allow to write an arbitrary matrix element of $\phi_{RR}$ in the diagonal form:
\begin{eqnarray*}
&& \hspace{-3 cm} \bra{S_{-I} \bar S_{- \bar I} w^{\pm}_{\Delta_3, \bar \Delta_3}}
 \phi_2 \ket{S_{-J} \bar S_{- \bar J} w^{\pm}_{\Delta_1, \bar \Delta_1}} \\[4pt]
&=& C^{(+)}_{\scriptscriptstyle [3]2[1]}
\rho^{(+)}_{\scriptscriptstyle \rm RR, e}(S_{-I}w^+_3, \nu_2, S_{-J} w^+_1) \,
\bar \rho^{(+)}_{\scriptscriptstyle \rm RR, e}(S_{-\bar I}w^+_3, \nu_2, S_{-\bar J} w^+_1)
\\[4pt]
&+&
C^{(-)}_{\scriptscriptstyle [3]2[1]} \rho^{(-)}_{\scriptscriptstyle \rm RR, e}(S_{-I}w^+_3, \nu_2, S_{-J} w^+_1) \,
\bar \rho^{(-)}_{\scriptscriptstyle \rm RR, e}(S_{-\bar I}w^+_3, \nu_2, S_{-\bar J} w^+_1)
\end{eqnarray*}
for $ \# I+\, \# J = 2\mathbb{N}$   and
 \begin{eqnarray*}
&& \hspace{-2 cm}
\bra{S_{-I} \bar S_{- \bar I} w^{\pm}_{\Delta_3, \bar \Delta_3}} \phi_2
\ket{S_{-J} \bar S_{- \bar J} w^{\pm}_{\Delta_1, \bar \Delta_1}}
\\[4pt]
&=& -i \, (-1)^{\# J}
\Big\lbrace  C^{(+)}_{\scriptscriptstyle [3]2[1]}
\rho^{(+)}_{\scriptscriptstyle \rm RR, o}(S_{-I}w^+_3, \nu_2, S_{-J} w^+_1) \,
\bar \rho^{(+)}_{\scriptscriptstyle \rm RR, o}(S_{-\bar I}w^+_3, \nu_2, S_{-\bar J} w^+_1)
\\ 
&& \hspace{1,5 cm} +
C^{(-)}_{\scriptscriptstyle [3]2[1]} \rho^{(-)}_{\scriptscriptstyle \rm RR, o}(S_{-I}w^+_3, \nu_2, S_{-J} w^+_1) \,
\bar \rho^{(-)}_{\scriptscriptstyle \rm RR, o}(S_{-\bar I}w^+_3, \nu_2, S_{-\bar J} w^+_1) \Big\rbrace
\end{eqnarray*}
 for $ \# I +\, \# J = 2\mathbb{N} +1 $.
  The basis for the 3-point blocks has the following form:
 \begin{eqnarray}\label{rhoRR}
 \nonumber
 \rho^{(\pm)}_{\scriptscriptstyle \rm RR, e}
	= \rho^{++}_{\scriptscriptstyle RR,e} \pm \rho^{--}_{\scriptscriptstyle RR,e}, \, \
&\qquad&
 \bar \rho^{(\pm)}_{\scriptscriptstyle \rm RR, e}
	= \bar \rho^{++}_{\scriptscriptstyle RR,e} \pm \bar \rho^{--}_{\scriptscriptstyle RR,e} ,
\\[-4pt]
\\[-4pt] \nonumber
 \rho^{(\pm)}_{\scriptscriptstyle \rm RR, o}
	= \rho^{+-}_{\scriptscriptstyle RR,o} \pm i\, \rho^{-+}_{\scriptscriptstyle RR,o},
&\qquad &
 \bar \rho^{(\pm)}_{\scriptscriptstyle \rm RR, o}
	= \bar \rho^{+-}_{\scriptscriptstyle RR,o} \mp i \bar \rho^{-+}_{\scriptscriptstyle RR,o}
\end{eqnarray}
  Introducing the normalized chiral vertex operators
\begin{eqnarray*}
 \bra{\eta_3} V^{(\pm)}_{\scriptscriptstyle \rm RR, f}(\nu|z) \ket{\eta_1}
&=& \rho^{(\pm)}_{\scriptscriptstyle \rm RR, f}(\eta_3,\nu,\eta_1|z)
\end{eqnarray*}
the superprimary field can be written in the diagonal form similar to (\ref{R_NRvertex}):
\begin{eqnarray}\label{phi_RR}
\phi_{\scriptscriptstyle \rm RR}= \,
{C^{(+)} \over \sqrt{2}}
  \left( V_{\scriptscriptstyle \rm RRe}^{(+)} \otimes
  \bar V_{\scriptscriptstyle \rm RRe}^{(+)}
            \,-\, i \,
V_{\scriptscriptstyle \rm RRo}^{(+)} \otimes
\bar V_{\scriptscriptstyle \rm RRo}^{(+)} \right)
+
{C^{(-)} \over \sqrt{2}}
\left( V_{\scriptscriptstyle \rm RRe}^{(-)} \otimes
\bar V_{\scriptscriptstyle \rm RRe}^{(-)}
            \,-\, i \,
V_{\scriptscriptstyle \rm RRo}^{(-)} \otimes
\bar V_{\scriptscriptstyle \rm RRo}^{(-)}\right)
\end{eqnarray}

Let us note that for the 3-point blocks (\ref{rhoRR}) the relations (\ref{e-e,o-o:p})-(\ref{e-o}) imply :
 \begin{eqnarray}\label{rhoRRo-e}
\rho^{(\pm)}_{\scriptscriptstyle \rm RR,o}(S_{J} w^-, \nu,  w^+) &=& \pm i \,
 \rho^{(\pm)}_{\scriptscriptstyle \rm RR,e}(S_{J} w^+, \nu,  w^+)
\\
 \rho^{(\pm)}_{\scriptscriptstyle \rm RR,o}(w^+, \nu, S_{J} w^-) &=&
 \rho^{(\pm)}_{\scriptscriptstyle \rm RR,e}(w^+, \nu, S_{J} w^+)
  \nonumber
  \end{eqnarray}
The similar identities are fulfilled by the $\rho^{(\pm)}_{\scriptscriptstyle \rm NR,f}$,  $\rho^{(\pm)}_{\scriptscriptstyle \rm RN,f}$ \cite{R}:
\begin{eqnarray}\label{rhoNRo-e}
 \rho^{(\pm)}_{\scriptscriptstyle \rm NR,o}(\nu, w^+, S_{J} w^-) &=&
 \rho^{(\pm)}_{\scriptscriptstyle \rm NR,e}(\nu, w^+,  S_{J} w^+),
 \\ \nonumber
 \rho^{(\pm)}_{\scriptscriptstyle \rm RN,o}(S_{J} w^-, w^+,  \nu) &=&
 \rho^{(\pm)}_{\scriptscriptstyle \rm RN,e}(S_{J} w^+, w^+,  \nu) ,
  \end{eqnarray}
Finally, we note that the blocks are functions of $\beta_i$ rather than the Ramond weights.
It follows from (\ref{e-e,o-o:p})-(\ref{e-o}) that the blocks with opposite signs of $\beta_i$
are related:
\begin{eqnarray}\label{3pktBetaRR}
\rho^{(\pm)}_{\scriptscriptstyle \rm RR,e}(S_{J}w^+_4, \nu_3 , S_{K}w^+_{-\beta})
&=& \rho^{(\mp)}_{\scriptscriptstyle \rm RR,e}(S_{J}w^+_4, \nu_3, S_{K}w^+_{\beta}),
\\ \nonumber
\rho^{(\pm)}_{\scriptscriptstyle \rm RR,e}( S_{J}w^+_{-\beta}, \nu_2 ,S_{K}w^+_1)
&=& \rho^{(\mp)}_{\scriptscriptstyle \rm RR,e}(S_{J}w^+_{\beta}, \nu_2, S_{K}w^+_1),
\end{eqnarray}
and  \cite{R}:
\begin{eqnarray}\label{3pktBetaNR}
\rho^{(\pm)}_{\scriptscriptstyle \rm NR,e/o}(S_{J}\nu_4 ,w^+_3, S_{K}w^+_{-\beta})
&=& \rho^{(\mp)}_{\scriptscriptstyle \rm NR,e/o}(S_{J}\nu_4 ,w^+_3, S_{K}w^+_{\beta}),
\\ \nonumber
\rho^{(\pm)}_{\scriptscriptstyle \rm RN,e/o}( S_{K}w^+_{-\beta},w^+_2 ,S_{J}\nu_1)
&=& \rho^{(\mp)}_{\scriptscriptstyle \rm RN,e/o}( S_{K}w^+_{\beta},w^+_2 ,S_{J}\nu_1)
\end{eqnarray}

\section{The 4-point blocks}
\subsection{Definitions}

We shall write the 4-point correlation functions (\ref{RRRRcs1})-(\ref{NNRRcs2}) in terms of the structure constants
and the 4-point superconformal blocks.
Let us start by defining  the 4-point blocks corresponding to the correlators factorized on NS states.
There are four even
\begin{eqnarray}
\label{evenblock4R}
\mathcal{F}^e_{\Delta}
\left[^{\pm \beta_3 \; \pm \beta_2}_{\;\;\;\beta_4
\,\;\;\;\beta_1} \right]
(z)
    &=&
z^{\Delta - \Delta_2 - \Delta_1} \left( 1 +
\sum_{m\in \mathbb{N}} z^m
    F^m_{c, \Delta}
    \left[^{\pm \beta_3 \; \pm \beta_2}_{\;\;\;\beta_4
\,\;\;\;\beta_1} \right]
     \right),
\end{eqnarray}
and four odd
\begin{eqnarray}
\label{oddblock4R}
\mathcal{F}^{o}_{\Delta}
\left[^{\pm \beta_3 \; \pm \beta_2}_{\;\;\;\beta_4
\,\;\;\;\beta_1} \right]
(z)
    &=&
z^{\Delta +{1\over 2}- \Delta_2 - \Delta_1 }
\sum_{k\in \mathbb{N}- \frac{1}{2}} z^{k-{1\over 2}}
        F^k_{c, \Delta}
       \left[^{\pm \beta_3 \; \pm \beta_2}_{\;\;\;\beta_4
\,\;\;\;\beta_1} \right],
\end{eqnarray}
conformal blocks related to correlators of four R primary fields.
The  coefficients are defined in terms of matrix elements of chiral vertex operators
$V^{(\pm)}_{\scriptscriptstyle \rm RN,f}$ and $V^{(\pm)}_{\scriptscriptstyle \rm NR,f}$
\footnote{We have corrected the definition from \cite{R} where the blocks coefficients were written in terms of
 conjugated 3-point blocks $\rho^{(\pm)}_{\scriptscriptstyle \rm NR,|f|}$}:
\begin{eqnarray*}
&&
F^{f}_{c, \Delta}
\left[^{\pm \beta_3 \; \pm \beta_2}_{\;\;\;\beta_4
\,\;\;\;\beta_1} \right] \; =
\hspace*{-20pt}
\begin{array}[t]{c}
{\displaystyle\sum} \\[2pt]
{\scriptstyle
|K|+|M| = |L|+|N| = f
}
\end{array}
\hspace*{-20pt}
\rho^{(\pm)}_{\scriptscriptstyle \mathrm{RN},|f|} (w^+_4 , w^+_{\beta_3} ,\nu_{\Delta,KM})
\ \left[G^{f}_{c, \Delta}\right]^{KM,LN}  \rho^{(\pm)}_{\scriptscriptstyle {\rm NR},|f|}
   (\nu_{\Delta,LN},  w^+_{\beta_2} , w^+_1 ).
\end{eqnarray*}
where ${|f|}={\rm e}$ for $f\in \mathbb{N}$, ${|f|}={\rm o}$ for $f\in \mathbb{N}-{1\over 2}$,
 $\nu_{\Delta,KM}$ is the standard basis in the NS supermodule ${\cal V}_{\Delta}$, and
$\left[G^{f}_{c, \Delta}\right]^{KM,LN}$ denotes the inverse NS Gram matrix.
In order to clarify the notation let us stress that the $\pm$ signs in front of $\beta_2, \beta_3$
in the 4-point blocks denote four different functions of the external parameters $\beta_i$ and are not
related to the actual sings of these arguments. This remark concerns all other types of blocks introduced in this
section.

A correlation function of two NS superprimaries and two R primary fields factorized on the
NS states will be expressed by two even blocks:
\begin{eqnarray*}
\mathcal{F}^{e}_{\Delta}
\left[^{\Delta_3 \; \pm \beta_2}_{\Delta_4 \; \;\;\;\beta_1} \right]
(z)
    &=&
z^{\Delta - \Delta_2 - \Delta_1} \left( 1 +
\sum_{m\in \mathbb{N}} z^m
    F^m_{c, \Delta}
  \left[^{\Delta_3 \; \pm \beta_2}_{\Delta_4 \; \;\;\;\beta_1} \right]
       \right),
\end{eqnarray*}
and two odd blocks
\begin{eqnarray*}
\mathcal{F}^{o}_{\Delta}
\left[^{\Delta_3 \; \pm \beta_2}_{\Delta_4 \; \;\;\;\beta_1} \right](z)
    &=&
z^{\Delta +{1\over 2}- \Delta_2 - \Delta_1 }
\sum_{l\in \mathbb{N}- \frac{1}{2}} z^{l-{1\over 2}}
        F^l_{c, \Delta}
       \left[^{\Delta_3 \; \pm \beta_2}_{\Delta_4 \; \;\;\;\beta_1} \right]
\end{eqnarray*}
where the coefficients are given by:
\begin{eqnarray} \label{4blockRRNN}
F^{f}_{c, \Delta}
\left[^{\Delta_3 \; \pm \beta_2}_{\Delta_4 \; \;\;\;\beta_1} \right]
= \hspace*{-20pt}
\begin{array}[t]{c}
{\displaystyle\sum} \\[2pt]
{\scriptstyle
|K|+|M| = |L|+|N| = f
}
\end{array}
\hspace*{-30pt}
\rho_{\scriptscriptstyle {\rm NN},|f|}(\nu_4 ,\nu_3,\nu_{\Delta,LN} )
\ \left[G^{f}_{c, \Delta}\right]^{KM,LN}
\rho^{(\pm)}_{\scriptscriptstyle {\rm NR},|f|} (\nu_{\Delta,KM}, w^+_{\beta_2}, w^+_1).
\end{eqnarray}

Let us now turn to the blocks with R intermediate states.
There are four even and four odd 4-point blocks contributing to the
correlation function $ \left\langle  \phi R R \phi \right\rangle$  factorized on R states:
\begin{eqnarray*}
\mathcal{F}^{\rm f}_{\beta}
\left[_{\hspace{4pt}\Delta_4 \; \hspace{4pt}\Delta_1}^{\pm\beta_3 \;\pm\beta_2} \right]\!(z)
    &=&
z^{\Delta - \Delta_2 - \Delta_1} \left( 1 +
\sum_{n\in \mathbb{N}} z^n
    F^{n,\mathrm{f}}_{c, \beta}
  \left[_{\hspace{4pt}\Delta_4 \; \hspace{4pt}\Delta_1}^{\pm\beta_3 \;\pm\beta_2} \right]
       \right),
\end{eqnarray*}
where $\mathrm{f} = e$ and $\mathrm{f} = o$ denote blocks with even and odd intermediate states, respectively.
The even and odd coefficients are defined in terms of the 3-point blocks with the corresponding parity:
\begin{eqnarray*} 
&& F^{n,\mathrm{f}}_{c, \beta}
\left[_{\hspace{4pt}\Delta_4 \; \hspace{4pt}\Delta_1}^{\pm\beta_3 \;\pm\beta_2} \right]
= \hspace*{-20pt}
 \begin{array}[t]{c}
{\displaystyle\sum} \\[2pt]
{\scriptstyle
|K|+|M| = |L|+|N| = n}
\end{array}
\hspace*{-30pt}
\rho^{(\pm)}_{\scriptscriptstyle \rm NR,f} (\nu_4,w^+_{\beta_3}, w^+_{\beta,KM})
\ \left[G^{\mathrm{f}, n}_{c, \Delta}\right]^{KM,LN}
\rho^{(\pm)}_{\scriptscriptstyle \rm RN,f}(w^+_{\beta,LN},w^+_{\beta_2}, \nu_1 ),
\end{eqnarray*}
where $w^{+}_{\beta,KM}$ is the standard basis in the R supermodule ${\cal W}_{\Delta}$,
${\textstyle \left[G^{\mathrm{f}, n}_{c, \Delta}\right]^{KM,LN}}$ is the inverse of the Gram matrix
\cite{Meurman:1986gr, Dorrzapf:1999nr}  in the subspaces of the
$ \mathrm{f} $  parity. Each parity subspace can be decomposed into the direct sum of the subspace
spanned by the basis vectors containing $S_0$ and the subspace spanned by all other basis vectors.
Blocks of  the Gram  matrices with respect to these decompositions are related to each other.
 This in order implies some relations between blocks of inverse  Gram  matrices.
Using these relations and the 3-point blocks  property (\ref{rhoNRo-e}) one can show that
the odd and the even 4-point blocks are equal:
$$
\mathcal{F}^{o}_{\beta}
\left[_{\hspace{4pt}\Delta_4 \; \hspace{4pt}\Delta_1}^{\pm\beta_3 \;\pm\beta_2} \right]\!(z)
    =
    \mathcal{F}^{e}_{\beta}
\left[_{\hspace{4pt}\Delta_4 \; \hspace{4pt}\Delta_1}^{\pm\beta_3 \;\pm\beta_2} \right]\!(z).
$$

We shall need two more types of 4-point blocks corresponding to the correlators factorized on
R states \footnote{
Due to clarity we have skipped definition of the rest of 4-point blocks corresponding to correlator of two
NS superprimaries and two R primary fields \emph{i.e.}
${\textstyle F^{n,\mathrm{f}}_{c, \beta}
\left[_{\pm\beta_4\; \hspace{4pt}\Delta_1}^{\hspace{4pt}\Delta_3  \;\pm\beta_2} \right],
F^{n,\mathrm{f}}_{c, \beta}
\left[_{\pm\beta_4\; \Delta_1}^{\pm\beta_3 \; \Delta_2 } \right]}$.
 The recursion representation of these blocks can be found within the universal method discussed in this section.
}:
\begin{eqnarray*} 
&& F^{n,\mathrm{f}}_{c, \beta}
\left[^{\pm \beta_3 \;  \hspace{4pt}\Delta_2}_{ \hspace{4pt}\Delta_4 \; \pm \beta_1} \right]
= \hspace*{-20pt}
 \begin{array}[t]{c}
{\displaystyle\sum} \\[2pt]
{\scriptstyle
|K|+|M| = |L|+|N| = n }
\end{array}
\hspace*{-30pt}
\rho^{(\pm)}_{\scriptscriptstyle \rm NR,f} (\nu_4, w^+_{\beta_3}, w^+_{\beta,KM})
\ \left[G^{\rm{f}, n}_{c, \Delta}\right]^{KM,LN}
 \rho^{(\pm)}_{\scriptscriptstyle \rm RR,f}(w^+_{\beta,LN}, \nu_2 ,w^+_{\beta_1} ).
\\[6pt]
&& F^{n,\mathrm{f}}_{c, \beta}
\left[^{\hspace{4pt}\Delta_3 \;  \hspace{4pt}\Delta_2}_{ \pm \beta_4  \; \pm \beta_1} \right]
= \hspace*{-20pt}
 \begin{array}[t]{c}
{\displaystyle\sum} \\[2pt]
{\scriptstyle
|K|+|M| = |L|+|N| = n }
\end{array}
\hspace*{-30pt}
\rho^{(\pm)}_{\scriptscriptstyle \rm RR,f} ( w^+_{\beta_4},\nu_3, w^+_{\beta,KM})
\ \left[G^{\rm{f}, n}_{c, \Delta}\right]^{KM,LN}
 \rho^{(\pm)}_{\scriptscriptstyle \rm RR,f}(w^+_{\beta,LN}, \nu_2 ,w^+_{\beta_1} ).
\end{eqnarray*}
As in the previous case, due to the properties of 3-point blocks (\ref{rhoNRo-e}) and (\ref{rhoRRo-e}),
 the odd coefficients are proportional to the even ones:
$$
 F^{n,o}_{c, \beta}
\left[^{\pm \beta_3 \;  \hspace{4pt}\Delta_2}_{ \hspace{4pt}\Delta_4 \; \epsilon_1 \beta_1} \right]
    =\, \epsilon_1 \, i \,   F^{n,e}_{c, \beta}
\left[^{\pm \beta_3 \;  \hspace{4pt}\Delta_2}_{ \hspace{4pt}\Delta_4 \; \epsilon_1 \beta_1} \right],
\qquad
 F^{n,o}_{c, \beta}
 \left[^{\hspace{4pt}\Delta_3 \;  \hspace{4pt}\Delta_2}_{ \pm \beta_4  \; \epsilon_1 \beta_1} \right]
 = \epsilon_1 i \,F^{n,e}_{c, \beta}
 \left[^{\hspace{4pt}\Delta_3 \;  \hspace{4pt}\Delta_2}_{ \pm \beta_4  \; \epsilon_1 \beta_1} \right]
  $$
Thus there are four independent even blocks of each type:
\begin{eqnarray*}
\mathcal{F}^{e}_{\beta}
\left[^{\pm \beta_3 \;  \hspace{4pt}\Delta_2}_{ \hspace{4pt}\Delta_4 \; \pm \beta_1} \right]\!(z)
    &=&
z^{\Delta - \Delta_2 - \Delta_1} \left( 1 +
\sum_{n\in \mathbb{N}} z^n
    F^{n, e}_{c, \beta}
 \left[^{\pm \beta_3 \;  \hspace{4pt}\Delta_2}_{ \hspace{4pt}\Delta_4 \; \pm \beta_1} \right]
       \right),
\\[6pt]
\mathcal{F}^{e}_{\beta}
\left[^{\hspace{4pt}\Delta_3 \;  \hspace{4pt}\Delta_2}_{ \pm \beta_4  \; \pm\beta_1} \right]\!(z)
    &=&
z^{\Delta - \Delta_2 - \Delta_1}  \left( 1 +
\sum_{n\in \mathbb{N}} z^n
    F^{n, e}_{c, \beta}
 \left[^{\hspace{4pt}\Delta_3 \;  \hspace{4pt}\Delta_2}_{ \pm \beta_4  \; \pm \beta_1} \right]
       \right).
\end{eqnarray*}

The representations of primary fields through the vertex operators (\ref{phiNN_vertex}), (\ref{R_NRvertex}),
 (\ref{phi_RR}) imply
 the following decompositions of the 4-point correlation functions on the 4-point blocks:
{\small
\begin{eqnarray}
\label{4ptRRRR-block}
&& \hspace{-2cm}
\left\langle
R^+_4(\infty,\infty) R^+_3(1,1) R^+_{2}(z, \bar z) R^+_1(0,0)
\right\rangle
\\ \nonumber
&=& \frac12 \int \mathrm{d}p \sum_{\epsilon_1,\epsilon_2=\pm}\Bigg\{
C^{(\epsilon_1)}_{\scriptscriptstyle [4][3]p} C^{(\epsilon_2)}_{\scriptscriptstyle -p[2][1]}
\left(
    \left|\mathcal{F}^e_{\Delta_p}
    \left[^{ \epsilon_1\beta_3 \; \epsilon_2\beta_2}_{\hspace{7pt}\beta_4 \ \hspace{7pt} \beta_1} \right](z)
    \right|^2
    +
    \left|\mathcal{F}^{o}_{\Delta_p}
    \left[^{ \epsilon_1\beta_3 \; \epsilon_2\beta_2}_{\hspace{7pt}\beta_4 \ \hspace{7pt} \beta_1} \right](z)
    \right|^2
    \right)
\Bigg\}
\\[6pt]
\label{4ptNNRR-block}
&& \hspace{-2cm}
\left\langle
\phi_4(\infty,\infty) \phi_3(1,1) R^+_{2}(z, \bar z) R^+_1(0,0)
\right\rangle
 \\ \nonumber
& =&\int \mathrm{d}p \sum_{\epsilon=\pm}\Bigg\{
C_{\scriptscriptstyle 43p} C^{(\epsilon)}_{\scriptscriptstyle -p[2][1]}
    \left|\mathcal{F}^{e}_{\Delta_p}
    \left[^{\Delta_3 \; \epsilon\beta_2 }_{\Delta_4 \; \;\beta_1 } \right](q)\right|^2
- i \,
\widetilde C_{\scriptscriptstyle 43p}  C^{(\epsilon)}_{\scriptscriptstyle -p[2][1]}
    \left|\mathcal{F}^{o}_{\Delta_p}
    \left[^{\Delta_3 \; \epsilon\beta_2 }_{\Delta_4 \; \;\beta_1 } \right](q)\right|^2
\Bigg\}
\end{eqnarray}
}
 and
{\small
\begin{eqnarray}\label{4ptNRRN-block}
\left\langle
\phi_4(\infty,\infty) R^+_3(1,1) R^+_{2}(z, \bar z) \phi_1(0,0)
\right\rangle
&=&
 \int \mathrm{d}p \sum_{\epsilon_1,\epsilon_2=\pm}
C^{(\epsilon_1)}_{\scriptscriptstyle 4[3][p]} C^{(\epsilon_2)}_{\scriptscriptstyle [-p][2]1}
    \left|\mathcal{F}^{e}_{\beta_p}
    \left[_{\hspace{5pt} \Delta_4 \; \hspace{10pt} \Delta_1}^{\epsilon_1\beta_3 \; -\epsilon_2\beta_2} \right](z)
    \right|^2
\\[6pt] \label{4ptNRNR-block}
\left\langle
\phi_4(\infty,\infty) R^+_3(1,1) \phi_{2}(z, \bar z) R^+_1(0,0)
\right\rangle
&=&
\int \mathrm{d}p \sum_{\epsilon_1,\epsilon_2=\pm}
C^{(\epsilon_1)}_{\scriptscriptstyle 4[3][p]} C^{(\epsilon_2)}_{\scriptscriptstyle [-p]2[1]}
\,
    \left|\mathcal{F}^e_{\beta_p}
    \left[^{ \epsilon_1\beta_3 \; \hspace{5pt} \Delta_2}_{\hspace{5pt}\Delta_4 \; -\epsilon_2\beta_1} \right](z)
    \right|^2
\\[6pt] \label{4ptRNNR-block}
\left\langle
R^+_4(\infty,\infty) \phi_3(1,1)  \phi_{2}(z, \bar z) R^+_1(0,0)
\right\rangle
&=&
\int \mathrm{d}p \sum_{\epsilon_1,\epsilon_2=\pm}
C^{(\epsilon_1)}_{\scriptscriptstyle [4]3[p]} C^{(\epsilon_2)}_{\scriptscriptstyle [-p]2[1]}
\,
    \left|\mathcal{F}^e_{\beta_p}
    \left[^{ \hspace{5pt} \Delta_3\;  \hspace{10pt}\Delta_2}_{\epsilon_1\beta_4 \ -\epsilon_2\beta_1} \right](z)
    \right|^2
\end{eqnarray}
}
Since the 4-point blocks with R intermediate states are functions of  $\beta$ (\ref{beta}),(\ref{momentum}),
 the momentum reflection  should be taken into account (\ref{3pktBetaRR}),(\ref{3pktBetaNR}).

\subsection{Classical limit of the 4-point blocks}

Let us now investigate the behavior of the 4-point blocks in the classical limit
$
b\to 0, \,
2 \pi \mu b^2 \to m=\mathrm{const}.
$
A correlation function of two NS superprimary fields (\ref{superprimary}) and two R primaries  (\ref{Rprimary})
is defined by the path integral with the action (\ref{SLFT}):
\begin{equation}\label{cor}
\langle R^+_{4} \phi_{3}  \phi_{2} R^+_{1} \rangle =
\int\limits {\cal D}\phi_{\scriptscriptstyle SL} {\cal D}\psi_{\scriptscriptstyle SL}
{\cal D}\bar\psi_{\scriptscriptstyle SL}\;
{\rm e}^{-{\cal S}_{\rm\scriptscriptstyle SL}[\phi_{\scriptscriptstyle SL},\psi_{\scriptscriptstyle SL}]}
\sigma^+ \, {\rm e}^{a_4\phi_{\scriptscriptstyle SL}}  {\rm e}^{a_3\phi_{\scriptscriptstyle SL}}
 {\rm e}^{a_2\phi_{\scriptscriptstyle SL}} \sigma^+ \, {\rm e}^{a_1\phi_{\scriptscriptstyle SL}}
\end{equation}
The twist fields have light weights and thus they do not contribute to the classical limit.
If all the exponential operators have heavy weights $\Delta_{a_i}$:
$$ a_i = \frac{Q}{2}(1-\lambda_i), \quad b a_i \to \frac{1-\lambda_i}{2},
\qquad 2b^2 \Delta_i \to \delta_i = \frac{1-\lambda_i^2}{4},
$$
 the correlator has the same asymptotic behavior as the bosonic 4-point function of primary fields:
 $$
 \langle R^+_{4} \phi_{3}  \phi_{2} R^+_{1} \rangle \sim {\rm e}^{-{1\over 2 b^2}
    S_{\rm\scriptscriptstyle cl}[\delta_4,\delta_3,\delta_2,\delta_1]},
 $$
 where $S_{\rm\scriptscriptstyle cl}[\delta_4,\delta_3,\delta_2,\delta_1]$ is the bosonic  Liouville
action
$$
S[\phi]=\frac{1}{2\pi}\int
\left(\left|\partial\phi\right|^2
+
 m^2{\rm e}^{2\phi}\right) d^2z,
$$
calculated on the classical configuration $\varphi$ satisfying the Liouville equation with sources
$$
\partial\bar\partial \varphi - {m^2}{\rm e}^{2\phi}=\sum\limits_1^4 {1-\lambda_i\over 4}\delta(z-z_i).
$$
The asymptotic behavior of the 3-point structure constants reads:
\begin{eqnarray}\label{3ptClass}
 \langle \phi_{3} \phi_{2} \phi_{1} \rangle
 &\sim& \mathrm{e}^{-\frac{1}{2b^2} S_{cl}[\delta_3,\delta_2,\delta_1]},
 \qquad
 \langle \phi_{3} \widetilde \phi_{2} \phi_{1} \rangle
 \,\sim\, \frac{1}{b^2} \mathrm{e}^{-\frac{1}{2b^2} S_{cl}[\delta_4,\delta_3,\delta_p]},
 \\\nonumber
\langle \phi_{3} R^{\pm}_{2} R^{\pm}_{1} \rangle &\sim& {\rm e}^{-{1\over 2 b^2}
    S_{\rm\scriptscriptstyle cl}[\delta_3,\delta_2,\delta_1]},
\end{eqnarray}
where $S_{\rm\scriptscriptstyle cl}[\delta,\delta_2,\delta_1]$ is the 3-point classical bosonic Liouville action.
The $b^{-2}$ coefficient in the second relation arises due to the fermionic contribution from $\widetilde \phi_{2}$ field (\ref{NSprimaries}) \cite{Hadasz:2007nt}.

A similar reasoning applied to the 4-point correlator (\ref{cor}) projected on
an even-even  subspace of  ${\cal W}_\Delta \otimes \bar{\cal W}_{\bar\Delta}$ yields:
$$
\langle R^+_{4} \phi_{3} \mid_{\Delta}^{ee} \phi_{2}R^+_{1} \rangle \sim {\rm e}^{-{1\over 2 b^2}
    S_{\rm\scriptscriptstyle cl}[\delta_4,\delta_3,\delta_2,\delta_1|\delta]},
$$
where the "$\Delta$-projected" classical action is given by
$$
S_{\rm\scriptscriptstyle cl}[\delta_4,\delta_3,\delta_2,\delta_1|\delta]=
S_{\rm\scriptscriptstyle cl}[\delta_4,\delta_3,\delta]
+S_{\rm\scriptscriptstyle cl}[\delta,\delta_2,\delta_1]
-
 f_{\delta}
\!\left[_{\delta_{4}\;\delta_{1}}^{\delta_{3}\;\delta_{2}}\right]
\!(z)
- \bar f_{\delta}
\!\left[_{\delta_{4}\;\delta_{1}}^{\delta_{3}\;\delta_{2}}\right]
\!(\bar z)
$$
 and $f_{\delta}
\!\left[_{\delta_{4}\;\delta_{1}}^{\delta_{3}\;\delta_{2}}\right]
\!(z)$
is the classical conformal block \cite{Zamolodchikov:3}.
The 4-point correlator projected on an even-even (or odd-odd) subspace of
 ${\cal V}_\Delta \otimes \bar{\cal V}_{\bar\Delta}$ has the same behavior:
$$
\langle \phi_{4} \phi_{3} \mid_{\Delta}^{ee} R^+_{2} R^+_{1} \rangle \sim {\rm e}^{-{1\over 2 b^2}
    S_{\rm\scriptscriptstyle cl}[\delta_4,\delta_3,\delta_2,\delta_1|\delta]},
\qquad \langle \phi_{4} \phi_{3} \mid_{\Delta}^{oo} R^+_{2} R^+_{1} \rangle \sim {\rm e}^{-{1\over 2 b^2}
    S_{\rm\scriptscriptstyle cl}[\delta_4,\delta_3,\delta_2,\delta_1|\delta]}.
$$
Thus the equations (\ref{4ptRRRR-block})-(\ref{4ptRNNR-block}) together with  (\ref{3ptClass})
lead to the following
asymptotical behavior of the 4-point blocks:
\begin{eqnarray}
\label{classRNNR}
 && \mathcal{F}^{1}_{\beta}
    \left[_{\hspace{5pt} \Delta_4 \; \hspace{5pt} \Delta_1}^{\pm\beta_3 \ \pm\beta_2} \right]\!(z)
\sim  \mathrm{e}^{\frac{1}{2b^2} f_{\delta}
    \left[^{\delta_3 \; \delta_2}_{\delta_4 \;\delta_1} \right](z)}, \qquad
\mathcal{F}^{1}_{\beta}
    \left[^{ \pm\beta_3 \; \hspace{5pt} \Delta_2}_{\hspace{5pt}\Delta_4 \ \pm\beta_1} \right]\!(z)
\sim  \mathrm{e}^{\frac{1}{2b^2} f_{\delta}
    \left[^{\delta_3 \; \delta_2}_{\delta_4 \;\delta_1} \right](z)},
\\ \nonumber
&&
\mathcal{F}^{1}_{\beta}
    \left[^{ \hspace{5pt}\Delta_4 \; \hspace{5pt} \Delta_2}_{\pm\beta_3 \ \pm\beta_1} \right]\!(z)
\sim  \mathrm{e}^{\frac{1}{2b^2} f_{\delta}
    \left[^{\delta_3 \; \delta_2}_{\delta_4 \;\delta_1} \right](z)},
\\[4pt] \label{classRRNN}
&&\mathcal{F}^1_{\Delta}
    \left[^{\Delta_3 \; \pm \beta_2}_{\Delta_4 \; \;\;\;\beta_1} \right](z)
\sim  \mathrm{e}^{\frac{1}{2b^2} f_{\delta}
    \left[^{\delta_3 \; \delta_2}_{\delta_4 \;\delta_1} \right](z)}, \qquad
\mathcal{F}^{\frac12}_{\Delta}
    \left[^{\Delta_3 \; \pm \beta_2}_{\Delta_4 \; \;\;\;\beta_1} \right](z)
\sim  \textstyle{b \over \delta} \,  \mathrm{e}^{\frac{1}{2b^2} f_{\delta}
    \left[^{\delta_3 \; \delta_2}_{\delta_4 \;\delta_1} \right](z)}.
 \end{eqnarray}
The coefficient in the last term can be derived analyzing the leading
 $\Delta$ dependence of the odd 3-point blocks with an arbitrary NS state from level $k$  \cite{Hadasz:2007nt,R}
$$
\rho_{\scriptscriptstyle \rm NN,o}(\nu_4 ,\nu_3,\nu_{\Delta,LN} )
\sim \beta_i \Delta^{k-\frac12}+ \ldots, \qquad
\rho^{(\pm)}_{\scriptscriptstyle \rm NR,o} (\nu_{\Delta,KM}, w^+_{\beta_2}, w^+_1)
 \sim \Delta^{k-\frac12}+ \ldots
$$

\subsection{Elliptic recurrence for the blocks with an NS intermediate weight}

It follows from the definitions that the coefficients of any type of 4-point blocks with an NS intermediate
weight are polynomials in the
external weights $\Delta_i$ or $\beta_i$ and rational functions of $\Delta$ and $c$.
The properties of the inverse Gram matrix imply that any coefficient can be  expressed as a sum over
simple poles in $\Delta$ or in the central charge.
In particular, for the coefficients (\ref{4blockRRNN}) one has:
\begin{eqnarray*}
F^{k}_{c, \Delta}\!
\left[^{\Delta_3 \; \pm\beta_2 }_{\Delta_4 \; \;\;\;\beta_1} \right]
&=&
h^{k}_{c, \Delta}
\left[^{\Delta_3 \; \pm \beta_2}_{\Delta_4 \; \;\;\;\beta_1} \right] +
 \begin{array}[t]{c}
{\displaystyle\sum} \\[2pt]
{\scriptstyle
1 \leq rs \leq 2k} \\[-4pt]
{\scriptstyle r+ s \in 2 \mathbb{N}
}
\end{array}
\frac{R^{k}_{c,rs}
\left[^{\Delta_3 \; \pm \beta_2}_{\Delta_4 \; \;\;\;\beta_1} \right]}{\Delta-\Delta_{rs}(c)}
\end{eqnarray*}
 with $\Delta_{rs}(c)$ given by the Kac determinant formula for NS Verma modules:
\begin{eqnarray*}
\Delta_{rs}(c)
& = &
-\frac{rs-1}{4} + \frac{1-r^2}{8}b^2 + \frac{1-s^2}{8}\frac{1}{b^2}\,,\hskip 10mm
c=\frac{3}{2} +3\left(b+{1\over b}\right)^2.
\end{eqnarray*}
In order to calculate the residues it is convenient to choose a specific basis in the NS module \cite{Hadasz:2007nt}.
Let us remind its construction. First, one introduces the states:
\begin{eqnarray*}
\chi_{rs}^{\Delta} &=& \sum_{|K|+|M|=\frac{rs}{2}} \chi_{rs}^{KM} S_{-K} L_{-M} \nu_{\Delta}
\end{eqnarray*}
where
$\chi_{rs}^{KM} $  are the coefficients of the singular vector
in the standard basis of ${\cal V}_{\Delta_{rs}}$:
\begin{eqnarray*}
  \chi_{rs} &=& \sum_{|K|+|M|=\frac{rs}{2}} \chi_{rs}^{KM} S_{-K}L_{-M} \nu_{\Delta_{rs}}
\qquad \rm{for} \ r+s \in 2 \mathbb{N},
\end{eqnarray*}
The family of states $\left\{S_{-L}L_{-N} \chi_{rs}^{\Delta}\right\}_{|L|+|N|=n-\frac{rs}{2}}$
can be completed to a full basis in the NS module ${\cal V}_{\Delta}$ at the level $n>\frac{rs}{2}$.
Working in such a basis in the NS module one obtains:
\begin{eqnarray}
\label{res:RRNN}
&& \hspace{-0.5cm}
{\mathcal R}^{{k}}_{c,\,rs}\!
\left[^{\Delta_3 \; \pm \beta_2}_{\Delta_4 \; \;\;\;\beta_1} \right]
\,=\, \lim_{\Delta \to \Delta_{rs}} (\Delta - \Delta_{rs}(c)) \, F^{k}_{c, \Delta}\!
\left[^{\Delta_3 \; \pm \beta_2}_{\Delta_4 \; \;\;\;\beta_1} \right]
=  A_{rs}(c) \, \times\\ [4pt] \nonumber
&&\times \hspace*{-20pt}
\begin{array}[t]{c}
{\displaystyle\sum} \\[2pt]
{\scriptstyle
|K|+|M| = |L|+|N| = k-\frac{rs}{2}
}
\end{array}
\hspace*{-30pt}
   \rho_{\scriptscriptstyle \rm NN,|k|}(\nu_4 , \nu_3, S_{-L}L_{-N}\chi_{rs} )
\,
 \left[G^{{k}-\frac{rs}{2}}_{c, \Delta_{rs}+\frac{rs}{2}}\right]^{KM,LN} \!
  \rho^{(\pm)}_{\scriptscriptstyle \rm NR,|k|}(S_{-K}L_{-M}\chi_{rs},w^+_2, w^+_1 ) ,
\end{eqnarray}
where the coefficient $A_{rs}(c)$ is given by:
\begin{eqnarray*}
A_{rs}(c)
\; = \;
\lim_{\Delta\to\Delta_{rs}}
\left(\frac{\left\langle\chi_{rs}^\Delta|\chi_{rs}^\Delta\right\rangle}{\Delta - \Delta_{rs}(c)}
\right)^{-1}.
\end{eqnarray*}
The exact formula for $A_{rs}(c)$ is due to A.~Belavin and Al.~Zamolodchikov \cite{Heqn}:
\begin{eqnarray*}
A_{rs}(c)
\; = \;
2^{rs-2}
\prod_{m=1-r}^r
\prod_{n=1-s}^s
\left(p b + \frac{q}{b}\right)^{-1}
\end{eqnarray*}
where $m+n \in 2{\mathbb Z}, \; (m,n) \neq (0,0),(r,s).$
It was proposed on the basis of  higher equations of motion in $N=1$ SLFT.
The corresponding formula in the bosonic case was recently proved  in \cite{Yanagida:2010qm}.

Due to the factorization property of the 3-point blocks (\ref{3ptFactorNS}) the residue (\ref{res:RRNN}) is proportional to a
coefficient of the same  block
\begin{eqnarray}\label{res:RRNNe}
{\mathcal R}^{{k}}_{c,\,rs}\!
\left[^{\Delta_3 \; \pm \beta_2}_{\Delta_4 \; \;\;\;\beta_1} \right]
=  A_{rs}(c) \
F^{k-\frac{rs}{2}}_{c, \Delta_{rs}+\frac{rs}{2}}\!
\left[^{\Delta_3 \; \pm \beta_2}_{\Delta_4 \; \;\;\;\beta_1} \right]
\, P^{rs}_{c}\!\left[^{\pm\beta_2}_{\;\;\; \beta_1} \right]
\, \times \,  \left\{
\begin{array}{rcl}
 && \hspace{-25pt} P^{rs}_c\!\left[^{\Delta_3}_{\Delta_4} \right]
 \qquad k \in \mathbb{N},
 \\ [4pt]
 && \hspace{-25pt} P^{rs}_c\!\left[^{\ast\Delta_3}_{\ \Delta_4} \right]
 \qquad k \in \mathbb{N}-\frac12,
  \end{array}
\right.
\end{eqnarray}
for $\frac{rs}{2} \in \mathbb{N}$, and to a coefficient of another block
\begin{eqnarray}\label{res:RRNNo}
 {\mathcal R}^{{k}}_{c,\,rs}\!
\left[^{\Delta_3 \; \pm \beta_2}_{\Delta_4 \; \;\;\;\beta_1} \right]
=  A_{rs}(c) \
F^{k-\frac{rs}{2}}_{c, \Delta_{rs}+\frac{rs}{2}}\!
\left[^{\Delta_3 \; \mp \beta_2}_{\Delta_4 \; \;\;\;\beta_1} \right]
\, P^{rs}_{c}\!\left[^{\pm\beta_2}_{\;\;\; \beta_1} \right]
\, \times \,
 \left\{
\begin{array}{rcl}
 && \hspace{-25pt} {\rm e}^{  i \frac{\pi}{4}} \
P^{rs}_c\!\left[^{\Delta_3}_{\Delta_4} \right]
 \qquad k \in \mathbb{N},
 \\ [4pt]
 && \hspace{-25pt} {\rm e}^{ -i \frac{\pi}{4}}
 \ P^{rs}_c\!\left[^{\ast\Delta_3}_{\ \Delta_4} \right]
 \qquad k \in \mathbb{N}-\frac12,
 \end{array}
\right.
\end{eqnarray}
for $\frac{rs}{2} \in \mathbb{N} -\frac12 $. The fusion polynomials
$ P^{rs}_c\!\left[^{\Delta_\imath}_{\Delta_\jmath} \right],
P^{rs}_c\!\left[^{\ast\Delta_\imath}_{\ \Delta_\jmath} \right],
 P^{rs}_{c}\!\left[^{\pm\beta_\imath}_{\;\;\; \beta_\jmath} \right]
$ are  given by formulae (\ref{fusionNS}).

In order to derive a closed elliptic recurrence for blocks' coefficients one has to investigate
the large $\Delta$ asymptotic.
According to Zamolodchikov's reasoning \cite{Zamolodchikov:2, Zamolodchikov:3}
the $\Delta_i$ and $c$ dependence of the first two terms in the $\frac{1}{\Delta}$
 expansion can be read from the classical limit of the superconformal blocks.
 From the path-integral arguments it follows that in the classical limit
 the bosonic classical block occurs (\ref{classRRNN}).
  It yields  the large $\Delta$ asymptotic  in the form:
\begin{eqnarray}
\label{asymptoticG1}
&& \hskip -1cm \ln
\mathcal{F}_{\Delta}^{e}\!
        \left[^{\Delta_3 \; \pm \beta_2}_{\Delta_4 \; \;\;\;\beta_1} \right]
         (z)
         =
\pi \tau \left( \Delta - \frac{c}{24} \right)
+ \left( \frac{c}{8} - \Delta_1 - \Delta_2 - \Delta_3 - \Delta_4\right) \, \ln{K^2(z)}
\\[6pt] \nonumber
  &&
+ \left( \frac{c}{24} - \Delta_2 - \Delta_3\right)  \, \ln(1-z)
+  \left( \frac{c}{24} - \Delta_1 - \ \Delta_2\right)  \, \ln(z) + f^{\pm}(z) +{\cal O}\left({1\over \Delta}\right),
\end{eqnarray}
for the even blocks, and:
\begin{eqnarray*}
&& \hskip -1cm \ln
\mathcal{F}_{\Delta}^{o}\!
        \left[^{\Delta_3 \; \pm \beta_2}_{\Delta_4 \; \;\;\;\beta_1} \right]
         (z)
         = -\ln \Delta
+\pi \tau \left( \Delta - \frac{c}{24} \right)
+ \left( \frac{c}{8} - \Delta_1 - \Delta_2 - \Delta_3 - \Delta_4\right) \, \ln{K^2(z)} \\
 [6pt] &&
+ \left( \frac{c}{24} - \Delta_2 - \Delta_3\right)  \, \ln(1-z)
+  \left( \frac{c}{24} - \Delta_1 - \ \Delta_2\right)  \, \ln(z) +{\cal O}\left({1\over \Delta}\right),
\end{eqnarray*}
for the odd blocks.
 $f^{\pm}(z)$ are functions specific for each type of block and independent of $\Delta_i$ and $c$.

Introducing the multiplicative factor which captures  all the $\Delta_i$ and $c$ dependence
of non-singular terms one defines the  elliptic blocks:
\begin{eqnarray}\label{elliptic_NNRR}
\nonumber
 \mathcal{F}^{e,o}_{\Delta}\!
\left[^{\Delta_3 \; \pm \beta_2}_{\Delta_4 \; \;\;\;\beta_1} \right]\!(z)
 \!& =&\!
(16q)^{\Delta - \frac{c-3/2}{24}}\ z^{\frac{c-3/2}{24} - \Delta_1 -\Delta_2} \
 (1- z)^{\frac{c-3/2}{24} - \Delta_2 - \Delta_3+\frac{1}{16}}\
\\
 \!& \times & \! \theta_3^{\frac{c - 3/2}{2}
- 4 (\Delta_1 +\Delta_2 +\Delta_3 + \Delta_4)+ \frac12 }  \
 \mathcal{H}^{e,o}_{\Delta}\! \left[^{\Delta_3 \; \pm \beta_2}_{\Delta_4 \; \;\;\;\beta_1} \right]\!(q).
\end{eqnarray}
The elliptic blocks
$\mathcal{H}^{e,o}_{\Delta}\! \left[^{\Delta_3 \; \pm \beta_2}_{\Delta_4 \; \;\;\;\beta_1} \right]\!(q)$
   can be written as sums over simple poles in $\Delta$.
The residues are given by the corresponding residues of the superconformal blocks (\ref{res:RRNNe}),(\ref{res:RRNNo}):
\begin{eqnarray}\label{H_NNRRe_rek}
\nonumber
 \mathcal{H}^{e}_{\Delta}\! \left[^{\Delta_3 \; \pm \beta_2}_{\Delta_4 \; \;\;\;\beta_1} \right]\!(q)
=
 g_{\pm}
 \! &+&\!
 \begin{array}[t]{c}
{\displaystyle\sum} \\[-5pt]
{\scriptscriptstyle
r,s>0}
\\[-7pt]
{\scriptscriptstyle
r,s\in 2{\mathbb N}
}
\end{array}
 \hspace{-7pt} (16q)^{\frac{rs}{2}}
 \frac{ A_{rs}(c)
P^{rs}_{c}\!\left[^{\Delta_3}_{\Delta_4}\right]
P^{rs}_{c}\!\left[^{\pm\beta_2}_{\;\;\; \beta_1} \right]
}{\Delta - \Delta_{rs}} \,
\mathcal{H}^{e}_{\Delta_{rs}+\frac{rs}{2}}\!
\left[^{\Delta_3 \; \pm \beta_2}_{\Delta_4 \; \;\;\;\beta_1} \right]\!(q)
 \\[-6pt]
 \\[-4pt] \nonumber
&&  + \hspace{-9pt} \begin{array}[t]{c}
{\displaystyle\sum} \\[-5pt]
{\scriptscriptstyle
r,s>0}
\\[-7pt]
{\scriptscriptstyle
r,s\in 2{\mathbb N}+1
}
\end{array}
\hspace{-12pt} (16q)^{\frac{rs}{2}}
\frac{  {\rm e}^{ i \frac{\pi}{4}} \,  A_{rs}(c) \,
P^{rs}_{c}\!\left[^{\Delta_3}_{\Delta_4}\right]
P^{rs}_{c}\!\left[^{\pm\beta_2}_{\;\;\; \beta_1} \right]}{\Delta - \Delta_{rs}} \,
\mathcal{H}^{o}_{\Delta_{rs}+\frac{rs}{2}}\!
\left[^{\Delta_3 \; \mp \beta_2}_{\Delta_4 \; \;\;\;\beta_1} \right]\!(q)
\end{eqnarray}
and
\begin{eqnarray}\label{H_NNRRo_rek}
\nonumber
 \mathcal{H}^{o}_{\Delta}\! \left[^{\Delta_3 \; \pm \beta_2}_{\Delta_4 \; \;\;\;\beta_1} \right]\!(q)
&=&
 \begin{array}[t]{c}
{\displaystyle\sum} \\[-5pt]
{\scriptscriptstyle
r,s>0}
\\[-7pt]
{\scriptscriptstyle
r,s\in 2{\mathbb N}
}
\end{array}
 \hspace{-7pt} (16q)^{\frac{rs}{2}}
 \frac{ A_{rs}(c)
P^{rs}_{c}\!\left[^{\ast\Delta_3}_{\ \Delta_4}\right]
P^{rs}_{c}\!\left[^{\pm\beta_2}_{\;\;\; \beta_1} \right]}{\Delta - \Delta_{rs}} \,
\mathcal{H}^{o}_{\Delta_{rs}+\frac{rs}{2}}\!
\left[^{\Delta_3 \; \pm \beta_2}_{\Delta_4 \; \;\;\;\beta_1} \right]\!(q)
  \\[-6pt]
 \\[-4pt] \nonumber
& +& \hspace{-9pt} \begin{array}[t]{c}
{\displaystyle\sum} \\[-5pt]
{\scriptscriptstyle
r,s>0}
\\[-7pt]
{\scriptscriptstyle
r,s\in 2{\mathbb N}+1
}
\end{array}
\hspace{-12pt} (16q)^{\frac{rs}{2}}
\frac{   {\rm e}^{  - i \frac{\pi}{4}} \, A_{rs}(c) \,
P^{rs}_{c}\!\left[^{\ast\Delta_3}_{\ \Delta_4}\right]
P^{rs}_{c}\!\left[^{\pm\beta_2}_{\;\;\; \beta_1} \right]}{\Delta - \Delta_{rs}} \,
\mathcal{H}^{e}_{\Delta_{rs}+\frac{rs}{2}}\!
\left[^{\Delta_3 \; \mp \beta_2}_{\Delta_4 \; \;\;\;\beta_1} \right]\!(q)
\end{eqnarray}
Since the functions $g_{\pm} $ are independent of $\Delta_i$ and central charge they can be read off
from the special $c=\frac32$ blocks. The explicit expressions for the $c=\frac32$ blocks
with the NS external weights $\Delta_1 = \Delta_2 = \frac18$ and the R external weights
$\Delta_3 = \Delta_4 = \frac{1}{16}$ can be calculated using the techniques
of the chiral superscalar model \cite{Hadasz:2007ns}. In the present case it yields
$$
\lim_{\beta_0\to 0}\, \lim_{2\alpha \to i} \,
 \lim_{b\to i}\, \mathcal{F}^{e}_{\Delta}\!
\left[^{ \Delta_a  \; \pm\beta_0}_{\Delta_a\; \;\;\beta_0 } \right]\!(z)
= (16q)^{\Delta} \left[z(1-z)\right]^{-\frac18} \theta_3^{-1}(q),
$$
which gives
$$
g_{\pm} = 1.
$$

The recursive representation of the 4-point blocks  corresponding to a correlator of four R fields
(\ref{evenblock4R}),(\ref{oddblock4R})
can be derived in the same way and it reads \footnote{We have corrected the corresponding formulae of \cite{R}}:
\begin{eqnarray}
\label{elliptic-RRRR}
  \mathcal{F}^{e/o}_{\Delta}\!
\left[^{\pm \beta_3 \; \pm \beta_2}_{\;\;\;\beta_4 \,\;\;\;\beta_1} \right] (z)
 & =&
(16q)^{\Delta - \frac{c-3/2}{24}}\ z^{\frac{c-3/2}{24} - \Delta_1 -\Delta_2} \
 (1- z)^{\frac{c-3/2}{24} - \Delta_2 - \Delta_3}\
\\
\nonumber
 & \times &\theta_3^{\frac{c - 3/2}{2}
- 4 (\Delta_1 +\Delta_2 +\Delta_3 + \Delta_4) }  \
 \mathcal{H}^{e/o}_{\Delta}\! \left[^{\pm \beta_3 \; \pm \beta_2}_{\;\;\;\beta_4 \,\;\;\;\beta_1} \right]\!(z),
\end{eqnarray}
\begin{eqnarray}\label{H_RRRR_rek}
\nonumber
 \mathcal{H}^{e/o}_{\Delta}\! \left[^{\pm \beta_3 \; \pm \beta_2}_{ \;\;\; \beta_4 \; \;\;\;\beta_1} \right]\!(q)
=
g^{e/o}
 \! &+&\!
 \begin{array}[t]{c}
{\displaystyle\sum} \\[-5pt]
{\scriptscriptstyle
r,s>0}
\\[-7pt]
{\scriptscriptstyle
r,s\in 2{\mathbb N}
}
\end{array}
 \hspace{-7pt} (16q)^{\frac{rs}{2}}
 \frac{ A_{rs}(c)
P^{rs}_{c}\!\left[^{\pm\beta_3}_{\;\;\; \beta_4}\right]
P^{rs}_{c}\!\left[^{\pm\beta_2}_{\;\;\; \beta_1} \right]
}{\Delta - \Delta_{rs}} \,
\mathcal{H}^{e/o}_{\Delta_{rs}+\frac{rs}{2}}\!
\left[^{\pm \beta_3 \; \pm \beta_2}_{ \;\;\; \beta_4 \; \;\;\;\beta_1} \right]\!(q)
 \\[-6pt]
 \\[-6pt] \nonumber
\! &-&\! \hspace{-9pt} \begin{array}[t]{c}
{\displaystyle\sum} \\[-5pt]
{\scriptscriptstyle
r,s>0}
\\[-7pt]
{\scriptscriptstyle
r,s\in 2{\mathbb N}+1
}
\end{array}
\hspace{-7pt} (16q)^{\frac{rs}{2}}
\frac{   A_{rs}(c) \,
P^{rs}_{c}\!\left[^{\pm\beta_3}_{\;\;\; \beta_4}\right]
P^{rs}_{c}\!\left[^{\pm\beta_2}_{\;\;\; \beta_1} \right]}{\Delta - \Delta_{rs}} \,
\mathcal{H}^{o/e}_{\Delta_{rs}+\frac{rs}{2}}\!
\left[^{\mp \beta_3 \; \mp \beta_2}_{ \;\;\; \beta_4 \; \;\;\;\beta_1} \right]\!(q)
\end{eqnarray}
where
$$ g^{e} = 1, \qquad g^{o}=0.
$$

\subsection{Elliptic recurrence for the blocks with  R intermediate states}

The coefficients of the 4-point blocks with R intermediate states
are polynomials in the external weights $\Delta_i$ or $\beta_i$ and rational functions of $c$
and the intermediate $\beta$. The inverse Gram matrix as a function of the intermediate weight has simple poles at:
\begin{eqnarray*}
\Delta_{rs}(c)
& = &
 \frac{c}{24} - \beta_{rs}^2(c), \qquad \beta_{rs}(c) = \frac{1}{2\sqrt{2}} \left(r b+ \frac{s}{b} \right)
\end{eqnarray*}
where  $r,s\in \mathbb{N}$ and the sum $r+s$ must be \emph{odd}.
Thus the 4-point blocks can be expressed as the following sums over simple poles in $\beta$:
\begin{eqnarray*}
F^{n,e}_{c, \beta}
\left[_{\hspace{4pt}\Delta_4 \; \hspace{4pt}\Delta_1}^{\pm\beta_3 \;\pm\beta_2} \right]
&=&
h^{n,e}_{c, \beta}
\left[_{\hspace{4pt}\Delta_4 \; \hspace{4pt}\Delta_1}^{\pm\beta_3 \;\pm\beta_2} \right]
 +
 \begin{array}[t]{c}
{\displaystyle\sum} \\[2pt]
{\scriptstyle
1< rs \leq 2n} \\[-4pt]
{\scriptstyle r+ s \in 2 \mathbb{N} +1
}
\end{array}
\frac{\mathcal{R}^{n,+}_{rs}
\left[_{\hspace{4pt}\Delta_4 \; \hspace{4pt}\Delta_1}^{\pm\beta_3 \;\pm\beta_2} \right] }{\beta-\beta_{rs}(c)}
+ \frac{\mathcal{R}^{n,-}_{rs}
\left[_{\hspace{4pt}\Delta_4 \; \hspace{4pt}\Delta_1}^{\pm\beta_3 \;\pm\beta_2} \right] }{\beta+\beta_{rs}(c)}
\\ [4pt]
F^{n,e}_{c, \beta}
\left[^{\pm \beta_3 \;  \hspace{4pt}\Delta_2}_{ \hspace{4pt}\Delta_4 \; \pm \beta_1} \right]
&=&
h^{n,e}_{c, \beta}
\left[^{\pm \beta_3 \;  \hspace{4pt}\Delta_2}_{ \hspace{4pt}\Delta_4 \; \pm \beta_1} \right]
 +
 \begin{array}[t]{c}
{\displaystyle\sum} \\[2pt]
{\scriptstyle
1< rs \leq 2n} \\[-4pt]
{\scriptstyle r+ s \in 2 \mathbb{N} +1
}
\end{array}
\frac{\mathcal{R}^{n,+}_{rs}
\left[^{\pm \beta_3 \;  \hspace{4pt}\Delta_2}_{ \hspace{4pt}\Delta_4 \; \pm \beta_1} \right]}{\beta-\beta_{rs}(c)}
+\frac{\mathcal{R}^{n,-}_{rs}
\left[^{\pm \beta_3 \;  \hspace{4pt}\Delta_2}_{ \hspace{4pt}\Delta_4 \; \pm \beta_1} \right]}{\beta+\beta_{rs}(c)}
\\[4pt]
F^{n,e}_{c, \beta}
\left[^{\hspace{4pt}\Delta_3 \; \hspace{4pt}\Delta_2}_{\pm\beta_4 \;\pm\beta_1} \right]
&=&
h^{n,e}_{c, \beta}
\left[^{\hspace{4pt}\Delta_3 \; \hspace{4pt}\Delta_2}_{\pm\beta_4 \;\pm\beta_1} \right]
 +
 \begin{array}[t]{c}
{\displaystyle\sum} \\[2pt]
{\scriptstyle
1< rs \leq 2n} \\[-4pt]
{\scriptstyle r+ s \in 2 \mathbb{N} +1
}
\end{array}
\frac{\mathcal{R}^{n,+}_{rs}
\left[^{\hspace{4pt}\Delta_3 \; \hspace{4pt}\Delta_2}_{\pm\beta_4 \;\pm\beta_1} \right] }{\beta-\beta_{rs}(c)}
+ \frac{\mathcal{R}^{n,-}_{rs}
\left[^{\hspace{4pt}\Delta_3 \; \hspace{4pt}\Delta_2}_{\pm\beta_4 \;\pm\beta_1} \right]] }{\beta+\beta_{rs}(c)}
\end{eqnarray*}
In order to determine the residues at $\beta_{rs}$ and $-\beta_{rs}$ we need to choose
a specific basis in the R module.
Let us introduce the states:
\begin{eqnarray*}
\chi_{rs}^{\beta} &=& \sum_{|K|+|M|=\frac{rs}{2}} \chi_{rs}^{+,KM} S_{-K} L_{-M} w^+_{\beta}
\qquad \rm{for} \ r+s \in 2 \mathbb{N}+1,
\end{eqnarray*}
where
 $\chi_{rs}^{+,KM}$  are the coefficients of the singular vector
in the standard basis of ${\cal W}_{\Delta_{rs}}$:
\begin{eqnarray*}
\chi^+_{rs} &=& \sum_{|K|+|M|=\frac{rs}{2}} \chi_{rs}^{+,KM} S_{-K} L_{-M} w^+_{\beta_{rs}}.
\end{eqnarray*}
The family of states
$\left\{S_{-L}L_{-N} \chi_{rs}^{\beta}\right\}_{|L|+|N|=n-\frac{rs}{2}}$   can be completed to a full basis in the
R module ${\cal W}_{\Delta}$ at the level $n>\frac{rs}{2}$.
Working in such a basis one can compute the residues:
\begin{eqnarray}
\label{res:NRRN+}
&& \hspace{-0.5cm}
\mathcal{R}^{n,+}_{rs}
\left[_{\hspace{4pt}\Delta_4 \; \hspace{4pt}\Delta_1}^{\pm\beta_3 \;\pm\beta_2} \right]
= \lim_{\beta \to \beta_{rs}} (\beta - \beta_{rs}(c)) \,
F^{n,e}_{c, \beta}
\left[_{\hspace{4pt}\Delta_4 \; \hspace{4pt}\Delta_1}^{\pm\beta_3 \;\pm\beta_2} \right]
\,=\,   A^R_{rs}(c) \, \times\\ [4pt] \nonumber
&&\times \hspace*{-20pt}
\begin{array}[t]{c}
{\displaystyle\sum} \\[2pt]
{\scriptstyle
|K|+|M| = |L|+|N| = n-\frac{rs}{2}
}
\end{array}
\hspace*{-30pt}
     \rho^{(\pm)}_{\scriptscriptstyle \rm NR,e}(\nu_4 ,w^+_3, S_{-K}L_{-M}\chi^+_{rs})
 \left[G^{e,\, {n}-\frac{rs}{2}}_{c, \Delta_{rs}+\frac{rs}{2}}\right]^{KM,LN} \!
   \rho^{(\pm)}_{\scriptscriptstyle \rm RN,e}
   (S_{-L}L_{-N}\chi^+_{rs},w^+_2 ,\nu_1 ),
\end{eqnarray}
and
\begin{eqnarray}
\label{res:NRRN-}
&& \hspace{-0.5cm}
\mathcal{R}^{n,-}_{\beta,rs}
\left[_{\hspace{4pt}\Delta_4 \; \hspace{4pt}\Delta_1}^{\pm\beta_3 \;\pm\beta_2} \right]
= \lim_{\beta \to -\beta_{rs}} (\beta + \beta_{rs}(c)) \,
F^{n,e}_{c, \beta}
\left[_{\hspace{4pt}\Delta_4 \; \hspace{4pt}\Delta_1}^{\pm\beta_3 \;\pm\beta_2} \right]
\,=\, - \,  A^R_{rs}(c) \, \times\\ [4pt] \nonumber
&&\times \hspace*{-20pt}
\begin{array}[t]{c}
{\displaystyle\sum} \\[2pt]
{\scriptstyle
|K|+|M| = |L|+|N| = n-\frac{rs}{2}
}
\end{array}
\hspace*{-30pt}
     \rho^{(\mp)}_{\scriptscriptstyle \rm NR,e}(\nu_4 ,w^+_3, S_{-K}L_{-M}\chi^+_{rs})
 \left[G^{e,\, {n}-\frac{rs}{2}}_{c, \Delta_{rs}+\frac{rs}{2}}\right]^{KM,LN} \!
   \rho^{(\mp)}_{\scriptscriptstyle \rm RN,e}
   (S_{-L}L_{-N}\chi^+_{rs},w^+_2 ,\nu_1 )
\end{eqnarray}
where we have used the relation between 3-point blocks
with opposite signs of $\beta_{rs}$ (\ref{3pktBetaNR}).
The coefficient $A^R_{rs}(c)$:
\begin{eqnarray*}
A^R_{rs}(c)
\; = \; -{1 \over 2\beta_{rs}} \,
\lim_{\Delta\to\Delta_{rs}}
\left(\frac{\left\langle\chi_{rs}^{+\,\Delta}|\chi_{rs}^{+\, \Delta}\right\rangle}{\Delta - \Delta_{rs}(c)}
\right)^{-1}.
\end{eqnarray*}
is of the form \cite{Heqn}:
\begin{eqnarray*}
A^R_{rs}(c)
\; = \;
- \, 2^{rs-\frac32}
\prod_{m=1-r}^r
\prod_{n=1-s}^s
\left(p b + \frac{q}{b}\right)^{-1}
\end{eqnarray*}
where $m+n \in 2{\mathbb Z}+1, \; (m,n) \neq (0,0).$
The equations (\ref{res:NRRN+}),(\ref{res:NRRN-}) together with factorization formulae
for 3-point blocks (\ref{faktorRRchi+})
lead to the expressions for the residues:
\begin{eqnarray}
\label{resNRRN}
\mathcal{R}^{n,\epsilon}_{rs}
\left[_{\hspace{4pt}\Delta_4 \; \hspace{4pt}\Delta_1}^{\pm\beta_3 \;\pm\beta_2} \right]
&=& \epsilon \,  A^R_{rs}(c)
  P^{rs}_c\!\left[_{\ \Delta_4}^{\pm \epsilon\beta_3} \right]
\, P^{rs}_c\!\left[_{\ \Delta_1}^{\pm \epsilon\beta_2} \right] \,
 F^{{m-\frac{rs}{2}},e}_{c, \beta_{rs}^{\prime}}\!
\left[_{\hspace{4pt}\Delta_4 \; \hspace{4pt}\Delta_1}^{\pm\epsilon\beta_3 \;\pm\epsilon\beta_2} \right],
\end{eqnarray}
where
\begin{equation}\label{betaPrime}
  \beta_{rs}^{\,\prime} =  {(-1)^s  \over 2\sqrt{2}}\left(rb - \frac{s}{b}\right)
\end{equation}
corresponds to the shifted weight $\Delta_{rs}+\frac{rs}{2}$ and
$P^{rs}_c\!\left[_{\ \Delta_\imath}^{\pm \epsilon\beta_\jmath} \right] $
are the fusion polynomials (\ref{fusionR}).
By similar calculation one can determine all the other residues:
\begin{eqnarray}
\label{resNRNR}
\mathcal{R}^{n,\epsilon}_{rs}
\left[^{\pm \beta_3 \;  \hspace{4pt}\Delta_2}_{ \hspace{4pt}\Delta_4 \; \pm \beta_1} \right]
&=& \epsilon \, (-1)^{rs \over 2}\, A^R_{rs}(c)
  P^{rs}_c\!\left[_{\ \Delta_4}^{\pm \epsilon\beta_3} \right]
\, P^{rs}_c\!\left[_{\ \Delta_2}^{\pm \epsilon\beta_1} \right] \,
 F^{{m-\frac{rs}{2}},e}_{c, \beta_{rs}^\prime}\!
\left[^{\pm \epsilon \beta_3 \;  \hspace{4pt}\Delta_2}_{ \hspace{4pt}\Delta_4 \; \pm \epsilon \beta_1} \right]
\\[4pt] \nonumber
\mathcal{R}^{n,\epsilon}_{rs}
\left[^{\hspace{4pt}\Delta_3 \; \hspace{4pt}\Delta_2}_{\pm\beta_4 \;\pm\beta_1} \right]
&=& \epsilon \,  A^R_{rs}(c)
  P^{rs}_c\!\left[_{\ \Delta_3}^{\pm \epsilon\beta_4} \right]
\, P^{rs}_c\!\left[_{\ \Delta_2}^{\pm \epsilon\beta_1} \right] \,
 F^{{m-\frac{rs}{2}},e}_{c, \beta_{rs}^\prime}\!
\left[^{\hspace{4pt}\Delta_3 \; \hspace{4pt}\Delta_2}_{\pm\epsilon\beta_4 \;\pm\epsilon\beta_1} \right]
\end{eqnarray}

In order to find a closed recurrence for the 4-point blocks with R intermediate states
we will define the corresponding elliptic blocks. The first step is to determine large $\beta$ asymptotic
of the 4-point blocks.
It follows from Zamolodchikov's reasoning \cite{Zamolodchikov:3},\cite{Hadasz:2007nt} that
 the $\Delta_i$,$\beta_i$ and c dependence of the first three terms in the
 large $\beta$ expansion of the block is given by the classical limit.
 Since in the classical limit the bosonic classical block occurs (\ref{classRNNR}), the linear in $\beta$ terms in
 the large $\beta$ asymptotic of the 4-point blocks have to be zero. The asymptotic takes the same form as
 in the case of the even block $ \mathcal{F}^{e}_{\Delta}\!
\left[^{\Delta_3 \; \pm \beta_2}_{\Delta_4 \; \;\;\;\beta_1} \right]\!(z)$ (\ref{asymptoticG1}). This  suggests
 the following definition of the elliptic blocks:
\begin{eqnarray}\label{elliptic_NRRN}
  \mathcal{F}^{e}_{\beta}\!
\left[_{\hspace{4pt}\Delta_4 \; \hspace{4pt}\Delta_1}^{\pm\beta_3 \;\pm\beta_2} \right]\!(z)
 \!& =&\!
(16q)^{\Delta - \frac{c-3/2}{24}-\frac{1}{16}}\ z^{\frac{c-3/2}{24} - \Delta_1 -\Delta_2 + \frac{1}{16}} \
 (1- z)^{\frac{c-3/2}{24} - \Delta_2 - \Delta_3}\
\\ \nonumber
 \!& \times & \! \theta_3^{\frac{c - 3/2}{2}
- 4 (\Delta_1 +\Delta_2 +\Delta_3 + \Delta_4) +\frac12 }  \
 \mathcal{H}^{e}_{\beta}\!
\left[_{\hspace{4pt}\Delta_4 \; \hspace{4pt}\Delta_1}^{\pm\beta_3 \;\pm\beta_2} \right]\!(q),
\\[6pt]\label{elliptic_NRNR}
 \mathcal{F}^{e}_{\beta}\!
  \left[^{ \pm\beta_3 \; \hspace{5pt} \Delta_2}_{\hspace{5pt}\Delta_4 \ \pm\beta_1} \right]\!(z)
 \!& =&\!
(16q)^{\Delta - \frac{c-3/2}{24}-\frac{1}{16}}\ z^{\frac{c-3/2}{24} - \Delta_1 -\Delta_2+\frac{1}{16}} \
 (1- z)^{\frac{c-3/2}{24} - \Delta_2 - \Delta_3 + \frac{1}{16}}\
\\ \nonumber
 \!& \times & \! \theta_3^{\frac{c - 3/2}{2}
- 4 (\Delta_1 +\Delta_2 +\Delta_3 + \Delta_4)+\frac12 }  \
 \mathcal{H}^{e}_{\beta}\!
  \left[^{ \pm\beta_3 \; \hspace{5pt} \Delta_2}_{\hspace{5pt}\Delta_4 \ \pm\beta_1} \right]\!(q)
\\[6pt]\label{elliptic_RNNR}
 \mathcal{F}^{e}_{\beta}\!
  \left[^{ \hspace{5pt}\Delta_3 \; \hspace{5pt} \Delta_2}_{\pm\beta_4 \ \pm\beta_1} \right]\!(z)
 \!& =&\!
(16q)^{\Delta - \frac{c-3/2}{24}-\frac{1}{16}}\ z^{\frac{c-3/2}{24} - \Delta_1 -\Delta_2+\frac{1}{16}} \
 (1- z)^{\frac{c-3/2}{24} - \Delta_2 - \Delta_3 }\
\\ \nonumber
 \!& \times & \! \theta_3^{\frac{c - 3/2}{2}
- 4 (\Delta_1 +\Delta_2 +\Delta_3 + \Delta_4)+\frac12 }  \
 \mathcal{H}^{e}_{\beta}\!
  \left[^{ \hspace{5pt}\Delta_3 \; \hspace{5pt} \Delta_2}_{\pm\beta_4 \ \pm\beta_1} \right]\!(q).
\end{eqnarray}
The elliptic blocks satisfy recursive relations with the coefficients at residues given by (\ref{resNRRN}),
(\ref{resNRNR}):
\begin{eqnarray*}
 \mathcal{H}^{e}_{\beta}\!
 \left[_{\hspace{4pt}\Delta_4 \; \hspace{4pt}\Delta_1}^{\pm\beta_3 \;\pm\beta_2} \right]\!(q)
=
 f_{\pm\pm}
 &+& \hspace{-3pt}
 \begin{array}[t]{c}
{\displaystyle\sum} \\[-5pt]
{\scriptscriptstyle
r,s>0}
\\[-7pt]
{\scriptscriptstyle
r+s\in 2{\mathbb N}+1
}
\end{array}
 \hspace{-5pt} (16q)^{\frac{rs}{2}}
 \frac{  A^R_{rs}(c)
  P^{rs}_c\!\left[_{\ \Delta_4}^{\pm \beta_3} \right]
\, P^{rs}_c\!\left[_{\ \Delta_1}^{\pm \beta_2} \right]}{\beta-\beta_{rs}} \,
\mathcal{H}^{e}_{\beta_{rs}^\prime}\!
\left[_{\hspace{4pt}\Delta_4 \; \hspace{4pt}\Delta_1}^{\pm\beta_3 \;\pm\beta_2} \right]\!(q)
\\
&&  \hspace{20pt} - (16q)^{\frac{rs}{2}}
 \frac{  A^R_{rs}(c)
  P^{rs}_c\!\left[_{\ \Delta_4}^{\mp \beta_3} \right]
\, P^{rs}_c\!\left[_{\ \Delta_1}^{\mp \beta_2} \right]}{\beta+\beta_{rs}} \,
\mathcal{H}^{e}_{\beta_{rs}^\prime}\!
\left[_{\hspace{4pt}\Delta_4 \; \hspace{4pt}\Delta_1}^{\mp\beta_3 \;\mp\beta_2} \right]\!(q),
\end{eqnarray*}
\begin{eqnarray}\label{H_NRNR_rek}
\nonumber
 \mathcal{H}^{e}_{\beta}\!
 \left[^{ \pm\beta_3 \; \hspace{5pt} \Delta_2}_{\hspace{5pt}\Delta_4 \ \pm\beta_1} \right]\!(q)
=
 g_{\pm\pm}
 &+& \hspace{-7pt}
 \begin{array}[t]{c}
{\displaystyle\sum} \\[-5pt]
{\scriptscriptstyle
r,s>0}
\\[-7pt]
{\scriptscriptstyle
r+s\in 2{\mathbb N}+1
}
\end{array}
 \hspace{-4pt} (16q)^{\frac{rs}{2}}
 \frac{  (-1)^{rs \over 2} A^R_{rs}(c)
  P^{rs}_c\!\left[_{\ \Delta_4}^{\pm \beta_3} \right]
\, P^{rs}_c\!\left[_{\ \Delta_2}^{\pm \beta_1} \right]}{\beta-\beta_{rs}} \,
\mathcal{H}^{e}_{\beta_{rs}^\prime}\!
\left[^{ \pm\beta_3 \; \hspace{5pt} \Delta_2}_{\hspace{5pt}\Delta_4 \ \pm\beta_1} \right]\!(q)
\\[-10pt]
\\[-6pt] \nonumber
&&  \hspace{15pt} -(16q)^{\frac{rs}{2}}
 \frac{  (-1)^{rs \over 2} A^R_{rs}(c)
  P^{rs}_c\!\left[_{\ \Delta_4}^{\mp \beta_3} \right]
\, P^{rs}_c\!\left[_{\ \Delta_2}^{\mp \beta_1} \right]}{\beta+\beta_{rs}} \,
\mathcal{H}^{e}_{\beta_{rs}^\prime}\!
\left[^{ \mp\beta_3 \; \hspace{5pt} \Delta_2}_{\hspace{5pt}\Delta_4 \ \mp\beta_1} \right]\!(q),
\end{eqnarray}
\begin{eqnarray}\label{H_RNNR_rek}
 \nonumber
 \mathcal{H}^{e}_{\beta}\!
 \left[^{ \hspace{5pt}\Delta_3 \; \hspace{5pt} \Delta_2}_{\pm\beta_4 \ \pm\beta_1} \right]\!(q)
=
 h_{\pm\pm}
 &+&\hspace{-7pt}
 \begin{array}[t]{c}
{\displaystyle\sum} \\[-5pt]
{\scriptscriptstyle
r,s>0}
\\[-7pt]
{\scriptscriptstyle
r+s\in 2{\mathbb N}+1
}
\end{array}
 \hspace{-4pt}  (16q)^{\frac{rs}{2}}
 \frac{ A^R_{rs}(c)
  P^{rs}_c\!\left[_{\ \Delta_3}^{\pm \beta_4} \right]
\, P^{rs}_c\!\left[_{\ \Delta_2}^{\pm \beta_1} \right]}{\beta-\beta_{rs}} \,
\mathcal{H}^{e}_{\beta_{rs}^\prime}\!
\left[^{ \hspace{5pt}\Delta_3 \; \hspace{5pt} \Delta_2}_{\pm\beta_4 \ \pm\beta_1} \right]\!(q)
\\[-10pt]
\\[-10pt] \nonumber
&&  \hspace{15pt} -(16q)^{\frac{rs}{2}}
 \frac{   A^R_{rs}(c)
  P^{rs}_c\!\left[_{\ \Delta_3}^{\mp \beta_4} \right]
\, P^{rs}_c\!\left[_{\ \Delta_2}^{\mp \beta_1} \right]}{\beta+\beta_{rs}} \,
\mathcal{H}^{e}_{\beta_{rs}^\prime}\!
\left[^{ \hspace{5pt}\Delta_3 \; \hspace{5pt} \Delta_2}_{\mp\beta_4 \ \mp\beta_1} \right]\!(q).
\end{eqnarray}
The independent of $\Delta_i$ and $c$ functions $f_{\pm\pm}, g_{\pm\pm}, h_{\pm\pm} $ can be deduced
from the explicit expressions for the specific  blocks with $c=\frac32$:
\begin{eqnarray*}
\lim_{\beta\to 0} \lim_{2\alpha \to i}
 \lim_{b\to i} \mathcal{F}^{e}_{\beta}\!
\left[_{ \Delta_a \  \Delta_a}^{\pm\beta \ \pm\beta} \right]\!(z)
\!&=& \!
\lim_{\beta\to 0} \lim_{2\alpha \to i}
 \lim_{b\to i} \mathcal{F}^{e}_{\Delta}\!
\left[^{ \pm\beta \; \; \Delta_a}_{ \Delta_a \; \pm\beta} \right]\!(z) =
(16q)^{\Delta-\frac{1}{16}} \, \left[z(1-z)\right]^{-\frac18}\, \theta_3^{-1}(q),
\\
\lim_{\beta\to 0} \lim_{2\alpha \to i}
 \lim_{b\to i} \mathcal{F}^{e}_{\Delta}\!
\left[^{\Delta_a \; \; \Delta_a}_{ \pm\beta \ \pm\beta} \right]\!(z)
\!&=& \! (16q)^{\Delta-\frac{1}{16}}\,  z^{-\frac18}(1-z)^{-\frac14}\, \theta_3^{-1}(q).
\end{eqnarray*}
Comparing the formulae  above with definitions (\ref{elliptic_NRRN}), (\ref{elliptic_NRNR}), (\ref{elliptic_RNNR})
 one gets:
$$
f_{\pm\pm} \,=\, g_{\pm\pm} \,=\,  h_{\pm\pm}=1,
$$

\section{Numerical check of bootstrap equations}

\subsection{Four R fields}

The correlation function of 4 R primary fields expressed in terms of elliptic blocks (\ref{4ptRRRR-block}),
(\ref{elliptic-RRRR}) take the following form:
\begin{eqnarray*}
&&  \hspace{-20pt}
\left\langle
R^+_4(\infty,\infty) R^+_3(1,1) R^+_{2}(z, \bar z) R^+_1(0,0)
\right\rangle
\\[4pt]
&&= \frac12 \ (z\bar z)^{\frac{c-3/2}{24} - \Delta_1 -\Delta_2} \
 \left[(1- z)(1-\bar z)\right]^{\frac{c-3/2}{24} - \Delta_2 - \Delta_3}
\left[\theta_3(q)\theta_3(\bar q)\right]^{\frac{c - 3/2}{2}
- 4 \sum_{i}\Delta_i } \times f_{\scriptscriptstyle [4][3][2][1]}(\tau, \bar \tau),
\\[4pt]
&& \hspace{-20pt}
f_{\scriptscriptstyle [4][3][2][1]}(\tau, \bar \tau) =\hspace{-2pt}
\int \mathrm{d}P |16q|^{P^2} \hspace{-3pt} \sum_{\epsilon_1,\epsilon_2=\pm}
 \hspace{-3pt} C^{(\epsilon_1)}_{\scriptscriptstyle [4][3]P} C^{(\epsilon_2)}_{\scriptscriptstyle -P[2][1]}
\left(
    \left|\mathcal{H}^e_{\Delta_P}\!
    \left[^{ \epsilon_1\beta_3 \; \epsilon_2\beta_2}_{\hspace{7pt}\beta_4 \ \hspace{7pt} \beta_1} \right]\!(q)
    \right|^2
    +
    \left|\mathcal{H}^{o}_{\Delta_P}\!
    \left[^{ \epsilon_1\beta_3 \; \epsilon_2\beta_2}_{\hspace{7pt}\beta_4 \ \hspace{7pt} \beta_1} \right]\!(q)
    \right|^2
    \right),
\end{eqnarray*}
where $ q(z)=\mathrm{e}^{i \pi \tau}$ and $\tau$ is defined by the complete elliptic integral of the first kind $K(z)$:
$$\tau = i {K(1-z)\over K(z)}.$$
The crossing symmetry conditions for the 4-point function (\ref{RRRRcs1}),(\ref{RRRRcs2}) read:
\begin{eqnarray}\label{ft_cs4R}
\nonumber
f_{\scriptscriptstyle [4][3][2][1]}(\tau, \bar \tau) &=& f_{\scriptscriptstyle [3][4][2][1]}(\tau+1, \bar \tau+1), \\
f_{\scriptscriptstyle [4][3][2][1]}(\tau, \bar \tau)
&=& (\tau \bar \tau)^{\frac{c - 3/2}{4} - 2 \sum_{i}\Delta_i }
 f_{\scriptscriptstyle [4][1][2][3]}(-{1 \over \tau}, -{1 \over \bar \tau}).
\end{eqnarray}
 The first equation can be verified analytically due to the relations:
 \begin{eqnarray*}
 \mathcal{H}^e_{\Delta}
    \left[^{ \pm\beta_3 \; \pm\beta_2}_{\hspace{7pt}\beta_4 \ \hspace{5pt} \beta_1} \right](-q)
   & =& \mathcal{H}^e_{\Delta}
    \left[^{ \pm\beta_4 \; \pm\beta_2}_{\hspace{7pt}\beta_3 \ \hspace{5pt} \beta_1} \right](q), \\
 \mathcal{H}^o_{\Delta}
    \left[^{ \epsilon\beta_3 \; \pm\beta_2}_{\hspace{3pt}\beta_4 \ \hspace{5pt} \beta_1} \right] (\mathrm{e}^{i\pi}q)
    &=& \mathrm{e}^{-  \epsilon { i\pi \over 2}} \mathcal{H}^o_{\Delta}
     \left[^{ \epsilon\beta_4 \; \pm\beta_2}_{\hspace{3pt}\beta_3 \ \hspace{5pt} \beta_1} \right](q).
\end{eqnarray*}
which follow from the recursive formula (\ref{H_RRRR_rek}) with the fusion polynomials satisfying (\ref{fuzPol_21_R}).

\begin{figure}
 \center
   \includegraphics[scale=0.7]{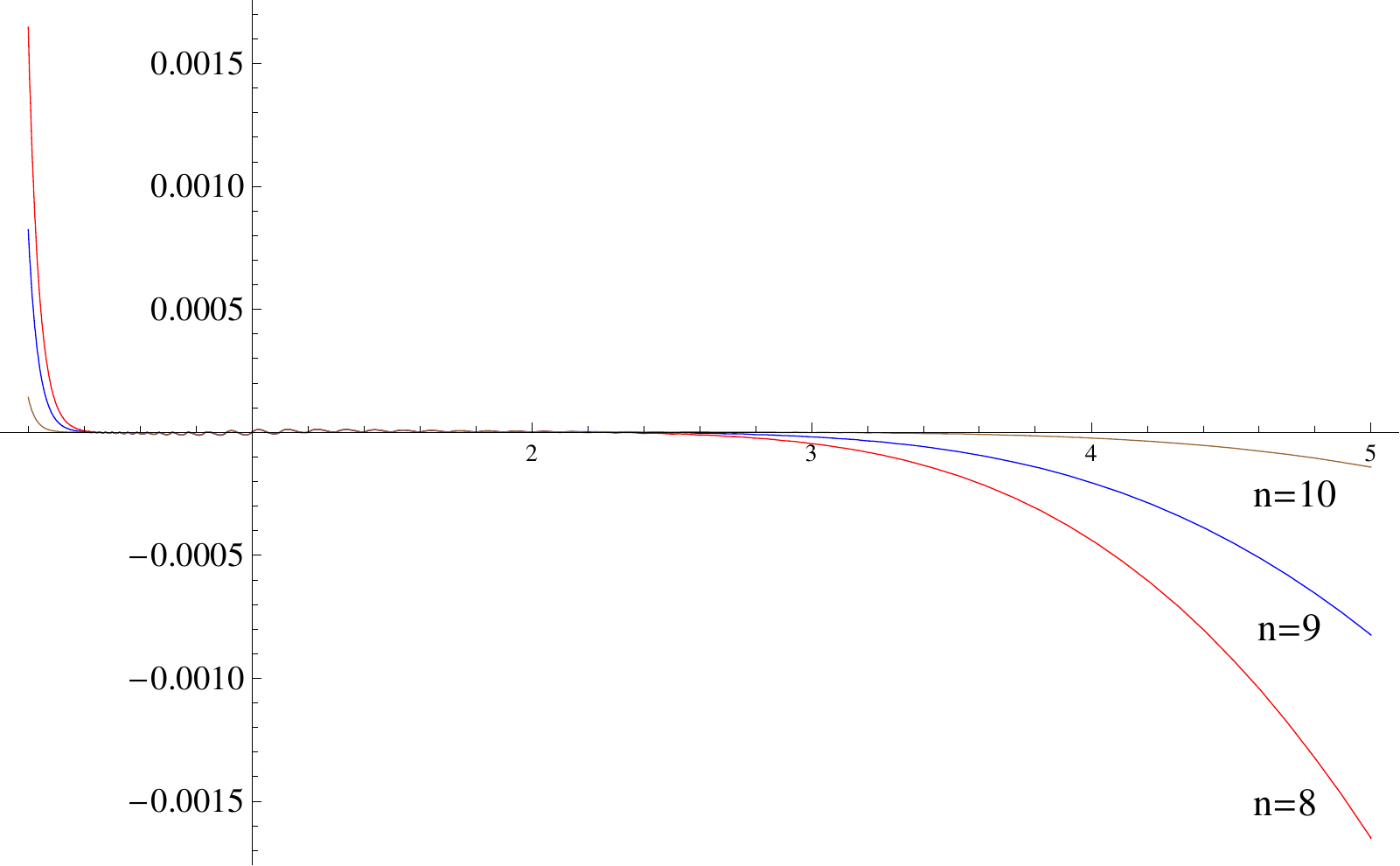}
  \caption{Numerical check of the bootstrap equation (\ref{boots4R}) }\label{plot4R}
\end{figure}

The second crossing symmetry condition (\ref{ft_cs4R}) can be checked numerically with the help of the
recursive relations for the blocks (\ref{H_RRRR_rek}).
In order to simplify calculations one can choose two different non-zero external weights:
$$
\beta_1=\beta_3= - {i p_1 \over \sqrt{2} }, \qquad \beta_2=-\beta_4= - {i p_2 \over \sqrt{2} }
$$
With such a choice none of blocks vanishes
which provides a non-trivial test of  the general recursive formulae.
The products of two SLFT structure constants (\ref{structureC_R}), (\ref{strCpm}) take the form:
\begin{eqnarray*}
C^{(\epsilon_1)}_{\scriptscriptstyle [-2][1]P} C^{(\epsilon_2)}_{\scriptscriptstyle -P[2][1]}
\!\! &=& \!\! \frac14 \left( { \pi \mu \over 2} \gamma\left({Qb \over 2}\right) b^{2-2b^2}\right)^{Q-2ip_1 \over b}
\!\! \Upsilon_{\scriptscriptstyle 0} \Upsilon_{\scriptscriptstyle \rm R}^2\!\left(Q+2ip_1\right)
 \Upsilon_{\scriptscriptstyle \rm R}\!\left(Q+2ip_2\right) \Upsilon_{\scriptscriptstyle \rm R}\!\left(Q-2ip_2\right)
\\[4pt]
 & \! \times  \! &
 r_1^{\epsilon_1,\epsilon_2}(P)
\end{eqnarray*}
with the $P$-dependent part:
{\small
\begin{eqnarray*}
&& \hspace{-20pt}
r_1^{\epsilon_1,\epsilon_2}(P) = P^2 \, \exp \int_{0}^{\infty} {\mathrm{d}t \over t}
\Bigg\{ \left( 4(p_1^2+p_2^2)-\frac{Q^2}{2} \right)\mathrm{e}^{-t}
+ \cos(Pt)\left(1-\coth\left(\frac{t}{2b}\right)\coth\left(\frac{b t}{2}\right)\right)
\\[4pt]
&& \hspace{40pt}
-{
6+\cos(Pt) \over
\sinh\left(\frac{t}{2b}\right)\sinh\left(\frac{b t}{2}\right)}
+{2 \cos\left(\frac{P t}{2}\right) \cos\left(\frac{p_1 t}{2}\right)\cos\left(\frac{p_2 t}{2}\right)
\over
\sinh\left(\frac{t}{4b}\right)\sinh\left(\frac{b t}{4}\right)
} + a_1^{\epsilon_1,\epsilon_2}\left(P,t\right)
\Bigg\}
\\[6pt]
&&  \hspace{-20pt}
a_1^{\pm\pm}\left(P,t\right) = 0, \qquad
a_1^{+-}\left(P,t\right) \,=\, - \, a_1^{-+}\left(P,t\right)
\,=\, {
-2 \sin\left(\frac{p_1 t}{2}\right)\sin\left(\frac{p_2 t}{2}\right)\cos\left(\frac{P t}{2}\right)
\over
 \cosh\left(\frac{t}{4b}\right)\cosh\left(\frac{b t}{4}\right)
}
\end{eqnarray*}
}
The form of the function above follows from the integral representation of $\Upsilon_b$ (\ref{Ups}).
The bootstrap equation (\ref{ft_cs4R}) reads:
{\small
\begin{eqnarray}\label{boots4R}
\nonumber
&& \hspace{-20pt}
\sum_{\epsilon_1,\epsilon_2=\pm} \int \mathrm{d}P |16q|^{P^2} \Bigg\{
r_1^{\epsilon_1,\epsilon_2}(P)
\left(
    \left|\mathcal{H}^e_{\Delta_P}
    \left[^{ \epsilon_1\beta_1 \; \epsilon_2\beta_2}_{-\beta_2 \ \hspace{5pt} \beta_1} \right](q)
    \right|^2
    +
    \left|\mathcal{H}^{o}_{\Delta_P}
    \left[^{ \epsilon_1\beta_1 \; \epsilon_2\beta_2}_{-\beta_2 \ \hspace{5pt} \beta_1} \right](q)
    \right|^2
    \right)\!\!
\Bigg\}
\\
&&= (\tau \bar \tau)^{\frac{c - 3/2}{4} - 4(\Delta_1+\Delta_2)} \,
\\ \nonumber
&& \times
\sum_{\epsilon_1,\epsilon_2=\pm} \int \mathrm{d}P |16q'|^{P^2} \Bigg\{
r_1^{\epsilon_1,\epsilon_2}(P)
\left(
    \left|\mathcal{H}^e_{\Delta_P}
    \left[^{ \epsilon_1\beta_1 \; \epsilon_2\beta_2}_{-\beta_2 \ \hspace{5pt} \beta_1} \right](q')
    \right|^2
    +
    \left|\mathcal{H}^{o}_{\Delta_P}
    \left[^{ \epsilon_1\beta_1 \; \epsilon_2\beta_2}_{-\beta_2 \ \hspace{5pt} \beta_1} \right](q')
    \right|^2
    \right)\!\!
\Bigg\}
\end{eqnarray}
}
where $q' = \mathrm{e}^{-{ i \pi \over \tau}}$.
Due to the highly oscillatory character of the integrant in the function  $r_1^{\epsilon_1,\epsilon_2}(P)$,
the numerical calculations should be carefully performed.
We present a check of this relation for $ p_1= 0.3,\, p_2 = 0.5$, $c= 3$
($b= \sqrt{-\frac34 + \frac{i \sqrt{7}}{4}}$)
 and for  $\tau$ along the imaginary axis in the range $[0.2 i, 5i]$.
On Figure \ref{plot4R} the relative difference of the left and the right side of (\ref{boots4R}) as a function
of $\tau$ is plotted. The three curves correspond to the expansions of the even and the odd elliptic blocks
 up to the terms $q^n$ and $q^{n+\frac12}$ for $ n=8,9,10$.

\subsection{R intermediate weights}

\begin{figure}
  \center
   \includegraphics[scale=0.55]{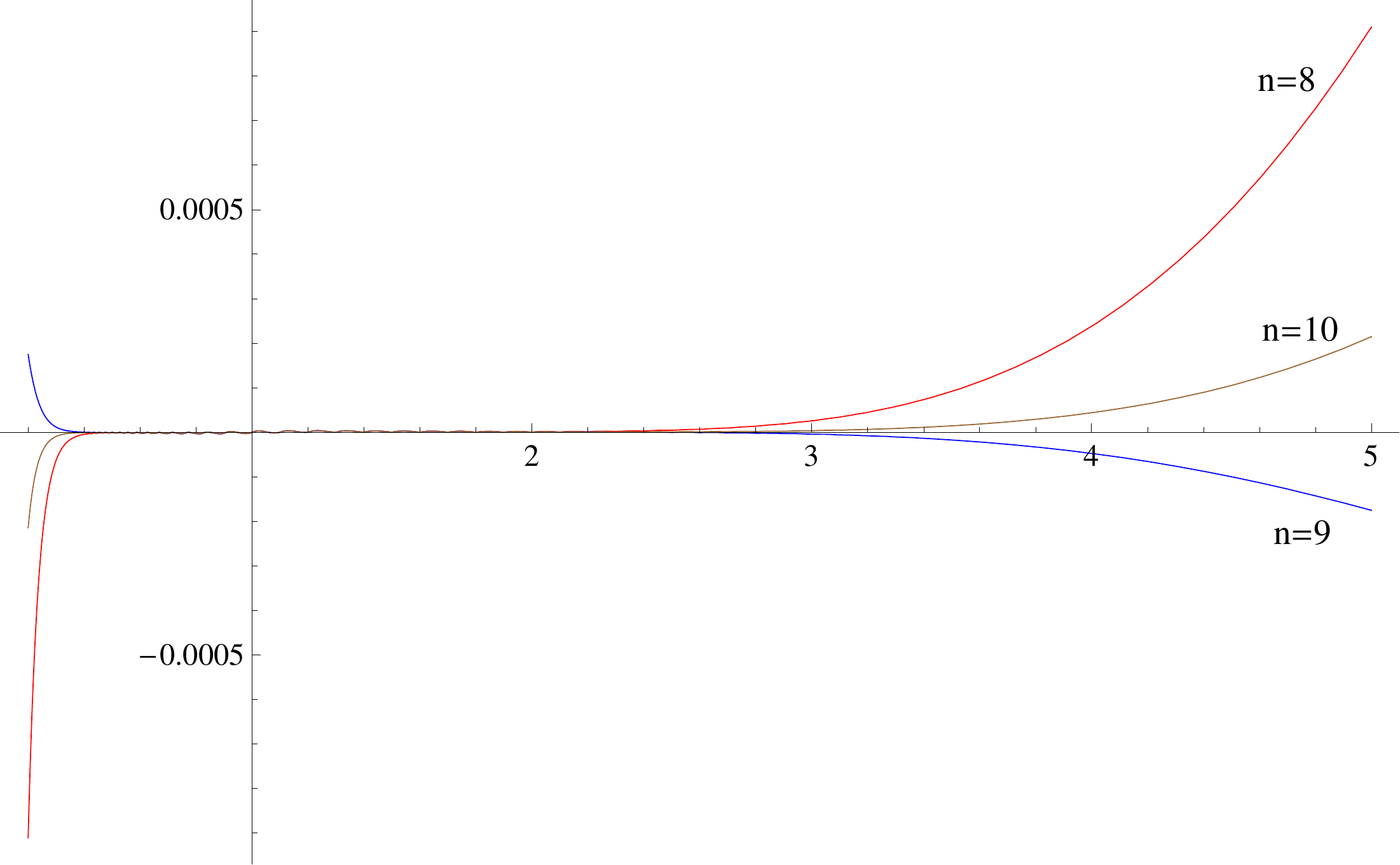}
  \caption{Numerical check of the bootstrap equation (\ref{bootsNRNR})}\label{plotNRNR}
\end{figure}

The correlation functions of 2 R primaries and 2 NS superprimary fields
(\ref{4ptNRNR-block}), (\ref{4ptRNNR-block}) written in terms of elliptic blocks
(\ref{elliptic_NRNR}), (\ref{elliptic_RNNR}) read:
\begin{eqnarray*}
&& \hspace{-60pt}
\left\langle
\phi_4(\infty,\infty) R^+_3(1,1) \phi_{2}(z, \bar z) R^+_1(0,0)
\right\rangle
=
\ (z\bar z)^{\frac{c-3/2}{24} - \Delta_1 -\Delta_2+\frac{1}{16}}
 \\[4pt] & \times&
  \left[(1- z)(1-\bar z)\right]^{\frac{c-3/2}{24} - \Delta_2 - \Delta_3+\frac{1}{16}} \,
\left[\theta_3(q)\theta_3(\bar q)\right]^{\frac{c - 3/2}{2}
- 4  \sum_{i}\Delta_i +\frac12 } \, f_{\scriptscriptstyle 4[3]2[1]}(\tau, \bar \tau)
\\[6pt]
&& \hspace{-60pt}
\left\langle
R^+_4(\infty,\infty) \phi_3(1,1)  \phi_{2}(z, \bar z) R^+_1(0,0)
\right\rangle
=
\ (z \bar z)^{\frac{c-3/2}{24} - \Delta_1 -\Delta_2+\frac{1}{16}} \
\\[4pt] & \times& \left[(1- z)(1-\bar z)\right]^{\frac{c-3/2}{24} - \Delta_2 - \Delta_3}
\, \left[\theta_3(q)\theta_3(\bar q)\right]^{\frac{c - 3/2}{2}
- 4  \sum_{i}\Delta_i +\frac12 } \times f_{\scriptscriptstyle [4]32[1]}(\tau, \bar \tau),
\end{eqnarray*}
\begin{eqnarray*}
f_{\scriptscriptstyle 4[3]2[1]}(\tau, \bar \tau)&=&
\int \mathrm{d}P |16q|^{P^2}\, \sum_{\epsilon_1,\epsilon_2=\pm}
C^{(\epsilon_1)}_{\scriptscriptstyle 4[3][P]} C^{(\epsilon_2)}_{\scriptscriptstyle [-P]2[1]}
\,
    \left|\mathcal{H}^e_{\beta_P}
    \left[^{ \epsilon_1\beta_3 \; \hspace{5pt} \Delta_2}_{\hspace{5pt}\Delta_4 \; -\epsilon_2\beta_1} \right](q)
    \right|^2
\\[6pt]
f_{\scriptscriptstyle [4]32[1]}(\tau, \bar \tau)&=&
\int \mathrm{d}P |16q|^{P^2} \, \sum_{\epsilon_1,\epsilon_2=\pm}
C^{(\epsilon_1)}_{\scriptscriptstyle [4]3[P]} C^{(\epsilon_2)}_{\scriptscriptstyle [-P]2[1]}
\,
    \left|\mathcal{H}^e_{\beta_P}
    \left[^{ \hspace{5pt} \Delta_3\;  \hspace{10pt}\Delta_2}_{\epsilon_1\beta_4 \ -\epsilon_2\beta_1} \right](q)
    \right|^2
\end{eqnarray*}
Crossing symmetry of the 4-point functions (\ref{NRNRcs1}),(\ref{NRNRcs2}) implies:
\begin{eqnarray*}
f_{\scriptscriptstyle 4[3]2[1]}(\tau, \bar \tau) &=& f_{\scriptscriptstyle [3]42[1]}(\tau+1, \bar \tau+1) \\
f_{\scriptscriptstyle 4[3]2[1]}(\tau, \bar \tau)
&=& (\tau \bar \tau)^{\frac{c - 3/2}{4} - 2 \sum_{i}\Delta_i + \frac14}
f_{\scriptscriptstyle 4[1]2[3]}(-{1 \over \tau}, -{1 \over \bar \tau})
\end{eqnarray*}
Since the elliptic blocks of the two types are related (\ref{H_NRNR_rek}, \ref{H_RNNR_rek}):
$$
\mathcal{H}^e_{\beta}
    \left[^{ \pm\beta_3 \; \hspace{5pt} \Delta_2}_{\hspace{5pt}\Delta_4 \; \pm\beta_1} \right](q)
    \,=\,
    \mathcal{H}^e_{\beta}
    \left[^{ \hspace{5pt} \Delta_4\;  \hspace{5pt}\Delta_2}_{\pm\beta_3 \ \pm\beta_1} \right](-q)
$$
the first condition is satisfied straightforwardly.
The second condition can be verified numerically. For the external weights:
$$
\beta_1=\beta_3= - {i p_1 \over \sqrt{2} }, \qquad \alpha_2= {Q \over 2} +i p_2, \qquad
 \alpha_4= {Q \over 2} -i p_2
$$
the products of the structure constants read:
\begin{eqnarray*}
C^{(\epsilon_1)}_{\scriptscriptstyle -2[1][P]} C^{(\epsilon_2)}_{\scriptscriptstyle -[P]2[1]}
\! &=& \! \frac14 \left( { \pi \mu \over 2} \gamma\left({Qb \over 2}\right) b^{2-2b^2}\right)^{Q-2ip_1 \over b}
\!\! \Upsilon_{\scriptscriptstyle 0} \Upsilon_{\scriptscriptstyle \rm R}^2\left(Q+2ip_1\right)
\\[4pt]
 & \! \times  \! &
   \Upsilon_{\scriptscriptstyle \rm NR}\left(Q+2ip_2\right) \Upsilon_{\scriptscriptstyle \rm NR}\left(Q-2ip_2\right)
    \,
 r_2^{\epsilon_1,\epsilon_2}(P)
\end{eqnarray*}
with the $P$-dependent part:
\begin{eqnarray*}
r_2^{\epsilon_1,\epsilon_2}(P) &=& \exp
\int_{0}^{\infty} {\mathrm{d}t \over t}
    \Bigg\{ \left( 1+ 4(p_1^2+p_2^2)-\frac{Q^2}{2}\right)\mathrm{e}^{-t}
    - {\cos\left( P t\right) \left( \cosh\left(\frac{t}{2b}\right)+ \cosh\left(\frac{b t}{2} \right) \right) +6
    \over \sinh\left(\frac{t}{2b}\right)\sinh\left(\frac{b t}{2}\right) }
\\
&& \hspace{30pt}
+
{2 \cos\left(\frac{P t}{2}\right) \cos\left(\frac{p_1 t}{2}\right)\cos\left(\frac{p_2 t}{2}\right)
\over
\sinh\left(\frac{t}{4b}\right)\sinh\left(\frac{b t}{4}\right)
}
+
a_2^{\epsilon_1,\epsilon_2}\left(P,t\right)
\Bigg\}
\\
a_2^{\pm\pm} &=&0, \qquad
 a_2^{+-}\left(P,t\right) =  \,-\, a_2^{-+}\left(P,t\right) =
 { 2 \sin\left(\frac{P t}{2}\right) \sin\left(\frac{p_1 t}{2}\right)\sin\left(\frac{p_2 t}{2}\right)
\over
\cosh\left(\frac{t}{4b}\right)\cosh\left(\frac{b t}{4}\right)
}
.
\end{eqnarray*}
Then the bootstrap equation takes the form:
\begin{eqnarray}\label{bootsNRNR}
&& \hspace{-40pt}
\sum_{\epsilon_1,\epsilon_2=\pm} \int \mathrm{d}P  |16q|^{P^2}\,
r_2^{\epsilon_1,\epsilon_2}(P)
\,
    \left|\mathcal{H}^e_{\beta_P}
    \left[^{ \epsilon_1\beta_1 \; \hspace{5pt} \Delta_2}_{\hspace{5pt}\Delta_2 \; -\epsilon_2\beta_1} \right](q)
    \right|^2
\\ \nonumber
&&= (\tau \bar \tau)^{\frac{c - 3/2}{4} - 4(\Delta_1+\Delta_2) + \frac14} \,
\sum_{\epsilon_1,\epsilon_2=\pm} \int \mathrm{d}P |16q'|^{P^2}\,
r_2^{\epsilon_1,\epsilon_2}(P)
\,
    \left|\mathcal{H}^e_{\beta_P}
    \left[^{ \epsilon_1\beta_1 \; \hspace{5pt} \Delta_2}_{\hspace{5pt}\Delta_2 \; -\epsilon_2\beta_1} \right](q')
    \right|^2
\end{eqnarray}
We present the sample  calculation for
 $ p_1= 0.3,\, p_2 = 0.4$, $c= 3$,
 and for  $\tau$ in the range $[0.2 i, 5i]$.
The relative difference of the both sides of (\ref{bootsNRNR}) is plotted on Figure \ref{plotNRNR}.
 The three curves correspond to the elliptic blocks expanded up to  $n=8,9,10$ power of $q$.

\subsection{Functions factorized on R and NS states}

\begin{figure}
  \center
   \includegraphics[scale=0.65]{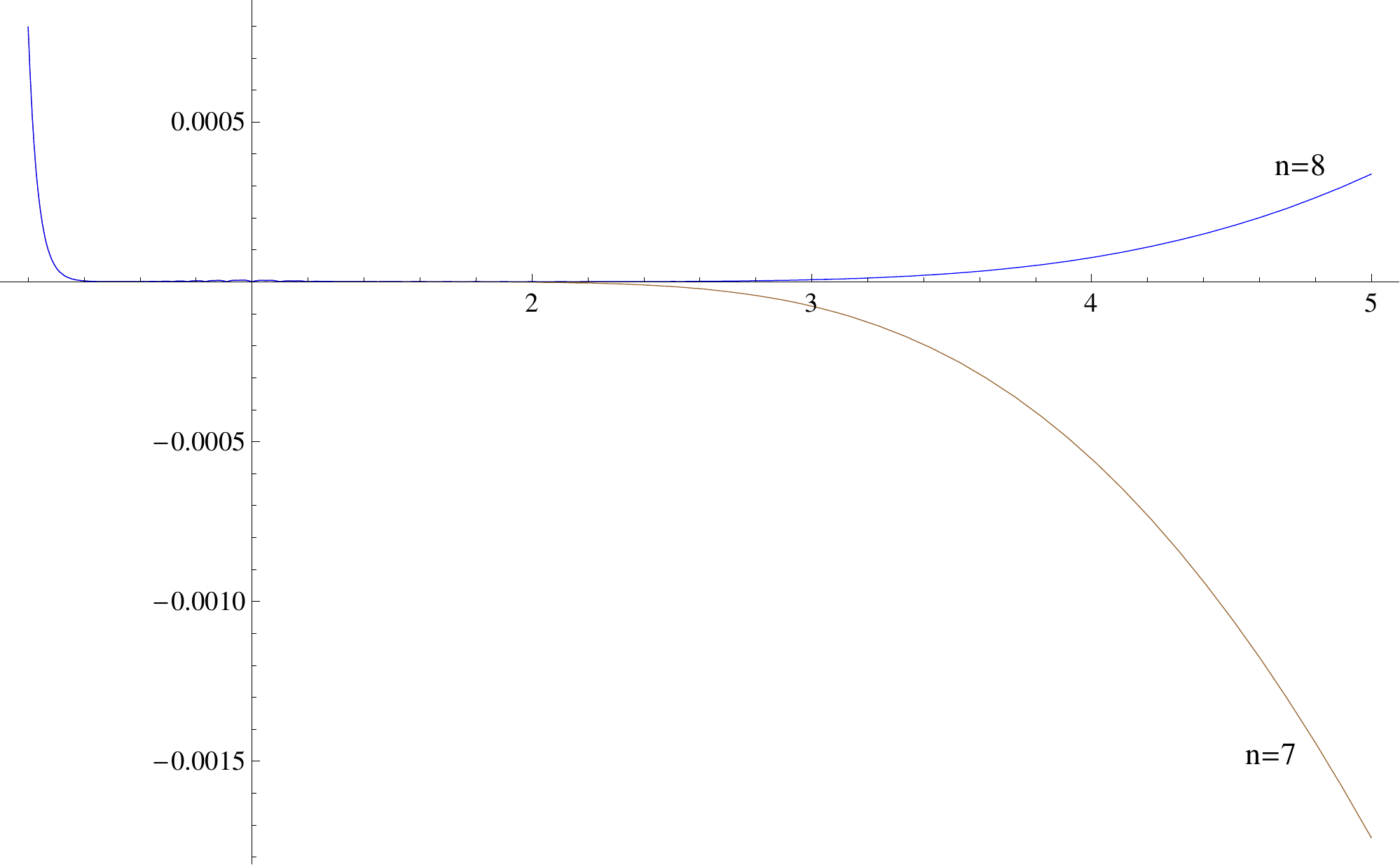}
  \caption{Numerical check of the bootstrap equation (\ref{bootsNNRR}) }\label{plotNNRR}
\end{figure}

Finally, we shall consider the bootstrap equation involving
4-point blocks from two different sectors and all the structure constants of $N=1$ SLFT.
Using formulae derived in the previous section
(\ref{4ptNNRR-block}), (\ref{4ptNRRN-block}), (\ref{elliptic_NNRR}), (\ref{elliptic_NRRN})  one can
express the correlation functions of 2 NS and 2 R fields
 in terms of elliptic blocks:
\begin{eqnarray*}
\left\langle
\phi_4(\infty,\infty) \phi_3(1,1) R^+_{2}(z, \bar z) R^+_1(0,0)
\right\rangle
&=&
\ (z \bar z)^{\frac{c-3/2}{24} - \Delta_1 -\Delta_2} \
 \left[(1- z)(1-\bar z)\right]^{\frac{c-3/2}{24} - \Delta_2 - \Delta_3+\frac{1}{16}}
 \\
&\times&
\left[\theta_3(q)\theta_3(\bar q)\right]^{\frac{c - 3/2}{2}
-  \sum_{i}\Delta_i +\frac12} \, f_{\scriptscriptstyle 43[2][1]}(\tau, \bar \tau)
\\[4pt]
\left\langle
\phi_4(\infty,\infty) R^+_3(1,1) R^+_{2}(z, \bar z) \phi_1(0,0)
\right\rangle
&=&
\ (z \bar z)^{\frac{c-3/2}{24} - \Delta_1 -\Delta_2+\frac{1}{16}} \
 \left[(1- z)(1-\bar z)\right]^{\frac{c-3/2}{24} - \Delta_2 - \Delta_3} \\
 &\times&
\left[\theta_3(q)\theta_3(\bar q)\right]^{\frac{c - 3/2}{2}
- 4  \sum_{i}\Delta_i +\frac12 } \, f_{\scriptscriptstyle 4[3][2]1}(\tau, \bar \tau),
\end{eqnarray*}
where
\begin{eqnarray*}
f_{\scriptscriptstyle 43[2][1]}(\tau, \bar \tau)
& =&\int \mathrm{d}P \, |16q|^{P^2}\,  \sum_{\epsilon=\pm}\Bigg\{
C_{\scriptscriptstyle 43P} C^{(\epsilon)}_{\scriptscriptstyle -P[2][1]}
    \left|\mathcal{H}^{e}_{\Delta_P}
    \left[^{\Delta_3 \; \epsilon\beta_2 }_{\Delta_4 \; \;\beta_1 } \right](q)\right|^2
\\
&& \hspace{70pt}
- i \,
\widetilde C_{\scriptscriptstyle 43P}  C^{(\epsilon)}_{\scriptscriptstyle -P[2][1]}
    \left|\mathcal{H}^{o}_{\Delta_P}
    \left[^{\Delta_3 \; \epsilon\beta_2 }_{\Delta_4 \; \;\beta_1 } \right](q)\right|^2
\Bigg\}
\\[4pt]
f_{\scriptscriptstyle 4[3][2]1}(\tau, \bar \tau)
&=&
 \int \mathrm{d}P |16q|^{P^2} \sum_{\epsilon_1,\epsilon_2=\pm}
C^{(\epsilon_1)}_{\scriptscriptstyle 4[3][P]} C^{(\epsilon_2)}_{\scriptscriptstyle [-P][2]1}
    \left|\mathcal{H}^{e}_{\beta_P}
    \left[_{\hspace{5pt} \Delta_4 \; \hspace{10pt} \Delta_1}^{\epsilon_1\beta_3 \; -\epsilon_2\beta_2} \right](q)
    \right|^2
\end{eqnarray*}
The crossing symmetry of the 4-point functions  (\ref{NNRRcs1}), (\ref{NNRRcs2}) yields:
\begin{eqnarray*}
f_{\scriptscriptstyle 43[2][1]}(\tau, \bar \tau) &=& f_{\scriptscriptstyle 34[2][1]}(\tau+1, \bar \tau+1) \\
f_{\scriptscriptstyle 43[2][1]}(\tau, \bar \tau)
&=& (\tau \bar \tau)^{\frac{c - 3/2}{4} - 2 \sum_{i}\Delta_i + \frac14}
f_{\scriptscriptstyle 4[1][2]3}(-{1 \over \tau}, -{1 \over \bar \tau})
\end{eqnarray*}
The first condition can be checked analytically using the relations:
\begin{eqnarray*}
\mathcal{H}^{e}_{\Delta_P}
    \left[^{\Delta_3 \; \epsilon\beta_2 }_{\Delta_4 \; \;\beta_1 } \right]\!(-q)&=&
    \mathcal{H}^{e}_{\Delta_P}
    \left[^{\Delta_4 \; \epsilon\beta_2 }_{\Delta_3 \; \;\beta_1 } \right]\!(q) \\
\mathcal{H}^{o}_{\Delta_P}
    \left[^{\Delta_3 \; \epsilon\beta_2 }_{\Delta_4 \; \;\beta_1 } \right]\!(\mathrm{e}^{i\pi}q)&=&
    \mathrm{e}^{i\pi \over 2} \mathcal{H}^{o}_{\Delta_P}
    \left[^{\Delta_4 \; \epsilon\beta_2 }_{\Delta_3 \; \;\beta_1 } \right]\!(q) ,
\end{eqnarray*}
which follow from recursions (\ref{H_NNRRe_rek}),(\ref{H_NNRRo_rek})
 and the properties of fusion polynomials (\ref{fuzPol_21_R}),(\ref{fuzPol_21_NS}).
 The second one can be numerically verified. We will consider a case of two different R fields and two
 equal non-zero NS weights:
$$
\beta_1= - {i p_1 \over \sqrt{2} }, \qquad
\beta_2= - {i p_2 \over \sqrt{2} }, \qquad \alpha_3= {Q \over 2} +i p_3, \qquad
 \alpha_4= {Q \over 2} -i p_3
$$
The structure constants with R intermediate states factorize onto the P-independent part:
\begin{eqnarray*}
 C^{(\epsilon_1)}_{\scriptscriptstyle -3[1][P]} C^{(\epsilon_2)}_{\scriptscriptstyle [-P][2]3}
&=&
r_3^{\epsilon_1,\epsilon_2}(P) \,\times\, g(p_i),
\end{eqnarray*}
\begin{eqnarray}\label{g}
g(p_i) &=& \frac14 \left( { \pi \mu \over 2} \gamma\left({Qb \over 2}\right) b^{2-2b^2}\right)^{Q-ip_2-ip_1 \over b}
\\[4pt] \nonumber
&\times& \Upsilon_{\scriptscriptstyle 0} \Upsilon_{\scriptscriptstyle \rm R}\!\left(Q+2ip_1\right)
\Upsilon_{\scriptscriptstyle \rm R}\!\left(Q+2ip_2\right)
 \Upsilon_{\scriptscriptstyle \rm NR}\!\left(Q+2ip_3\right) \Upsilon_{\scriptscriptstyle \rm NR}\!\left(Q-2ip_3\right)
\end{eqnarray}
and the nontrivial integral
{\small
\begin{eqnarray*}
&& \hspace{-20pt} r_3^{\epsilon_1,\epsilon_2}(P) =
\exp \int_{0}^{\infty} {\mathrm{d}t \over t}
    \Bigg\{ \left( 1+ 4 p_3^2 + 2(p_1^2+p_2^2)-\frac{Q^2}{2}\right)\mathrm{e}^{-t}
    - {\cos\left( P t\right) \left( \cosh\left(\frac{t}{2b}\right)+ \cosh\left(\frac{b t}{2} \right) \right) +6
    \over \sinh\left(\frac{t}{2b}\right)\sinh\left(\frac{b t}{2}\right) }
\\
&& \hspace{-10pt} +
  \cos\left( {p_3 t \over 2} \right)  \,
\left[
{ \cos\left(\frac{P t}{2}\right) \left(\cos\left(\frac{p_1 t}{2}\right)+ \cos\left(\frac{p_2 t}{2}\right) \right)
\over
\sinh\left(\frac{t}{4b}\right)\sinh\left(\frac{b t}{4}\right)
}
+ \epsilon_1 { \sin\left(\frac{P t}{2}\right) \left(
\sin\left(\frac{p_1 t}{2}\right) - (\epsilon_1 \epsilon_2) \sin\left(\frac{p_2 t}{2}\right) \right)
\over
\cosh\left(\frac{t}{4b}\right)\cosh\left(\frac{b t}{4}\right)
}
\right]
\Bigg\}
\end{eqnarray*}
}
The structure constants with NS intermediate states have the same P-independent part (\ref{g})
\begin{eqnarray*}
&& C_{\scriptscriptstyle -33P} C^{(\epsilon)}_{\scriptscriptstyle -P[2][1]}
= r_4^{\epsilon}(P) \,\times\, g(p_i),
 \qquad
-i\,\widetilde C_{\scriptscriptstyle -33P} C^{(\epsilon)}_{\scriptscriptstyle -P[2][1]}
 = \tilde r_4^{\epsilon}(P) \,\times\, g(p_i).
\end{eqnarray*}
The P-dependent parts read:
{\small
\begin{eqnarray*}
r_4^{\epsilon}(P)
    &=& P^2 \, \exp \int_{0}^{\infty} {\mathrm{d}t \over t}
    \left\{ \left( 4p_3^2 + 2(p_1^2+p_2^2)-\frac{Q^2}{2} \right)\mathrm{e}^{-t}
    + a_4^{+,\epsilon}\left(P,t\right)\right\}
\\
\tilde r_4^{\epsilon}(P)
    &=& P^2 \, \exp \int_{0}^{\infty} {\mathrm{d}t \over t}
    \left\{ \left( 1+ 4p_3^2 + 2(p_1^2+p_2^2)-\frac{Q^2}{2} \right)\mathrm{e}^{-t}
    + a_4^{-,\epsilon}\left(P,t\right)\right\}
\\
a_4^{\epsilon_1,\epsilon_2}\left(P,t\right)
&=&
 \cos(Pt)\left(1-\coth\left(\frac{t}{2b}\right)\coth\left(\frac{b t}{2}\right)\right)
-{
6+\cos(Pt) \over
\sinh\left(\frac{t}{2b}\right)\sinh\left(\frac{b t}{2}\right)}
\\
&+&\cos\left(\frac{P t}{2}\right) \,
\left[
{ \cos^2\left( {p_3 t \over 2} \right) + \cos\left(\frac{p_1 t}{2}\right)\cos\left(\frac{p_2 t}{2}\right)
\over
\sinh\left(\frac{t}{4b}\right)\sinh\left(\frac{b t}{4}\right)
}
+ \epsilon_1 { \cos^2\left( {p_3 t \over 2} \right) + \epsilon_2 \sin\left(\frac{p_1 t}{2}\right)\sin\left(\frac{p_2 t}{2}\right)
\over
\cosh\left(\frac{t}{4b}\right)\cosh\left(\frac{b t}{4}\right)
}
\right].
\end{eqnarray*}
}
The bootstrap equation in this case have the following form:
\begin{eqnarray}\label{bootsNNRR}
&& \hspace{-20pt}
\sum_{\epsilon=\pm}\int \mathrm{d}P \, |16q|^{P^2} \Bigg\{
r_4^{\epsilon}(P)
    \left|\mathcal{H}^{e}_{\Delta_P}
    \left[^{\Delta_3 \; \epsilon\beta_2 }_{\Delta_3 \; \;\beta_1 } \right](q)\right|^2
+
\tilde r_4^{\epsilon}(P)
    \left|\mathcal{H}^{o}_{\Delta_P}
    \left[^{\Delta_3 \; \epsilon\beta_2 }_{\Delta_3 \; \;\beta_1 } \right](q)\right|^2
\Bigg\}
\\ \nonumber
&& = (\tau \bar \tau)^{\frac{c - 3/2}{4} - 2(\Delta_1+\Delta_2+2\Delta_3) + \frac14} \,
\sum_{\epsilon_1,\epsilon_2=\pm} \int \mathrm{d}P \, |16q'|^{P^2} \,
r_3^{\epsilon_1,\epsilon_2}(P) \,
    \left|\mathcal{H}^{e}_{\beta_P}
    \left[_{\hspace{5pt} \Delta_3 \; \hspace{10pt} \Delta_3}^{\epsilon_1\beta_1 \; -\epsilon_2\beta_2} \right](q')
    \right|^2
\end{eqnarray}
We present a check of this relation for
 $ p_1= 0.2,\, p_2 = 0.4,\, p_3=0.3$, $c= 3$,
 and for  $\tau$  in the range $[0.2 i, 5i]$.
The relative difference of both sides of (\ref{bootsNNRR}) is plotted
on Figure \ref{plotNNRR}.
 The two curves correspond to the elliptic blocks with the NS intermediate states
  expanded up to $q^7,\,q^{\frac{13}{2}}$ and $q^8,\,q^{\frac{15}{2}}$, respectively.
  In both cases the elliptic blocks with the R intermediate states are expanded up to
   $q^8$.

\section*{Acknowledgements}

The author is grateful to L. Hadasz and Z. Jask\'{o}lski for valuable
discussions and the collaboration on \cite{Hadasz:2006sb, Hadasz:2007nt, R, Hadasz:2009db}.
I would also like to thank Z. Jask\'{o}lski for useful comments and careful reading of the manuscript.

\section*{Appendix}
In this Appendix we collect some useful properties of the 3-point blocks of different types.

\renewcommand{\theequation}{A.\arabic{equation}}
\setcounter{equation}{0}

\subsection*{3-point blocks with an R singular state}

Let us consider  the three point correlation functions with a degenerate field $R_{rs}^{\pm}$ within
the Feigin-Fuchs construction \cite{Bershadsky:1985dq}. In this approach
the NS superprimary and the R primary fields are represented by vertex operators in the free superscalar Hilbert space
\begin{eqnarray*}
\phi_{\lambda, \bar \lambda} (z,\bar z) &=&  {\rm e}^{a \phi(z)+\bar a \bar \phi(\bar z)} \\[4pt]
R^+_{\beta,\bar \beta} (z,\bar z)&=&  {\rm e}^{a \phi(z)+\bar a\bar \phi(\bar z)+i{\pi\over 4}}\sigma^+(z,\bar z)
\;\;\;,\;\;\;
R^-_{\beta,\bar \beta} (z,\bar z) \,=\,   {\rm e}^{a \phi(z)+\bar a\bar  \phi(\bar z)- i{\pi\over 4}}\sigma^-(z,\bar z)
\end{eqnarray*}
where $a = \frac{Q - \lambda}{2}$ or $ a = \frac{Q}{2} - \sqrt{2} \beta $ for the NS or  the R fields,  respectively.
 $\sigma^\pm$ are the twist operators of the fermionic sector:
$$
\psi(z)\sigma^\pm(z,\bar z) \sim {1\over \sqrt{2(z-w)}} \;\sigma^\mp(z,\bar z)\ .
$$
The left chiral screening charges are defined by:
\[
Q_b = \oint dz \, \psi(z) {\rm e}^{b \phi(z)}, \qquad
Q_{\frac{1}{b}} = \oint dz \, \psi(z) {\rm e}^{\frac{1}{b} \phi(z)},
\]
and analogously the ones in the right sector.
The 3-point functions with a degenerate R field and various number of left screening charges take the forms:
\begin{eqnarray*}
\begin{array}{llllll}
C^{\epsilon}_{[\beta_{rs},\delta], (\lambda_2,0),  [\beta_1,0]}
&=& \left\langle R_{rs}^{\epsilon} \phi_{\lambda_2} R_{\beta_1}^{\epsilon} \,
 Q^k_b \, Q^l_{\frac{1}{b}}  \right\rangle\ ,
&
k+l\in 2\mathbb{N}\ ,
& \textstyle \delta=-{1\over 2\sqrt{2}}({1\over b}+b)\; ;
\\
C^{\epsilon}_{[\beta_{rs},\delta], (\lambda_2,0),  [\beta_1,0]}
&=& \left\langle R_{rs}^{\epsilon} \phi_{\lambda_2} R_{\beta_1}^{\epsilon} \,
 Q^k_b \, Q^l_{\frac{1}{b}} \, \bar Q_b  \right\rangle \ ,
&k+l\in2\mathbb{N} +1 \ ,
& \textstyle \delta={1\over 2\sqrt{2}}({1\over b}-b).
 \end{array}
\end{eqnarray*}
The charge conservation implies  that the
 structure constants above are non-zero if and only if
the even fusion rules:
 \begin{equation}
\label{efusionRules} \beta_1 + \frac{1}{2 \sqrt{2}} \, \lambda_2 \; =
\; \frac{1}{2 \sqrt{2}} \,(1-r+2k)b + \frac{1}{2 \sqrt{2}} \, (1-s+2l)\frac{1}{b},
\qquad  k+l \in 2{\mathbb N} \cup \{0\}
\end{equation}
or the odd fusion rules:
\begin{equation}
\label{ofusionRules}
\beta_1 +  \frac{1}{2 \sqrt{2}} \, \lambda_2\; =
\; \frac{1}{2 \sqrt{2}} \,(1-r+2k)b + \frac{1}{2 \sqrt{2}} \, (1-s+2l)\frac{1}{b},
\qquad  k+l \in 2{\mathbb N} -1
\end{equation}
are satisfied ($k,l$ are integers in the range $ 0 \leq k \leq r-1,\ 0 \leq l \leq s-1 $).
Moreover, for any even integer $n\in 2\mathbb{N}$ one has \cite{R}:
\begin{eqnarray*}
\left\langle \psi(w_1) \ldots \psi(w_{n}) \sigma^-(1,1)  \sigma^-(0,0) \right\rangle
&=&-\left\langle \psi(w_1) \ldots \psi(w_{n})
) \sigma^+(1,1) \sigma^+(0,0)\right\rangle
 \\ \nonumber
 \left\langle \psi(w_1) \ldots \psi(w_{n-1}) \, \bar \psi(\bar w) \sigma^-(1,1)
 \sigma^-(0,0) \right\rangle
&=& \left\langle \psi(w_1) \ldots \psi(w_{n-1})\, \bar \psi(\bar w) \sigma^+(1,1) \sigma^+(0,0)\right\rangle
\end{eqnarray*}
 Thus if the even fusion rules (\ref{efusionRules}) are fulfilled,
 the structure constants are related
 $$
C^{+}_{(\alpha_{rs},\delta), (\beta_2,0), (\beta_1,0)} =
- C^{-}_{(\alpha_{rs},\delta), (\beta_2,0), (\beta_1,0)},
$$
what implies
$C^{(-)}_{(\alpha_{rs},\delta), (\beta_2,0), (\beta_1,0)}  \neq 0. $ Similarly, for the odd fusion
rules (\ref{ofusionRules}) the second constant does not vanish
\mbox{$C^{(+)}_{(\alpha_{rs},\delta), (\beta_2,0), (\beta_1,0)}  \neq 0 .$}

Consider now a 3-point function with the singular R field. Due to the Ward identities it can be written in terms of
the structure constants and the 3-point blocks (\ref{phi_RR}).
Since the correlator is identically equal to zero, the 3-point blocks have to vanish
\[ \rho_{\scriptscriptstyle \rm RRe}^{(-)}( \chi_{rs}^+, \nu_2, w_{1}^+)=0
 \]
 for even fusion rules and
  \[
\rho_{\scriptscriptstyle \rm RRe}^{(+)}( \chi_{rs}^+, \nu_2, w_{1}^+)=0
 \]
 for odd fusion rules.
An additional information on zeros of the 3-point blocks can be derived from the property of R fields
\cite{Poghosian:1996dw, R}:
$R_{-\beta}^\epsilon = \epsilon R_{\beta}^\epsilon
$
leading to formula
$$
C^{(\pm)}_{(\beta_{rs},\delta), (\lambda_2,0), (-\beta_1,0)}
=
C^{(\mp)}_{(\beta_{rs},\delta), (\lambda_2,0), (\beta_1,0)}.
$$
 The 3-point block $\rho_{\scriptscriptstyle \rm RRe}^{(+)}( \chi_{rs}^+, \nu_2, w_{1}^+)$
has to vanish for the even fusion rules (\ref{efusionRules})
and $\rho_{\scriptscriptstyle \rm RRe}^{(-)}( \chi_{rs}^+, \nu_2, w_{1}^+)$
 for the odd fusion rules (\ref{ofusionRules}) with the  opposite sign of $\beta_1$ in both cases.
This suggests the following definition of the fusion polynomials:
\begin{eqnarray}\label{fusionR}
P^{rs}_c\!\left[^{\pm\beta_1}_{\ \Delta_2} \right]
= \!
	\prod_{p=1-r}^{r-1} \prod_{q=1-s}^{s-1}\!\!\!
	\left({\lambda_2 \over 2 \sqrt2} \mp \beta_1 - \frac{ p b+ q b^{-1}}{2\sqrt2}\right) \!\!
	\prod_{p'=1-r}^{r-1} \prod_{q'=1-s}^{s-1}\!\!\!
	\left({\lambda_2 \over 2 \sqrt2}\pm \beta_1 - \frac{ p' b+q' b^{-1}}{2\sqrt2}\right)
\end{eqnarray}
where the products run over:
\begin{eqnarray}\label{p-q}
p = 1-r + 2k, \quad  q=1-s+2l, &\qquad& k+l \in 2 \mathbb{N} \cup \{0\} \\
p' = 1-r + 2k, \quad  q'=1-s+2l, &\qquad& k+l \in 2 \mathbb{N} +1 \nonumber
\end{eqnarray}
and $k,l$ are integers in the range $0 \leq k \leq r-1, \, 0\leq l \leq s-1$.

The 3-point blocks in terms of the fusion polynomials read:
\begin{equation}\label{rhoRR_P}
 \rho_{\scriptscriptstyle \rm RRe}^{(\pm)}( \chi_{rs}^+, \nu_2, w_{1}^+) =
 (-1)^{rs \over 2} \,P^{rs}_c\!\left[_{\ \Delta_2}^{\pm \beta_1} \right],
\end{equation}
where the proportionality coefficient is set by the singular vector normalization
 condition
 \footnote{The singular vectors are normalized such that $\chi^+_{rs} =  L_{-1}^{\frac{rs}{2}}w^+_{rs} + \ldots$}
 and the formula:
$$
\rho_{\scriptscriptstyle \rm RRe}^{(\pm)}( L_{-1}^{n}w^+, \nu_2, w_{1}^+) =
(\Delta + \Delta_2 - \Delta_1)_n, \qquad
(a)_n = {\Gamma(a+n) \over \Gamma(a)}
$$

A similar reasoning leads to the formulae for other types of 3-point blocks with R singular state:
\begin{eqnarray*}
\rho_{\scriptscriptstyle \rm RRe}^{(\pm)}(w_{4}^+ , \nu_3, \chi_{rs}^+) &=&
 (-1)^{rs \over 2} \,P^{rs}_c\!\left[_{\ \Delta_3}^{\pm \beta_4} \right],
\\
\rho_{\scriptscriptstyle \rm RNe}^{(\pm)}( \chi_{rs}^+,  w_{2}^+ ,\nu_1) &=&
  P^{rs}_c\!\left[_{\ \Delta_1}^{\pm \beta_2} \right],
\qquad \qquad
\rho_{\scriptscriptstyle \rm NRe}^{(\pm)}( \nu_4, w_{3}^+ ,\chi_{rs}^+) =
 P^{rs}_c\!\left[_{\ \Delta_4}^{\pm \beta_3} \right].
\end{eqnarray*}

\subsection*{Factorization over an NS singular vector}

 The fusion polynomials corresponding to 3-point blocks with the singular NS state were defined
in \cite{Hadasz:2006sb},\cite{R}:
 \begin{eqnarray}\label{fusionNS}
 \nonumber
 P^{rs}_c\!\left[^{\Delta_2}_{\Delta_1} \right]&=&
 \;
	\prod_{p=1-r}^{r-1} \prod_{q=1-s}^{s-1}
	\left(\frac{\lambda_1 + \lambda_2 -  p b- q b^{-1}}{2\sqrt2}\right)
	\left( \frac{\lambda_1 - \lambda_2 - p b- q b^{-1}}{2\sqrt2}\right)
\\[4pt]
 P^{rs}_c\!\left[^{\ast\Delta_2}_{\ \Delta_1} \right]&=&
  \;
 \prod_{p'=1-r}^{r-1} \prod_{q'=1-s}^{s-1}
 \left(\frac{\lambda_1 + \lambda_2 - p' b- q' b^{-1}}{2\sqrt2}\right)
  \left(\frac{\lambda_1 - \lambda_2 - p' b- q' b^{-1}}{2\sqrt2}\right)
  \\[4pt]
   \nonumber
P^{rs}_c\!\left[^{\pm\beta_2}_{\;\;\; \beta_1} \right]
&=&
	\;
	\prod_{p=1-r}^{r-1} \prod_{q=1-s}^{s-1}
	\left(\beta_1\mp \beta_2 - \frac{ p b+ q b^{-1}}{2\sqrt2}\right) \,
	\prod_{p'=1-r}^{r-1} \prod_{q'=1-s}^{s-1}
	\left(\beta_1\pm \beta_2 - \frac{ p' b+q' b^{-1}}{2\sqrt2}\right),
\end{eqnarray}
where the $(p,q)$ and $(p',q')$ correspond to the even and the odd fusion rules, respectively (\ref{p-q}).
The number of possible pairs in the products is given by
\begin{eqnarray*}
 \#(p,q)_{rs} &=&   \left\{
\begin{array}{rcl}
 && \hspace{-25pt} \left({rs+1 \over 2} \right),
 \\[4pt]
 && \hspace{-25pt}  \left({rs \over 2} \right),
 \end{array}
\right.
\qquad
\#(p',q')_{rs} \,=\, \left\{
\begin{array}{rcl}
 && \hspace{-25pt} \left({rs-1 \over 2} \right),
 \\[4pt]
 && \hspace{-25pt}  \left({rs \over 2} \right),
 \end{array}
\right.
\end{eqnarray*}
where $r,s\in 2\mathbb{N}+1$ for upper lines and
 $r,s\in 2\mathbb{N}$ for lower lines.
The fusion polynomials with interchanged weights are related to each other in the following way:
\begin{eqnarray}\label{fuzPol_21_R}
P^{rs}_c\!\left[^{\beta_2}_{ \beta_1} \right]
&=& (-1)^{ \#(p,q)_{rs}} \, P^{rs}_c\!\left[^{\beta_1}_{ \beta_2} \right], \qquad
P^{rs}_c\!\left[^{-\beta_2}_{\;\;\;  \beta_1} \right]
= (-1)^{\#(p',q')_{rs}} \, P^{rs}_c\!\left[^{-\beta_1}_{\;\;\;  \beta_2} \right],
\\
\label{fuzPol_21_NS}
 P^{rs}_c\!\left[^{\Delta_2}_{\Delta_1} \right]&=& (-1)^{ \#(p,q)_{rs}} \,
  P^{rs}_c\!\left[^{\Delta_1}_{\Delta_2} \right], \qquad
  P^{rs}_c\!\left[^{\ast\Delta_2}_{\ \Delta_1} \right] =
   (-1)^{\#(p',q')_{rs}} \,  P^{rs}_c\!\left[^{\ast\Delta_1}_{\ \Delta_2} \right].
\end{eqnarray}
The factorization of 3-point blocks over the NS singular state is given by the following formulae
\cite{Hadasz:2006sb},\cite{R}:
\begin{eqnarray}\label{3ptFactorNS}
\nonumber
\rho_{\scriptscriptstyle \rm NN,e}
  (\nu_4 , \nu_3, S_{-L}L_{-N}\chi_{rs} )
   &=&
   P^{rs}_c\!\left[^{\Delta_3}_{\Delta_4} \right]
\, \times \, \left\{
\begin{array}{rcl}
 && \hspace{-25pt}  \rho_{\scriptscriptstyle \rm NN,e}( \nu_4 , \nu_3,S_{-L}L_{-N}\nu_{\Delta_{rs}+\frac{rs}{2}} ),
 \\[4pt]
 && \hspace{-25pt} \rho_{\scriptscriptstyle \rm NN,o}( \nu_4 , \nu_3,S_{-L}L_{-N}\nu_{\Delta_{rs}+\frac{rs}{2}} ),
 \end{array}
\right.
\\[4pt]
\rho_{\scriptscriptstyle \rm NN,o}
   (\nu_4 , \nu_3, S_{-L}L_{-N}\chi_{rs} )
   &=&
  P^{rs}_c\!\left[^{\ast\Delta_3}_{\ \Delta_4} \right]
\, \times \, \left\{
\begin{array}{rcl}
 && \hspace{-25pt}  \rho_{\scriptscriptstyle \rm NN,o}( \nu_4 , \nu_3,S_{-L}L_{-N}\nu_{\Delta_{rs}+\frac{rs}{2}} ),
 \\[4pt]
 && \hspace{-25pt} \rho_{\scriptscriptstyle \rm NN,e}( \nu_4 , \nu_3,S_{-L}L_{-N}\nu_{\Delta_{rs}+\frac{rs}{2}} ),
 \end{array}
\right.
\\[6pt] \nonumber
\rho_{\scriptscriptstyle \rm NR, \rm e}^{(\pm)}( S_{-L}L_{-N}\chi_{rs},w_{2}^+, w_1^+)
&=& P^{rs}_{c}\!\left[^{\pm\beta_2}_{\;\;\; \beta_1} \right]  \times \left\{
\begin{array}{rcl}
 && \hspace{-25pt}
 \rho_{\scriptscriptstyle \rm NR,\rm e}^{(\pm)}(S_{-L}L_{-N}\nu_{\Delta_{rs}+\frac{rs}{2}},w_{2}^+, w_1^+),
 \\ [6pt]
&& \hspace{-25pt} {\rm e}^{ i \frac{\pi}{4}} \
\rho_{\scriptscriptstyle \rm NR \rm o}^{(\mp)}(S_{-L}L_{-N}\nu_{\Delta_{rs}+\frac{rs}{2}},w_{2}^+, w_1^+), \,
 \end{array}
\right.
\\[4pt] \nonumber
\rho_{\scriptscriptstyle \rm NR, \rm o}^{(\pm)}(S_{-L}L_{-N}\chi_{rs}, w_{2}^+, w_1^+ )
&=& P^{rs}_{c}\!\left[^{\pm\beta_2}_{\;\;\; \beta_1} \right] \times \left\{
\begin{array}{rcl}
 && \hspace{-25pt}
 \rho_{\scriptscriptstyle \rm NR,\rm o}^{(\pm)}(S_{-L}L_{-N}\nu_{\Delta_{rs}+\frac{rs}{2}},w_{2}^+, w_1^+),
 \\ [6pt]
&& \hspace{-25pt} {\rm e}^{- i \frac{\pi}{4}} \
\rho_{\scriptscriptstyle \rm NR \rm e}^{(\mp)}(S_{-L}L_{-N}\nu_{\Delta_{rs}+\frac{rs}{2}},w_{2}^+, w_1^+),
 \end{array}
\right.
\\[6pt] \nonumber
\rho_{\scriptscriptstyle \rm RN, \rm e}^{(\pm)}(w_{4}^+, w_3^+, S_{-L}L_{-N}\chi_{rs})
&=& P^{rs}_{c}\!\left[^{\pm\beta_3}_{\;\;\; \beta_4} \right]  \times \left\{
\begin{array}{rcl}
 && \hspace{-25pt}
 \rho_{\scriptscriptstyle \rm RN,\rm e}^{(\pm)}(w_{4}^+, w_3^+, S_{-L}L_{-N}\nu_{\Delta_{rs}+\frac{rs}{2}}),
 \\ [6pt]
&& \hspace{-25pt} -{\rm e}^{ - i \frac{\pi}{4}} \
\rho_{\scriptscriptstyle \rm RN \rm o}^{(\mp)}(w_{4}^+, w_3^+, S_{-L}L_{-N}\nu_{\Delta_{rs}+\frac{rs}{2}}), \,
 \end{array}
\right.
\\[4pt] \nonumber
\rho_{\scriptscriptstyle \rm RN, \rm o}^{(\pm)}(w_{4}^+, w_3^+, S_{-L}L_{-N}\chi_{rs})
&=& P^{rs}_{c}\!\left[^{\pm\beta_3}_{\;\;\; \beta_4} \right] \times \left\{
\begin{array}{rcl}
 && \hspace{-25pt}
 \rho_{\scriptscriptstyle \rm RN,\rm o}^{(\pm)}(w_{4}^+, w_3^+, S_{-L}L_{-N}\nu_{\Delta_{rs}+\frac{rs}{2}}),
 \\ [6pt]
&& \hspace{-25pt} - {\rm e}^{ i \frac{\pi}{4}} \
\rho_{\scriptscriptstyle \rm RN \rm e}^{(\mp)}(w_{4}^+, w_3^+, S_{-L}L_{-N}\nu_{\Delta_{rs}+\frac{rs}{2}}),
 \end{array}
 \right.
\end{eqnarray}
where the upper and lower lines correspond to ${rs\over 2} \in \mathbb{N}$ and
 ${rs\over 2} \in \mathbb{N} + \frac12$, respectively.

\subsection*{Factorization over an R singular vector}

There are two independent singular vectors $\chi^{\pm}_{rs}$ on the same level in
 the R module $\mathcal{W}_{\Delta_{rs}}$. Let us introduce the operator $D_{rs}$ generating
 the  even singular vector:
 $$
 \chi_{rs}^+ = D_{rs} \, w^+_{rs}
 $$
Then the odd singular vector is defined as
 $\chi_{rs}^{-} = {{\rm e}^{i{\pi\over 4}}\over i \beta^{\prime}}S_0 \chi_{rs}^+ $ with $\beta^{\prime}$
 given by (\ref{betaPrime}). There is however a second odd singular vector
$
 D_{rs} S_0 w^+_{rs}
$
on the same level. It follows from the general properties of R singular vectors discussed in \cite{Dorrzapf:1999nr}
 that these two odd vectors are proportional to each other:
$$
S_0 D_{rs}  w^+_{rs} = (-1)^s \, {rb^2 - s \over r b^2+ s} \, D_{rs} \, S_0 \, w^+_{rs}
$$
The proportionality coefficient can be easily recognized as the quotient
 $\beta^{\prime}_{rs}/\beta_{rs}$, what leads to the useful relation
\begin{equation}\label{chi-}
\chi_{rs}^{-} =  D_{rs} \, w^-_{rs}.
\end{equation}
With the help of this identity
one can compute the factorization of 3-point blocks  over an R singular state.
The  Ward identities for the 3-form $\varrho_{\scriptscriptstyle \rm RR,e}$
imply:
\begin{eqnarray*}
&& \varrho_{\scriptscriptstyle \rm RR}(S_{-L}L_{-N}\chi^+_{rs} ,  \nu_2 , w^+_1) \,=\,
\rho^{++}_{\scriptscriptstyle \rm RR}(S_{-L}L_{-N}w^{+}_{\Delta_{rs}+\frac{rs}{2}} ,  \nu_2 , w^+_1)
\varrho_{\scriptscriptstyle \rm RR}(\chi^+_{rs} ,  \nu_2 , w^+_1)
\\
&& \hspace{4cm} \ \ \,+\, \rho^{--}_{\scriptscriptstyle \rm RR}(S_{-L}L_{-N}w^{+}_{\Delta_{rs}+\frac{rs}{2}} ,  \nu_2 , w^+_1)
\varrho_{\scriptscriptstyle \rm RR}(\chi^-_{rs} ,  \nu_2 , w^-_1)
\\
&& \varrho_{\scriptscriptstyle \rm RR}(\chi^+_{rs} ,  \nu_2 , w^+_1)
 \,=\, \rho^{++}_{\scriptscriptstyle \rm RR}(\chi^+, \nu, w^+) \varrho_{\scriptscriptstyle \rm RR}(w^{+}_{rs}, \nu, w^+)
+\rho^{--}_{\scriptscriptstyle \rm RR}(\chi^+, \nu, w^+) \varrho_{\scriptscriptstyle \rm RR}(w^{-}_{rs}, \nu, w^-)
\\
&& \varrho_{\scriptscriptstyle \rm RR}(\chi^-_{rs} ,  \nu_2 , w^-_1)
 \,=\,\rho^{++}_{\scriptscriptstyle \rm RR}(\chi^-, \nu, w^-) \varrho_{\scriptscriptstyle \rm RR}(w^{+}_{rs}, \nu, w^+)
+ \rho^{--}_{\scriptscriptstyle \rm RR}(\chi^-, \nu, w^-) \varrho_{\scriptscriptstyle \rm RR}(w^{-}_{rs}, \nu, w^-)
\end{eqnarray*}
Using the identity (\ref{chi-})
and the properties of the 3-point blocks $\rho^{\imath\jmath}_{\scriptscriptstyle \rm RR,e}$ (\ref{e-e,o-o:p})
one gets the factorization formula:
\begin{eqnarray*}
\rho^{(\pm)}_{\scriptscriptstyle \rm RR,e}(S_{-L}L_{-N}\chi^+_{rs} ,  \nu_2 , w^+_1 )
&=& \rho^{(\pm)}_{\scriptscriptstyle \rm RR,e}(S_{-L}L_{-N}w^{+}_{\Delta_{rs}+\frac{rs}{2}} ,  \nu_2 , w^+_1)
\, \rho^{(\pm)}_{\scriptscriptstyle \rm RR,e}(\chi^+_{rs} ,  \nu_2 , w^+_1 ).
\end{eqnarray*}
Expressing the 3-point blocks with the singular state in terms of the fusion polynomials  (\ref{rhoRR_P}) we obtain:
\begin{eqnarray*}
\rho^{(\pm)}_{\scriptscriptstyle \rm RR,e}(S_{-L}L_{-N}\chi^+_{rs} ,  \nu_2 , w^+_1 )
&=&
(-1)^{rs \over 2} \,P^{rs}_c\!\left[_{\ \Delta_2}^{\pm \beta_1} \right]
 \rho^{(\pm)}_{\scriptscriptstyle \rm RR,e}(S_{-L}L_{-N}w^{+}_{\Delta_{rs}+\frac{rs}{2}} ,  \nu_2 , w^+_1)
\end{eqnarray*}
and similarly:
\begin{eqnarray}
\label{faktorRRchi+}
\nonumber
\rho^{(\pm)}_{\scriptscriptstyle \rm RR,e}( w^+_4,  \nu_3 , S_{-L}L_{-N}\chi^+_{rs})
&=& (-1)^{rs \over 2} \,P^{rs}_c\!\left[_{\ \Delta_3}^{\pm \beta_4} \right]
\  \rho^{(\pm)}_{\scriptscriptstyle \rm RR,e}(w^+_4 ,  \nu_3 , S_{-L}L_{-N}w^+_{\Delta_{rs}+\frac{rs}{2}}  )
\\
\rho^{(\pm)}_{\scriptscriptstyle \rm NR,e}( \nu_4,  w^+_3 , S_{-L}L_{-N}\chi^+_{rs} )
&=&  P^{rs}_c\!\left[^{\pm \beta_3}_{\ \Delta_4} \right]
\  \rho^{(\pm)}_{\scriptscriptstyle \rm NR,e} (\nu_4, w^+_3 , S_{-L}L_{-N}w^+_{\Delta_{rs}+\frac{rs}{2}})
\\ \nonumber
\rho^{(\pm)}_{\scriptscriptstyle \rm RN,e}( S_{-L}L_{-N}\chi^+_{rs},w^+_2, \nu_1 )
&=& P^{rs}_c\!\left[_{\ \Delta_1}^{\pm \beta_2} \right]
\  \rho^{(\pm)}_{\scriptscriptstyle \rm RN,e} ( S_{-L}L_{-N}w^+_{\Delta_{rs}+\frac{rs}{2}},w^+_2, \nu_1).
\end{eqnarray}

\end{document}